# EFFICIENT AND SECURE ROUTING PROTOCOL FOR WIRELESS SENSOR NETWORK

**A THESIS**

*Submitted by*

**S. GANESH**
**[Reg.No. 2009192101]**

*in partial fulfillment for the award of the degree*

*of*

**DOCTOR OF PHILOSOPHY**

**FACULTY OF ELECTRONICS ENGINEERING**
**SATHYABAMA UNIVERSITY**
**JEPPIAAR NAGAR, CHENNAI-119, INDIA**

**AUGUST2014**

Covering letter

From

Dr.R.Amutha,

Professor,Deparftment of ECE,

SSN College of Engineering, SSN Nagar,

Kalavakkam-603110

To

Dr.B.Sheela Rani,

Vice Chancellor,

Sathyabama University, Chennai –600110.

Respected Sir/Madam

Sub: Submission of thesis for Viva Voce- Reg

Ref: Research Scholar Mr.S.Ganesh, Reg.N0: 2009192101.

The thesis copies of the above mentioned research scholar have been attached, after incorporating the necessary corrections suggested by Indian/Foreign Examiner/Foreign Examiner Nominee.

Thanking you

Yours

Dr.R.Amutha

**SATHYABAMA UNIVERSITY**
Jeppiaar Nagar, Chennai

Proceedings of the Ph.D Viva Voce Examination of Mr./~~Ms.~~ S. GANESH (Reg. No. 200919210l) held at 11 am/pm on (date) 11/06/2016 in Sathyabama University

The Ph.D Viva Voce Examination of Mr./~~Ms.~~ S. GANESH on his/her Ph.D Thesis entitled "Efficient and secure Routing protocol for wireles sensor Networks"

was conducted in Sathyabama University, Chennai – 119 at 11 am/pm on(date) 11/6/2016

The Following members of the Oral Examination Board were present:

1. Dr. Mrinal kanti Naskar Indian Examiner
2. Dr. V. Vaideki Expert member
3. Dr. R. Amutha Supervisor & Convener

The Research Scholar Mr./Ms. S. GANESH

Presented the salient features of his/her Ph.D work. This was followed by questions from the board members. The questions raised by the Indian Examiner and nominee of the foreign examiner were also put to the research scholar. The research scholar answered / ~~did not answer~~ the questions to the full satisfaction of the board members.

**Based on the research scholar's work, his/her presentation and also the clarification and answers by the research scholar to the questions, the board recommends/~~does not recommend~~ that Mr./~~Ms.~~ S. GANESH be awarded the Ph.D degree in the faculty of** Electronics Engineering.

| | | |
|---|---|---|
| [signature] 11.06.16. Dr. Mrinal kanti Naskar<br>Name & signature of the Indian Examiner | [signature] 11/6/16 Dr. V. Vaideki<br>Name & signature of the Expert Member | [signature] 11.6.16 Dr. R. Amutha<br>Name & signature of the Supervisor and Convener |

# BONAFIDE CERTIFICATE

Certified that this thesis titled "**EFFICIENT AND SECURE ROUTING PROTOCOL (ESRP) FOR WIRELESS SENSOR NETWORK**" is the bonafide work of **Mr. S.GANESH [Reg. No.2009192101]** who carried out the research under my supervision. Certified further, that to the best of my knowledge the work reported herein does not form part of any other thesis or dissertation on the basis of which a degree or award was conferred on an earlier occasion of thesis of any other candidate.

SUPERVISOR

Signature :

Name : **Dr.R.AMUTHA**

Academic Designation : Professor

Department : Electronics and Communication Engineering

College with Address : SSN College of Engineering<br>Rajiv Gandhi Salai, OMR Road,<br>SSN Nagar, Kalavakkam-603 110<br>Tamil Nadu, India.

# ABSTRACT

Advances in Wireless Sensor Network (WSN) have provided the availability of small and low-cost sensors with the capability of sensing various types of physical and environmental conditions, data processing, and wireless communication. Since WSN protocols are application specific, the focus has been given to the routing protocols that might differ depending on the application and network architecture. In this work, novel routing protocols have been proposed which is a cluster-based security protocol is named as Efficient and Secure Routing Protocol (ESRP) for WSN. The goal of ESRP is to provide an energy efficient routing solution with dynamic security features for clustered WSN.

During the network formation, a node which is connected to a Personal Computer (PC) has been selected as a sink node. Once the sensor nodes were deployed, the sink node logically segregates the other nodes in a cluster structure and subsequently creates a WSN. This centralized cluster formation method is used to reduce the node level processing burden and avoid multiple communications.

In order to ensure reliable data delivery, various security features have been incorporated in the proposed protocol such as Modified Zero-Knowledge Protocol (MZKP), Promiscuous hearing method, Trapping of adversaries and Mine detection. One of the unique features of this ESRP is that it can dynamically decide about the selection of these security methods, based on the residual energy of nodes.

Simulation of ESRP is carried out along with routing protocols, namely Lightweight Dependable Trust System (LDTS) for clustered WSN, Low Energy Adaptive Clustering Hierarchy (LEACH) based protocol known as Specification-based intrusion detection mechanism for LEACH (SLEACH), Security and Energy Efficient Disjoint Route (SEDR) for WSN, Secure and Energy-Efficient Routing Protocol (SERP) for WSN, to find out the evaluation performance such as energy consumption and network lifetime. From the performance metrics, it is consummated that ESRP is more secure and energy efficient than the other routing protocols.

# ACKNOWLEDGEMENT

I would like to express heartfelt thanks to **Col.Dr.JEPPIAAR M.A.,B.L., Ph.D.,** Founder and Chancellor, Sathyabama University, Chennai, for providing me the great academic infrastructure to carry out my research work.

I wish to express my sincere thanks to the Directors **Dr.MARIE JOHNSON,B.E, M.B.A, MPhil., Ph.D.,** and **Dr.MARIAZEENA JOHNSON**, **B.E, M.B.A, MPhil., Ph.D**., of Sathyabama University, Chennai, for their support to carry out my research work.

I express my deep gratitude to the Secretary and Correspondent **Dr.P.CHINNADURAI, M.A., Ph.D,** Panimalar Institute of Technology, for his kind words and motivation, which inspired me a lot in completing this thesis. I sincerely thank **Mr.C.SAKTHIKUMAR, M.E.,** Director, Panimalar Institute of Technology for providing the necessary facilities for completion of my research work.

I wish to express my deepest gratitude to the Vice Chancellor and Dean **Dr. B. SHEELA RANI, M.S., Ph.D**., of Sathyabama University, for giving me academic guidance to carry out my research work.

It is indeed a great privilege to express my deep sense of gratitude to my respected supervisor **Dr.R.AMUTHA, M.E., Ph.D.,** Professor, Department of ECE,SSN College of Engineering, who suggested this work. She has been a perennial fountainhead of kindness, an inspiring friend, a philosopher and guide to make this work a reality.

My sincere thanks to my Doctoral Committee members **Dr.C.GOMATHY, M.E., Ph.D.,** and **Dr.K.S.SHAJI, M.E., Ph.D.,** for their support, guidance and helpful suggestions. Their guidance has served me well and I owe them my heartfelt appreciation.

I also express my gratefulness to the Principal **Dr.T.JAYANTHY**, **M.E., Ph.D.,** of Panimalar Institute of Technology who helped me in the completion of my research work.

I wish to express my sincere thanks to the head of the ECE department **Dr. M.P.CHITRA**, **M.E., Ph.D.,** of Panimalar Institute of Technology, for the constant support and encouragement. I thank all the review committee members for reviewing my research work at regular intervals.

I express my heartfelt thanks and gratitude to my parents, and family members for their blessing, wishes and support.

**S. GANESH**

# TABLE OF CONTENTS

**CHAPTER NO. TITLE PAGE NO.**

| CHAPTER NO. | TITLE | PAGE NO. |
| --- | --- | --- |

# LIST OF TABLES

# LIST OF FIGURES

| FIGURE NO. | TITLE | PAGE NO. |
| --- | --- | --- |

| FIGURE NO. | TITLE | PAGE NO. |
|---|---|---|

| FIGURE NO. | TITLE | PAGE NO. |
|---|---|---|

# LIST OF SYMBOLS AND ABBREVATIONS

| | |
|---|---|
| AODV | Ad-hoc On-demand Distance Vector |
| CBR | Constant Bit Rate |
| DoS | Denial of Service |
| DSDV | Destination Sequenced Distance Vector |
| DSR | Dynamic Source Routing |
| ESRP | Efficient and Secure Routing Protocol |
| HSN | Heterogeneous Sensor Network |
| LEACH | Low-Energy Adaptive Clustering Hierarchy |
| MAC | Medium Access Control |
| PEGASIS | Power-Efficient GAthering in Sensor Information Systems |
| TDMA | Time Division Multiple Access |
| TCL | Tool Command Language |
| US | Umpiring System |
| WSN | Wireless Sensor Network |

# CHAPTER 1

# INTRODUCTION

## 1.1 APPLICATIONS OF WSN

The outstanding progress in natural philosophy, technology sealed the trail for the expansion of micro-electronics, thus facilitating the manufacture of tiny chips and small devices. The communication technology is undergoing transformation because of the planning and improvement of small devices and thus expedited the planning and advancement of WSN with low price and low energy consumption.

WSN has many applications in military, health and in different industrial sectors. Due to the characteristics of WSN, sensor nodes are typically attributed with restricted power, low information measure, low memory size and restricted energy. Due to the measurability and energy effectiveness options, investigators prompt many routing protocols for cluster-based WSN. Routing could be a method of determinative a path between supply and destination upon request of information transmission. In WSN, the network layer is employed to implement the routing of the incoming information.

It is illustrious that in multi-hop networks the sensing node cannot reach the sink directly. So, intermediate detector nodes have to be compelled to relay their packets. The implementation of routing tables offers the answer. These contain the lists of node possibility for any given packet's destination.

Numerous improved hierarchal routing protocols were suggested in many research papers. But, some requirements for the routing protocols are conflicting. Always selecting the shortest route towards the base station causes the intermediate nodes to deplete faster, which result in a decreased network lifetime. At the same time, always choosing the shortest path might result in lower energy consumption and lower network delay. Since the routing objectives are tailored by the application, different routing mechanisms have been proposed for different applications. These routing mechanisms primarily differ in terms of routing objectives and routing techniques.

The majority routing protocols are vulnerable to uncounted security risks. Attacks comprising Cluster Head (CH) are in the main harmful. Because of resource restrictions, the general public key based algorithms like Rivest Shamir Adelman (RSA) and Diffie-Hellman are terribly complicated and energy-consuming for WSN.

In several cases, multiple sensing element nodes are needed to beat environmental obstacles like obstructions and line of sight constraints. Also, the setting to be monitored doesn't have an associate degree of existing infrastructure for energy economical communication. Therefore, it becomes imperative for sensing element nodes to survive on tiny, finite sources of energy and communicate through a wireless communication. Security could be a crucial issue because of inherent limitations in WSN.

## 1.2 WSN MODEL

Unlike their ancestor ad hoc networks, WSN are resource limited, they are deployed densely, they are prone to failures, the number of nodes are several orders higher than that of ad hoc networks, their network topology is constantly changing, they use broadcast communication mediums and finally wireless sensor nodes does not have global identification tags.

Figure 1.1 shows the major components of a typical WSN.

- Sensor Field*:* A sensor field can be considered as the area in which the nodes are placed.
- Sensor Nodes*:* Sensors nodes are the heart of the network. They are in charge of collecting data and routing this information back to the sink.

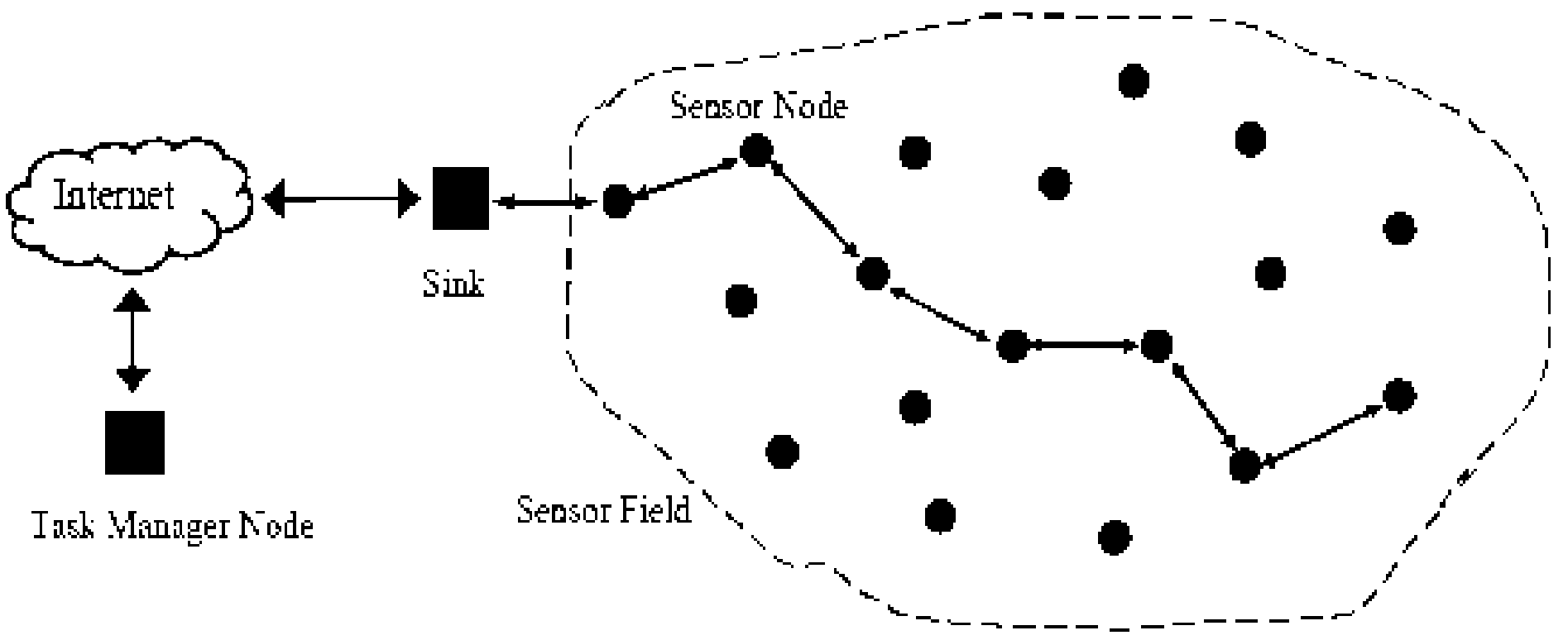

**Figure 1.1 Components of WSN**

- Sink*:* A sink is a sensor node with the specific task of receiving, processing and storing data from the other sensor nodes. They serve to reduce the total number of messages that need to be sent, hence reducing the overall energy requirements of the network.

- Task Manager*:* The task manager is the centralized point of control within the network, which extracts information from the network and disseminates control information back to the network. It also serves as a gateway to other networks, a powerful data processing and storage centre and also an access point for a human interface. This can be either a laptop or a workstation.

## 1.3 WSN PROTOCOL STACK

Nearly all communication protocols are based on the seven layers of the Open System Interconnection (OSI) model. Likewise the communication architecture of WSN can be classified into different layers. However, WSN does not adhere as closely to the layered architecture of the OSI model as other networks.

As shown in Figure 1.2, the protocol stack can be roughly broken into five major layers such as Physical, Data link, Network, Transport and Application layer. Nevertheless, the layered model is also useful in WSN for categorizing various protocols, attacks, and defenses.

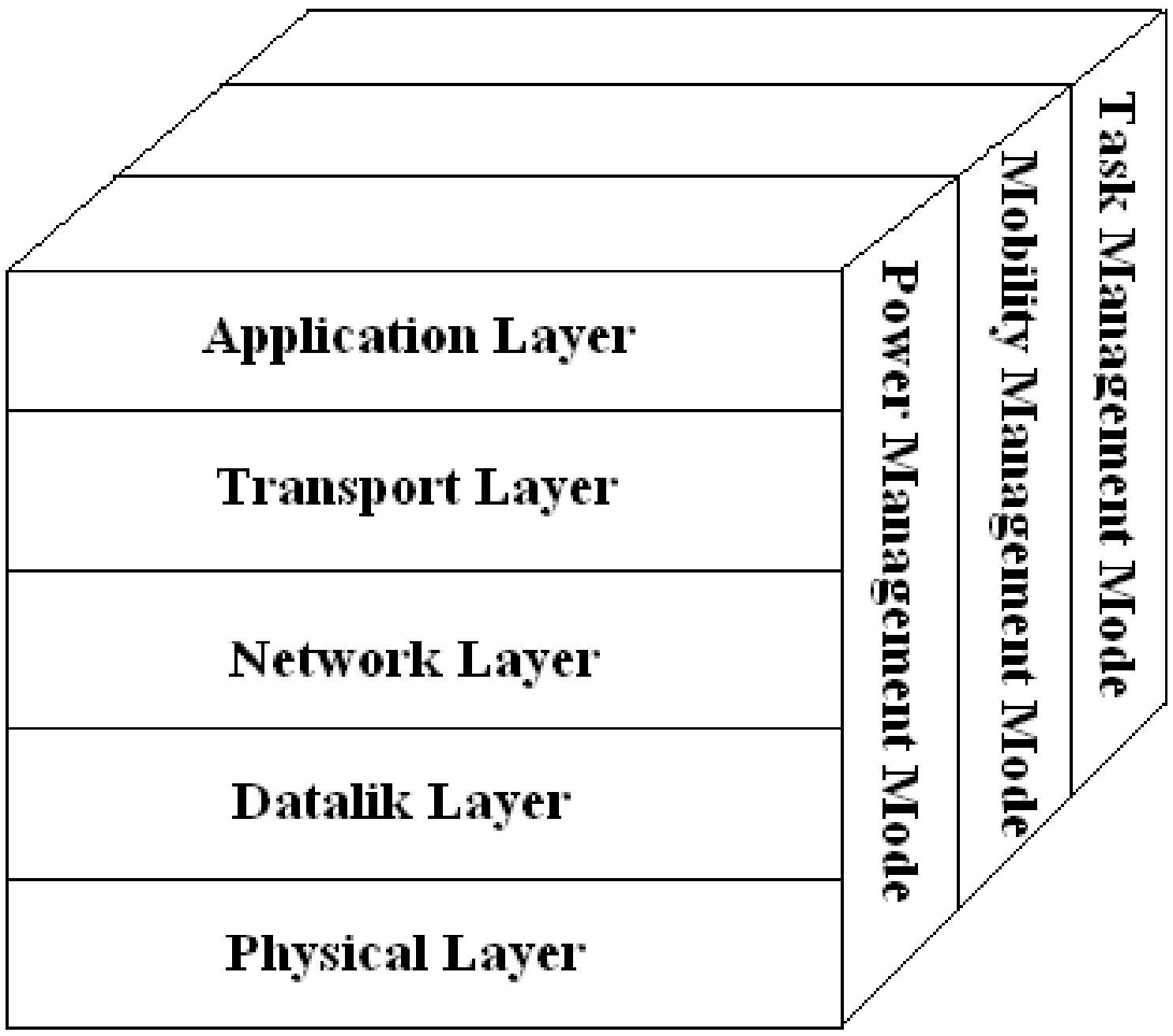

**Figure 1.2 WSN Protocol Stack**

Each layer provides services to the layer above and receives services from the layer below.

- The Physical layer should provide robust modulation, transmission, and receiving techniques.
- The Medium Access Control (MAC) convention at the Data connection layer must be power-mindful and ready to minimize impacts, subsequent to the remote channel is defenseless to clamor and sensor hubs may be versatile.
- The Network layer is responsible for routing of data, particularly selecting paths to send data in the network.

- Transport layer helps to maintain the data flow. This layer is especially needed when the system is planned to be accessed through Internet or other external networks. Unlike protocols such as Transmission Control Protocol (TCP) and User Datagram Protocol (UDP), the multi hop communication scheme of WSN is not based on global addressing. Therefore, new schemes that can split the end-to-end communication, probably at the sinks may be needed.

- Application layer defines a standard set of services and interface primitives. An example is the so called Sensor Network Services Platform (SNSP).

Orthogonal to these five layers defines Power, Mobility, and Task Management planes which are responsible for monitoring the power, movement, and task distribution among sensor nodes. These planes help sensor nodes to coordinate sensor tasks and reduces the overall power consumption.

## 1.4 WSN IMPLEMENTATION FRAME WORK

The literature and information on WSN applications, solutions, scientific and technological development are abundant. Internet and Mica (The platform is named Mica due to its resemblance with the final electronic implementation of silicate relative) mote based WSN implementation were briefed here.

### 1.4.1 Present Methodologies

Internet based WSN is a good substrate for WSN implementation. The altered TCP for WSN is good with a few arrangements accessible at the physical, data link, network, transport, and application layers of the OSI model, which rearranges WSN configuration and operation. The topologies and administration of the system are adaptable. Applications with diverse prerequisites can be obliged.

Mica motes are commonly planned in stackable layers. The center of a mote is a little, minimal effort, low-control PC. The PC screens, one or more sensors and unites with the outside world with a radio connection. Most of the time, they remained in a standby mode for power saving purpose. Several times in each second, the device flicks on its radio to check for incoming messages, but if there are none, the radio is shut off within milliseconds. Similarly, the sensors usually take their readings only once every few minutes. Data is transmitted only when the memory is full.

Motes Operating System (MOS) forces mote programs to shut down except when certain events that warrant action occur. The operating system is also highly modular. If a program needs only certain functions from MOS, the nonessential parts of the operating system are automatically removed from the mote.

### 1.4.2 Future Requirements

Applications requiring additional measurements and additional activity points are in demand. Necessary applications could need massive or extraordinarily massive mounted fixed or mobile networks, for which, WSN are the natural answer. However, many challenges at completely different levels have to be faced, specifically hardware, software, sensors and node sizes, power they will harvest or store, harsh environmental in operating conditions, node failures, quality of nodes, dynamic topology, communication failures, heterogeneousness of nodes, massive scale of deployment, unattended operation, schedule, period of time maximization, robustness, fault tolerance, self-configuration, and security.

**Power:** The current technologies already permit extended operation using small size, low Ampere-hour batteries. This is achieved with both low power consumption components as well as by keeping the nodes in stand-by as long as possible. However, some applications will surely require battery recharge. Sun, Wind vibration and Bio energy are some of the possible solutions already under study.

**Network programming and re-programming**: Programming of a wireless network with many nodes is not easy. New solutions have to be found. To replace the software on motes with updated versions, an idea based on the way viruses and worms are spread on the Internet, have been tried.

The new program is packaged into a special form and delivered to the root mote, which installs it and infects its neighbors with the package. The upgrade makes its way through the network like an epidemic, but it does so in a more controlled fashion that avoids redundant communications and adapts to the way that the motes are scattered in space.

**Node failure**: A WSN is unlikely to crash outright, but as some nodes die and others generate noisy or corrupted data, the measurements of the overall system may become biased or inconsistent. Work has been done on techniques to judge the health of a WSN by perturbing the system in a controlled way and observing how the sensor nodes respond.

**Privacy:** Over the next decade or so, wireless sensor nodes will probably evolve into a much less distinct and less visible form. Devices will gradually migrate out of their little boxes and will instead be incorporated directly into various materials and objects. Many of them will draw energy directly from the environment in which they operate.

To the extent that these kinds of WSN infiltrate homes, workplaces, farms, transportation terminals and shopping sites and are able to sense the presence, motion and even physiological states of individuals, they will raise substantial privacy concerns. Privacy issues are quite straightforward for many valuable applications, but in other domains, a careful balance must be struck to ensure that the technology properly empowers the individual.

The present technologies limit a WSN in hardware related aspects like sensed quantities, the size of nodes, and nodes autonomy, but it is at the software level where improvements and new solutions are required. All these developments must produce very low-cost devices so that the cost of a WSN with a large number of nodes is not prohibitive.

## 1.5 MOTIVATION

The long run goal of this analysis is to support high speed, energy economical knowledge delivery to the service-oriented specification in an urban setting that provides distinctive services to the user. There are several potentialities which might create the urban setting actually awake to the wants of the user. This level of responsiveness and interactivity between the user and, therefore, the urban setting take the client expertise to a replacement level.

Some of the applications facilitated by such a network layer protocol could be advertisements and offers based on the proximity to a shop, display of user-selected content on public displays, asking a user to switch off the phone in a silent zone (e.g. conference hall), enabling a user to change the Air Conditioner (AC) temperature or switch it off in a certain area and providing security messages in case the user goes in a restricted area. These applications and much more can be implemented at the application layer with the help of this work.

## 1.6 OBJECTIVE

The purpose of this work is to develop a protocol which should support energy efficiency and real-time traffic for environments like habitat monitoring or area surveillance.

This work focuses towards finding out suitable routing protocols that would be useful for various sensing applications with reduced energy consumption. Important goals of the proposed protocol were listed underneath.

- To consume minimum energy at node level.
- To reduce node level processing load.
- To reduced end to end delay.
- To identify and isolate malicious nodes.
- To deliver the data in a reliable manner.
- To improve the life time of a WSN.

## 1.7 STATE OF ART

So as to perceive the objective and prospective application of WSN, subtle and exceptionally competent communication protocols are necessary. WSN are application specific, therefore, routing protocol options are being differed from one application to another.

Nevertheless, routing protocols of all WSN, irrespective of the application, should attempt to capitalize on the network life span and reduce the energy utilization of the entire network. For these causes, the energy utilization parameter has elevated precedence than other features. At the network level, it is greatly advantageous to locate techniques for energy efficient route location and transmitting information from the sensor nodes to the base stations, in order to maximize the lifespan of the work.

Directing in WSN is exceptionally requesting inferable from the inbuilt elements which separate these systems from different remote systems, for example, cell, and portable specially appointed systems. Amid the larger part of use circumstances, nodes in WSNs are generally altered after arrangement with the exception of, maybe, a little number of versatile nodes. Inferable from such varieties, a few calculations, for example, LEACH, Power Efficient Gathering in Sensor Information Systems (PEGASIS) and Virtual Grid Architecture (VGA) have been proposed for the directing inconveniences in WSN.

In this work, the functioning of ESRP for WSN has been examined.

## 1.8 THESIS CONTRIBUTION

In this thesis, two techniques, namely ‘Centralized cluster formation’ and ‘Light weight security mechanism’ has been considered to provide energy efficient and reliable data delivery in a WSN. The distributed LEACH protocol has been customized to form clusters in a centralized manner.

The Zero Knowledge Protocol (ZKP), Triple Umpiring System (TUS) for WSN which offers security for routing and data forwarding functions by including certain features of Ad hoc On-demand Distance Vector (AODV) routing protocol and a few other methods, have been modified to implement the security mechanisms in the proposed protocol.

## 1.9 THESIS OUTLINE

This thesis is organized into five chapters. **Chapter 2** will offer comprehensive information connected to WSN literatures and also will deal with accessible research papers connected to routing in WSN.

**Chapter 3** describes the proposed frame work, namely ESRP for WSN. The selection of centralized cluster formation method has been justified by comparing it with its distributed counterparts. Further, it describes the light weight security methods and discusses their merits and influence on ESRP. Also, this chapter comprises particulars connected to the performance analysis with other similar protocols with the help of simulated results.

**Chapter 4** describes the hardware implementation of ESRP.

Lastly **Chapter 5** offers a summary of the whole thesis and scope for future researches and investigations in the same field.

# CHAPTER 2

# LITERATURE REVIEW

## 2.1 ROUTING IN WSN

One of the fundamental configuration objectives of WSN is to do information correspondence while attempting to draw out the lifetime of the system and forestall integration debasement by utilizing forceful vitality administration methods. The configuration of steering conventions is affected by numerous testing elements as given underneath:

### 2.1.1 Node Deployment

Node organization in WSN is application-particular and can be either manual (deterministic) or randomized. In manual sending, the sensors are physically put and information is directed through foreordained ways. On the other hand, in arbitrary arrangement, the nodes are scattered arbitrarily, making a specially appointed directing framework.

### 2.1.2 Energy Consumption

Sensor node lifetime depends enormously on battery lifetime. In the multi-bounce WSN, every node assumes a double part as information sender and information switch. The breaking down of some sensor nodes emerging out of force disappointment can bring about critical topological changes and may oblige re-directing of parcels and revamping of the system.

### 2.1.3 Fault Tolerance

Some sensor nodes may come up short or be obstructed because of absence of force, physical harm or natural impedance. The disappointment of sensor nodes ought not to influence the general functionalities of the sensor system. The steering convention needs to focus the other conceivable way to course the information to the sink node.

### 2.1.4 Scalability

The quantity of sensor nodes conveyed in the detecting zone may be in the request of hundreds or thousands, or considerably more. Any directing plan must be scaled up to handle steering, among the immense number of sensor nodes. By and large nodes in the sensor system are in rest mode and at whatever point an occasion is detected, the nodes are changed over to dynamic state.

### 2.1.5 Coverage

In WSN, every sensor nodes gets a certain perspective of the earth. A given sensor's perspective of the earth is constrained both in extent and exactness. It can just cover a constrained physical range of the earth. Subsequently, range scope is additionally a vital configuration parameter in WSN.

In a WSN, the network data between the nodes are traded when sending them. However the learning about node's area is needed in numerous sensor system applications. This area data empowers early expectation of the marvel, along these lines, minimizing the impact of unsafe fiasco. In directing, the area data of nodes aided in simple discovery of steering way between the source and the destination which thusly minimizes the inertness included in information transmission.

## 2.2 CLASSIFICATION OF WSN ROUTING PROTOCOLS

Routing protocols are often classified as Proactive, Reactive and Hybrid looking on the sort of communication routes processed at intervals the network for information transmission from the supply to sink.

In **Proactive routing protocols** all the routes are calculated before the sink makes associate initiation to speak with the nodes within the network, wherever as in **Reactive routing protocols**, the trail values square measure calculated only if needed. Whenever a sink needs to contact a specific node, the path values were calculated and therefore the best path is chosen for information transmission.

**Hybrid routing protocols,** as the name suggests, is a combination of both proactive and reactive routing protocols, which decides when to calculate the path from the sink to the source depending on the type of communication. Generally, it has been suggested that proactive routing protocols are better for static nodes. The reason is that a lot of energy can be saved compared to reactive routing protocols which depend on the discovery of the best route path for data transmission. In proactive routing it is not necessary to search for the nearest neighbors for every next hop when data is transmitted.As shown in Figure 2.1, the routing protocols in WSN can be coarsely divided into the following two categories:

- Based on Network Structure, which can be sub divided as Flat, Hierarchical and Location Based protocols.

- Based on Protocol Operation, which can be sub divided as Negotiation based, Multi-Path based, Query based, QoS based and Coherent based routing protocols.

Directing is one of the basic assignments in WSN. Restricted to conventional specially appointed systems, steering in WSN is additionally difficult as an after effect of its inalienable qualities. Firstly, assets are extraordinarily obliged as far as power supply, preparing ability and transmission data transfer capacity. Also, it is hard to plan a worldwide tending to plan like an Internet Protocol (IP).

Besides, IP can't be connected to WSN, since location redesigning in an expansive scale or element WSN can bring about overwhelming overhead. Thirdly, because of the constrained assets, it is hard for a directing convention to adapt to eccentric and regular topology changes, particularly in a versatile domain. Fourthly, information accumulation by numerous sensor nodes more often than not brings about a high likelihood of information excess. Fifthly, most uses of WSN require the normal correspondence plan of numerous to-one, i.e., from various sources to one specific sink, as opposed to multicast or shared.

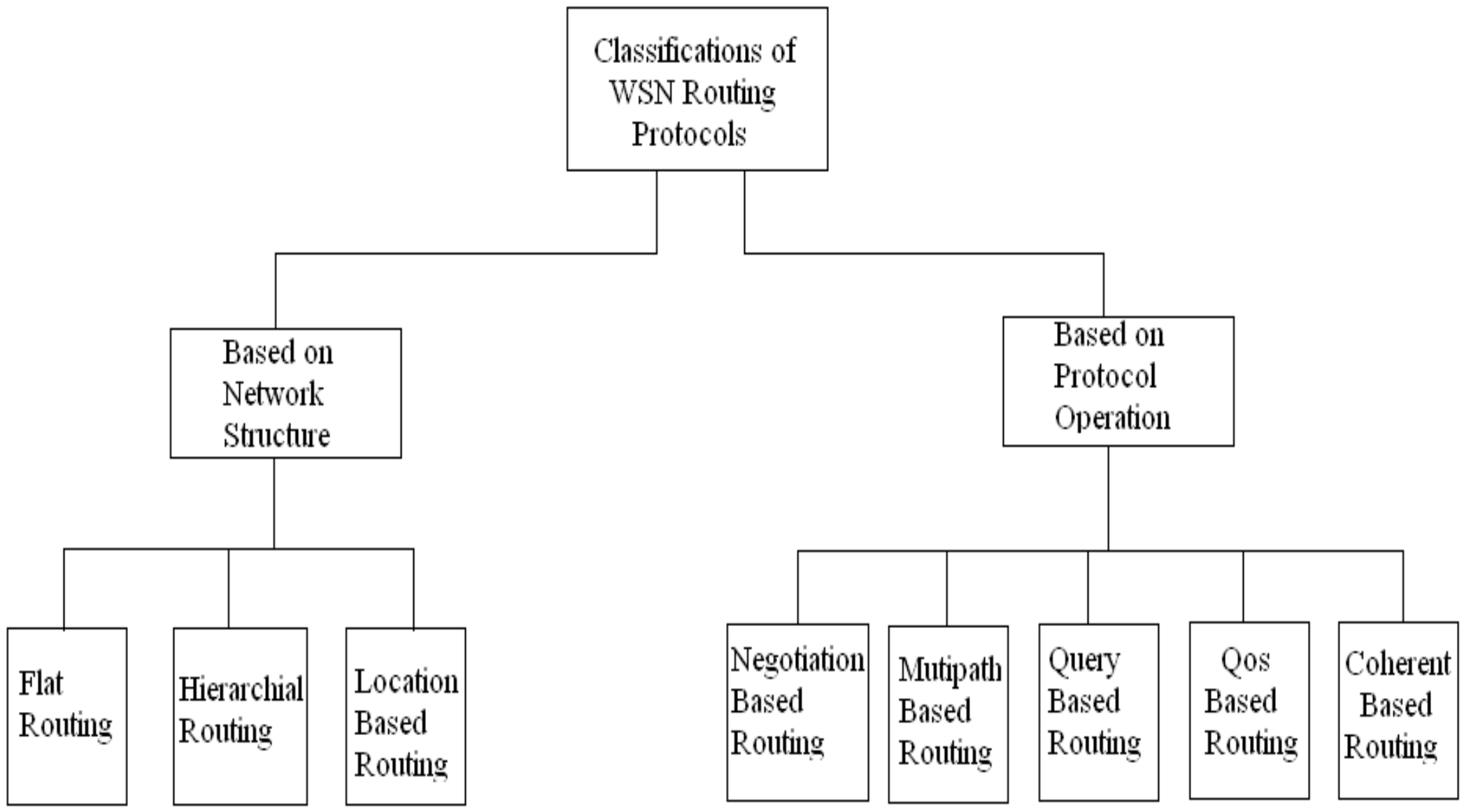

**Figure 2.1 A Taxonomy of WSN Routing Protocols**

Subsequently, limited inertness for information transmissions must be thought seriously about in this sort of uses. By the by, vitality protection is more essential than QoS in many applications, in that all sensor nodes are obliged with vitality, which is specifically identified with the system lifetime.

### 2.2.1 Classification based on Network Structure

In a level topology, all nodes perform the same errands and have the same functionalities in the system. Information transmission is performed jump by bounce normally utilizing the type of flooding.

#### *2.2.1.1 Flat routing protocols*

The common level steering convention in WSN incorporates Flooding and Gossiping, Directed Diffusion (DD), Greedy Perimeter Stateless Routing (GPSR), Trajectory Based Forwarding (TBF), Energy-Aware Routing (EAR), Gradient-Based Routing (GBR), Sequential Assignment Routing (SAR) and so on.

In little scale deals with level organizing customs are all around persuading. In any case, it is generally undesirable in huge scale structures due to constrained assets.

#### *2.2.1.2 Hierarchical routing protocols*

In a various levelled topology, nodes perform diverse undertakings in WSN and commonly are sorted out into heaps of groups as per particular necessities or measurements.

In Hierarchical conventions, nodes are assembled into groups. A node is chosen as CH. CH performs conglomeration of information transmitted by its group part utilizing standard data combination system. The totaled data is then transmitted to the base station in this manner diminishing vitality utilization. A few progressive conventions are accessible in the literary works. LEACH protocol was the first group based directing convention for WSN. It utilizes a stochastic model for group head determination. LEACH has persuaded the outline of a few different conventions which attempt to enhance the CH determination process.

Other typical cluster routing protocols in WSNs include LEACH, Hybrid Energy-Efficient Distributed (HEED) clustering protocol, Distributed Weight-based Energy-efficient Hierarchical Clustering (DWEHC) protocol, Position-based Aggregator Node Election (PANEL) protocol, Two-Level Hierarchy LEACH (TLH-LEACH), Unequal Clustering Size (UCS) model, Energy Efficient Clustering Scheme (EECS),Energy-Efficient Uneven Clustering (EEUC) algorithm, Base-station Controlled Dynamic Clustering Protocol (BCDCP), PEGASIS, Threshold sensitive Energy Efficient sensor Network (TEEN) protocol, Two-Tier Data Dissemination (TTDD) protocol, Concentric Clustering Scheme (CCS),Hierarchical Geographic Multicast Routing (HGMR) etc.

#### *2.2.1.3 Location based routing protocols*

In Location-based conventions, position data of nodes is used to transfer the information to the wanted destination. Power headway can be expert in Location-based directing conventions as stand out jump neighbors data are, obliged to course the information to the destination. By this the control overheads used can be minimized.

By and large area data are required to ascertain the separation between two specific nodes, so that vitality utilization can be evaluated. For the most part, two procedures are utilized to discover the area in which the first strategy is to locate the direction of the neighboring node and the second method is to utilize Global Positioning System (GPS).Since, there is no addressing scheme for WSN like IP-addresses and they are spatially deployed on a region, location information can be utilized in routing data in an energy efficient way.

### 2.2.2 Classification based on Protocol Operation

There are five different classifications as follows

#### *2.2.2.1 Negotiation based routing protocols*

These protocols use high-level descriptors so as to eliminate the redundant data transmissions. Flooding is used to disseminate data, due to the fact that flooding data are overlapped. But collisions occur during transmissions and also nodes receive duplicate copies of data during transmission. The same data content is sent or exchanged again and again between the same set of nodes, and a lot of energy is utilized during this process. Negotiation protocols like SPIN are used to suppress duplicate information and prevent redundant data from being sent to the next neighboring node or towards the base station by performing several negotiation messages on the real data that has to be transmitted.

#### *2.2.2.2 Multi-path based routing protocols*

Multipath directing conventions take a shot at the rule that higher execution can be accomplished by recording more than one plausible way. At the point when numerous courses are known, regardless of the fact that the essential way fizzles information sending can proceed continuously on the other most accessible ways without sitting tight for another course to be found. These protocols are efficient in handling multiple paths. Nodes send the collected data on multiple paths rather than using a single path. The reliability and fault tolerance of the network increases as there is, as long as it is possible, an alternative path exist, when the primary path fails.

#### 2.2.2.3 *Query based routing protocols*

Query based directing spreads the utilization of questions issued by the base station. The base station sends questions asking for certain data from the hubs in the system. A node, which is in charge of detecting and gathering information, peruses these inquiries and if there is a match with the information asked for in the inquiry, then it begins sending the information to the hub or the base station. This process is known as Directed Diffusion where the base station sends interest messages on to the network. These interest messages, which move in the network, create a path while passing through all the sensor nodes. Any sensor node, which has the data suitable to the interest message, sends the collected data along with the interest message towards the base station. Thus, less energy is consumed and data aggregation is performed on a route.

#### 2.2.2.4 *QoS based routing protocols*

In this type of routing protocol, both quality and energy have to be maintained within the network. Whenever a sink requests for data from the sensed nodes in the network, the transmission has to satisfy certain QoS parameters, such as, for example, bounded latency and bandwidth consumed.

Successive Assignment Routing (SAR) is one of the first directing conventions that utilize the idea of QoS in steering choices. Directing choice in SAR relies on upon three elements: Energy Consumption inside of the system by the sink and the hubs, QoS of every way in the system, and the Priority level of every bundle sent.

#### *2.2.2.5 Coherent based routing protocols*

In a WSN, sensor nodes collect data and send it to the nearest neighbors or the sink within the network. In this process, the processing of the collected data is the most important event. There are two types of data-processing techniques followed within the network structure: Coherent and Non-Coherent. All nodes within the WSN collect the data and process it before sending to the next nearest node for further processing. This technique is called Non-Coherent data process routing and the nodes that perform further processing on the data are called aggregators. In Coherent steering, after least handling, the information is sent to the aggregators. This base preparing incorporates capacities, like time stamping or copy concealment. This method is vitality effective since all the preparing is finished by the hubs, which decreases the aggregate time and vitality utilization.

## 2.3 ON-DEMAND ROUTING PROTOCOLS

On-demand directing is more suitable for the remote sensor system because of its successive topologically evolving environment. As these systems are primarily worked by batteries, the directing convention should be vitality mindful. Power enhancement can be accomplished by performing steering with area data of the hubs.

A few conventions have been proposed in the writing to address the aforementioned issues. This segment examines a portion of the on-interest and area based steering conventions proposed in WSN. Many routing protocols developed for WSN were proposed in this literature. Then again, it is hard to study them independently. Other than that, few of these conventions have basic attributes and can be gathered into classes.

Young Bae Ko and Nitin Vaidya (2000) have talked about Location-Aided Routing (LAR) for ad-hoc systems which execute course revelation through constrained flooding which brought about lessening in control overhead when contrasted with AODV. However, this technique accepts area to be figured through GPS beneficiary which does not adjust well to indoor based applications.

As a rule, non-geographic steering conventions like DSR and AODV experiences an enormous measure of control overhead to set up and keep up course owning to the regularly changing system topology. In sensor systems where a large number of nodes correspond with one another, telecast tempests may bring about noteworthy force utilization. Then again, geographic directing conventions require just neighborhood data and subsequently are more suitable for remote sensor systems. As just the area data of a node and its immediate neighbors are obliged to forward the packets, memory and transfer speed use likewise get minimized in this methodology.

Carman et al. (2000) have identified the following confidentiality parameters in sensor networks:

- A sensor centre point should not reveal its data to the neighbors. A valid example, in unstable military applications, an adversary has implanted a couple of toxic centre points into the framework, protection will piece them from becoming acquainted with information as for distinctive centre points.

- Building up and keeping up classifieds is discriminating when centre point characters and keys are being flowed to set up a sheltered relating channel among sensor centers.

Eschenauer and Gligor (2000) presented an essential key administration plan to address the key dissemination, renouncement, re-keying and flexibility to sensor hub catch issues. Key dissemination comprises of three stages, to be specific, key pre-appropriation, shared-key disclosure and way key foundation. It involves five offline steps:

- Generation of an extensive pool of ‘P’ keys and their key identifiers.
- Random withdrawal of ‘k’ keys from ‘P’ to build up key ring of every hub.
- Loading of the key ring of every hub into its memory.
- Saving of the key identifiers of a key ring and related sensor identifier on a trusted controller hub.
- For every hub stacking the key shared in the middle of it and the $i^{th}$ controller node.

The routing protocol is present in the network layer. The function of the wired routing protocol is merely switching over data and discovering practical routes, but in the wireless routing, there are further functions to meet the wireless setting that comprises of minimal available energy as well as network resources.

Karp and Kung (2000) proposed Greedy Perimeter Stateless Routing (GPSR), an area based directing convention that works in two modes: insatiable mode and edge mode. In the covetous mode/avaricious sending, every hub advances the parcel to the neighbor nearest to the destination. At the point when covetous sending is unrealistic, the parcel changes to border mode. The voracious directing comes up short at whatever point a deadlock issue is experienced. In the border directing (face steering) the bundles are steered around deadlocks until hubs closer to the destination are found. The drawback of this procedure is reduced framework lifetime. An uneven dissemination of movement in the system results because of the choice of hubs that are closer to the destination. As the lingering vitality of the hub and the connection quality are not considered in steering, the bundle conveyance proportion is all that much diminished. This system takes after an occasional beaconing technique that expends superfluous vitality and data transfer capacity.

Elizabeth Royer and Charles Perkins (2000) proposed an effective AODV steering convention for adhoc system. AODV comprises of three stages, to be specific, course disclosure, course repair and course support. In the course disclosure stage, the end-to-end way foundation in the middle of source and sink is finished by flooding the course asks for bundles in the system. This procedure results in noteworthy increment in vitality utilization which constrains its selection for remote sensor system.

Hubaux et al. (2001) proposed a declaration archive plan. These storehouses store the general population authentications that the hub themselves issue, and a choice set of declarations issued by the others. The execution is characterized by the likelihood that any hub can get and check people in general key of whatever other client, utilizing just the nearby declaration storehouses of the two clients. The situation is that an excess of testaments in a sensor hub would effectively surpass their ability.

Yu et al (2001) devised Geographic and Energy Aware Routing (GEAR) convention that uses remaining vitality and geological data of nodes for neighbor choice through which the bundles are steered towards the destination district. This is a proactive based approach and results in high bundle overhead and is not actualized in a sensible situation.

Young Bae Ko and Nitin Vaidya (2002) have suggested another kind of geographic routing known as Geo-casting which combines multicast with geographical routing. In this approach, the packets are delivered to nodes within certain geographic area.

Lu et al. (2002) have grown Real-time Architecture and Protocol (RAP) in view of speed. A neighboring hub with a high bundle speed to achieve the destination is chosen as the sending hub.

An effective confirmed key foundation convention is proposed by Huang et al (2003) in a self-arranging sensor system. This is taking into account a mix of Elliptic Curve Cryptography (ECC) and symmetric-key operations. The half and half plan lessens the high cost open key operations on the sensor side and replaces them with effective symmetric-key based operations. The ECC has speedier reckoning time, littler keys, and uses less memory and data transfer capacity than the Reed Solomon Algorithm (RSA). Both ECC and RSA can be quickened with committed co-processors.

The package movement in the remote sensor framework is examined in the arrangement given by Zhao and Govindan (2003). The best association quality is considered as the coordinating metric. The imperativeness usage in the framework is lessened by minimizing the amount of retransmission.

Pietro et al. (2003) presented an anticipating model of two traditions, to be particular, the quick and supportive traditions remembering the final objective to set up a secured pair canny correspondence channel between any two sensors. With a particular finished objective to ensure the viability of the tradition and adaptability to the coalition of polluted sensors they give

- Key organization plan, which portrays how the sensors are stacked with the keys.

- Key revelation methodology to figure the arrangements of keys for a given pair of sensors.

- Security versatile channel foundation strategy to empower a discretionary pair of sensors to concede to a typical key to be utilized to secure a channel.

They reasoned that utilizing pseudo-arbitrary era of the key files of every key ring utilizing the sensor ID as the seed has points of interest and burdens. Be that as it may, thusly of determining key rings lessens computational and correspondence overheads contrasted and utilizing a mystery seed as portrayed in Eschenauer and Gligor (2000) plan.

SrdjanKrco (2003) addressed the problem related to the behavior of Wireless Local Area Network (WLAN) 802.11b network cards in ad hoc mode. An improvement in the neighbour detection algorithm was proposed based on the differentiation of good and bad neighbours using Signal to Noise Ratio (SNR).

Zhu et al. (2003) conceived a plan called Localized Encryption and Authentication Protocol (LEAP) that would take into consideration information combination, in-system handling and detached support. Other than offering fundamental prerequisites like classifieds and affirmation, LEAP supports distinctive correspondence examples, including unicast (tending to a solitary hub), nearby show (tending to a gathering of hubs in an area) and worldwide telecast (tending to every one of the hubs in a WSN). In some cases WSN is conveyed in an enemy's enclosure, where more often than not traded off hubs are undetected. Jump gives survivability such that trading off of a few hubs does not surrender the whole system.

He et al. (2003) have proposed a Stateless convention for constant correspondence called SPEED. The bound on end-to-end correspondence postponement is achieved by implementing a uniform correspondence speed in each bounce in the system.

Gang Zhou et al. (2004) discussed the problem of radio irregularity which aroused due to different path loss and variation in the transmitted power. The impact of radio irregularity on routing protocols have been analysed with the help of a simulation model known as a Radio Irregularity Model (RIM) which was an extension of isotropic radio models. The simulation was carried out using Global Mobile Simulator (GloMoSim) and it has been concluded that location-based routing protocols were affected more compared to on-demand based routing protocols because of this irregularity.

Ian D. Chakeres (2004) have imagined an outline for the execution of AODV steering convention. The configuration portrayed the occasion triggers needed for AODV operation, the outline conceivable outcomes and the choices for AODV directing convention usage.

WSN has for the most part been utilizing symmetric key and other non-open key encryption plans nitty- gritty by Gura et al. (2004). A downside of these plans is that they are not as adaptable as open key plans. In any case, they are computationally speedier. With restricted memory, figuring and correspondence limit, and power supply, sensor hubs can't utilize complex cryptographic advancements, for example, normal open key cryptographs. The use of public key cryptography on WSN has not been tested sufficiently to rule it out completely. Through the use of the MICA2 mote and TinyOS, public-key schemes are tested to determine their performance.

Du et al. (2004) make an augmentation to the key administration plan created by Eschenauer and Gligor (2000). Their plan gives sending information if sensors are pre-orchestrated in a grouping before organization. Once dropped from a plane or helicopter, the sensors have a superior possibility of perceiving their neighbors through the pre-masterminded succession. This former organization information is valuable for key pre-circulation.

At the point when neighbor sensors are known, key pre-appropriation gets to be inconsequential and essentially obliges that, for every hub n, a couple astute keys in the middle of n and each of its neighboring hubs is created, and that these keys are spared in n's memory. This ensures the foundation of a protected channel with each of its neighbors after sending.

The transmission range modification plan (Hwang and Kim 2004) proposes sensor hubs to build their transmission extents amid the common key disclosure stage. Hubs come back to their unique ideal transmission extend once the keys are found. The thought is to decrease the correspondence trouble in way key foundation stage, and to spare vitality while as yet giving a decent key network.

Jamal Al Karaki and Ahmed Kamal (2004) have devised the accompanying steering difficulties in WSN. To start with, because of the moderately substantial number of hubs, it is impractical to assemble a worldwide tending to plot for the conveyed hubs, as it expands the overhead. Consequently a conventional IP-based convention may not be connected to WSN. Second, the hubs are firmly obliged as far as vitality, preparing and stockpiling limits.

Third, WSN are application particular. Fourth, position attention to sensor hubs is vital since information accumulation is ordinarily in view of area. At last, information gathered by numerous sensors in WSN is regularly taking into account basic wonders, so there is a high likelihood of excess. The directing conventions need to adventure this excess issue to enhance vitality and transmission capacity usage.

Chan and Perrig (2005) portrayed a system for building up a key between two sensor hubs taking into account the normal trust of a third hub some place inside of the sensor system. The hubs and their mutual keys are spread over the system such that for any two hubs ‘A’ and ‘B’, there is a hub ‘C’ that imparts a key to both ‘A’ and ‘B’. In this way, the key foundation convention in the middle of ‘A’ and B can be safely steered through ‘C’.

Geo-casting is an important communication paradigm (Navas and Imielinski 1997, Sanchez et al. 2006) in wireless sensor networks, as in some applications the objective is to communicate data to a set of destination nodes that are within a certain region. The challenge involved in this kind of paradigm is to determine the common set of nodes that yield better cost over progress ratio to the destination nodes.

Chakrabarti et al. (2006) presented a randomized square mixing methodology for key pre-dispersion in WSN. Arjan Durresi et al. (2006)

proposed a key pre dissemination plan for heterogeneous systems which comprise of hubs that are stationary separated from being very versatile. They utilize a different disjoint key pool to build up connections between the stationary and portable hubs of the system in light of the fact that, if the same key pool is utilized as a part of numerous systems, the trade-off of keys in one system would prompt bargain of keys in every one of the systems. Xiaobing Hou et al (2006) clarified a Gossip-based Sleep Protocol (GSP) with decreased end-to-end delay and less vitality utilization.

Mohamed et al. (2006) proposed a lightweight combinatorial advancement of the key organization arrangement for grouped WSN, called Scalable, Hierarchical, Efficient, Location-careful and Light-weight (SHELL) tradition. In SHELL, interest is diminished by using the physical regions of the centre points in figuring their keys. This arrangement uses a summon centre to speak to the entire framework. The request centre particularly compares with the portal centre points which are in charge of the individual groups. Centers can be added to this framework at whatever point.

The passage hubs are sufficiently capable to correspond with the order hub and do the obliged key administration capacities. Every entryway hub can correspond with no less than two other passage hubs in the system and has three sorts of keys. The main key sort is a preloaded one that permits the door to specifically correspond with the order hub. The second sort permits the distinctive door hubs to convey. The third key sort permits the passage to speak with the majority of the sensor hubs in its group. A percentage of the obligations of the summon hub are to go about as a key store, confirm entryway and sensor hubs, appropriate keys to every single other hub and it performs key reestablishment when required.

Eltoweissy et al. (2006) proposed a dynamic key administration framework, called Exclusion-Based System (EBS). A portion of the benefits of utilizing a dynamic key administration plan are enhanced system survivability and better backing for system development. The issue in making a dynamic key administration framework is the capacity to make it secure and productive. The EBS relegates k keys to every hub from a key pool. In the event that a hub catch is identified, rekeying happens all through the system. A burden of this EBS plan is that, if even a little number of hubs in the system are bargained, data about the whole system could be revealed by an enemy.

The primary utilization of the EBS plan was made with unknown hubs in the system. The hubs did not have IDs rather; they were recognized by their areas. This plan is heterogeneous and relies on upon a focal base station for key dispersion. This EBS plan is extremely effective, however it doesn't avoid agreement among hubs that are traded off. The Localized Combinatorial Keying (LOCK) is a dynamic key administration plan taking into account the EBS plan. This remote sensor system model comprises of a three level pecking order, base station, group heads and sensor hubs.

## 2.4 SECURITY GOALS OF WSN ROUTING PROTOCOLS

Security goals encompass both the customary networks and a WSN. The four security goals for a WSN are determined as Confidentiality, Integrity, Authentication and Availability. Confidentiality is the ability to conceal messages from a passive attacker so that any message communicated via the sensor network remains confidential. The following confidentiality parameters have been identified in WSN:

- A sensor node ought not to uncover its information to the neighbors, for instance, in delicate military applications, a foe has infused a few pernicious hubs into the system, and classifieds will block them from obtaining entrance to data with respect to different nodes.
- Setting up and keeping up privacy is amazingly critical when hub characters and keys are being appropriated to build up a safe corresponding channel among sensor hubs.

Information respectability in WSN is expected to guarantee the dependability of the information and alludes to the capacity to affirm that a message has not been messed around with, adjusted or changed while on the system. Regardless of the fact that the system has classifieds measures set up, there is still a possibility that the uprightness of the information has been traded off by the changes.

The integrity of the network will be in question if

- A malicious node present in the network injects bogus data.
- Turbulent conditions due to wireless channel cause damage or loss of data.

Verification guarantees the dependability of the message by distinguishing its inception. Assaults in WSN don't simply include the change of bundles yet can likewise infuse extra counterfeit parcels. Information confirmation checks the character of senders and can be accomplished through symmetric or uneven systems, where sending and getting nodes offer mystery keys to process the message validation code.

Due to the remote media and the unattended way of WSN, guaranteeing verification turn into a testing employment. The vitality and computational constraints of remote sensor hubs make the organization of complex cryptographic strategies illogical.

Availability determines whether a node has the ability to use the resources and whether the network is available for the messages to communicate. In an intrusion, a correspondence join in sensor systems gets lost or gets to be occupied. Illustrations of this kind of danger are hub catch, message debasement and insertion of vindictive code and so on. A block attempt means trade off in the sensor system by an enemy whereby the aggressor increases unapproved access to the sensor hub or information put away inside of it. A case of this is a hub catch assault.

Change implies an unapproved gathering gets to the information as well as messes around with it, by altering the information parcels being transmitted or bringing on a refusal of administration assault, for example, flooding the system with counterfeit information. In creation, an enemy infuses false information and bargains the dependability of the data handed-off.

The major assets in computing systems are the hardware, software and the data in WSN. The goal is to protect the network itself. The four classes of security in computing systems such as Interruption, Interception, Modification and Fabrication are explained below.

In an Interruption, a correspondence interface in a WSN gets lost or gets the chance to be involved. Instances of this sort of risk are message corruption, insertion of malicious codes etc. An Interception means deal in the WSN by an adversary whereby the attacker builds unapproved access to the sensor centre or data set away within it. An instance of this is a centre point get attack.

Modification suggests an unapproved assembling gets to the data and in addition disturbs it, by changing the data bundles being transmitted or bringing on a Denial of Service (DoS) attack, for instance, flooding the framework with sham data. In Fabrication, an adversary implants false data and deals the reliability of the information gave off.

## 2.5 SECURITY FRAMEWORK

Attacks against routing protocols in WSN fall into one of the following categories:

The **Spoofed Routing Information attack** includes defilement of the inside control data, for example, the routing tables. The **Selective forwarding attack**, specifically forward the packets that cross a vindictive node. In this attack, pernicious nodes may decline to forward specific messages and basically drops them, guaranteeing that they are not proliferated any further. These attacks are commonly more viable when the aggressor is expressly included on the way of an information stream.

The **Wormhole attack** includes making of a "wormhole" that catches the data in one area and replays them in another area. If the replayed information is tampered, it is known as **Sinkhole attack.** Further the creation of false control packets during the formation of the initial routes of the network is called as **Hello Flood Attack**, and the creation of false acknowledgement information is called as **Acknowledgment Spoofing.**

### 2.5.1 Attacking the external flash

Some applications might want to store valuable data on the external Electrically Erasable Programmable Read Only Memory (EEPROM) .Probably the simplest form of attack is eavesdropping on the conductor wires connecting the external memory chip to the microcontroller. Using a suitable logic analyzer makes it easy for the attacker to read all data that are being transferred to and from the external EEPROM while she is listening. In the event that a strategy was found to make the microcontroller read the whole outer memory, the aggressor would realize all memory substance. This sort of assault could be going on unnoticed for broadened times of time, as it doesn't impact typical operation of the sensor node. A more modern attack would join a second microcontroller to the Input/yield pins of the flash chip. On the off chance that the aggressor is fortunate, the mote microcontroller will not get to the information transport while the assault is in advancement, and the assault will be totally unnoticed.

On the off chance that the aggressor is dexterous, she can disjoin the immediate association between the bit microcontroller and the glimmer chip, and afterward interface the two to her own microcontroller. The attacker could then recreate the outside memory to the mote, making everything seem unsuspicious. Obviously, as opposed to utilizing her own particular chip, the aggressor could basically do a "mass delete" of the mote's microcontroller and put her own system on it to peruse the outer memory substance. This operation is even conceivable without learning of the bootstrap loader secret key. While this reasons "decimation" of the node from the networks perspective, in numerous situations this may not make any difference to the aggressor.

A conceivable countermeasure could be checking the area of the outside flash all things considered between times, putting a condition of constraint at the time the attacker is permitted to independent the microcontroller from the outer flash memory. These strikes could be assembled into three classes relying on the exertion basic.

- The class containing the “easy” attacks: Attacks in this class are able to influence sensor readings, and may be able to control the radio function of the node, including the ability to read, modify, delete, and create radio messages without, however, having access to the program or the memory of the sensor node. These attacks are termed “easy” because they can be mounted quickly with standard and relatively cheap equipment.

- The category containing the "medium" attacks: Attacks during this category permissible to be told in any event a rate of the substance of the memory of the within, either the Random Access Memory (RAM) on the microcontroller, its internal impact memory, or the external flash memory. This might offer the aggressor, e.g., cryptographic keys of the within. These strikes termed "medium" on the grounds that they need a non-standard examination centre gear, but allow lining up this equipment some spot else, i.e., not within the sensor failed.

- The class containing the “medium” attacks: Attacks in this class allowed to learn at least some of the contents of the memory of the node, either the Random Access Memory (RAM) on the microcontroller, its internal flash memory, or the external flash memory. This may give the attacker, e.g., cryptography keys of the node. These attacks are termed “medium” because they require non-standard laboratory equipment but allow to prepare this equipment elsewhere, i.e., not in the sensor field.

- The class containing the “hard” attacks: Using attacks in this class the adversary complete read/write access to the microcontroller. This gives the attacker the ability to analyze the program, learn the secret key material, and change the program to his own needs. These attacks are termed “hard” because they require the adversary to deploy non-standard laboratory equipment in the field.

An imperativeness gainful component key organization arrangement in WSN proposed by Huanzhao Wang et al (2006) presents a dynamic key administration plan of bunched sensor systems. By utilizing a pseudo-irregular capacity and the elliptic bend computerized mark calculation in this plan, vitality utilization can be diminished fundamentally in key foundation and upkeep stages. Examination demonstrates that this plan has a low overhead as far as processing, correspondence, and capacity.

Traynor et al. (2006) considered the hubs in the systems which are more effective and more secure than others. A hub that has a restricted memory and handling force is recognized as L1 and a hub that has more memory and additionally preparing force is distinguished as L2. L2 hubs go about as head hubs for the L1 hubs and have the obligation of directing bundles all through the framework. These L2 centers have section to entrance to portal servers which are associated with a wired system. There are three entering and trust models in these heterogeneous systems. The principal is called backhaul, in which the L1 hubs just send information to L2 hubs and do this just on the off chance that they both straightforwardly share a key. The second plan is known as the shared with constrained trust, in which two L1 hubs wish to trade information and they just trust one another or another L2 hub. In the event that the two L1 hubs can build up a pair wise key, they speak with each other. If they are not ready to set up a key, they swing to the L2 hub for help with setting up one. The third plan is distributed with liberal trust, which works like the second plan, however rather than L1 hubs just believing the other L1, it needs to speak with and other L2 hubs, the L1 hubs believe all other L1 hubs and L2 hubs in the system.

Chipara et al. (2006) have devised a Real-time Power-Aware Routing (RPAR) convention. The parcel speed of each one-jump nodes needed to achieve the destination figured by speed task arrangement is considered as the steering metric. The information is sent to the hub that offers high bundle speed. On the other hand, all the three conventions (RAP, SPEED and RPAR) neglect to give high throughput in remote correspondence as they depend just on the speed.

A steering convention taking into account join quality is proposed by Mohammed F.Younis et al (2006). The Expected Transmission tally (ETX) is produced as a metric to choose sending node. It discovers all the conceivable ways with the base expected number of transmission (counting retransmissions) needed to convey a bundle to the destination. The way offering high throughput is chosen as the sending way. On the other hand, ETX does not consider the remaining power and end-to-end due date.

Jing Feng (2006) had proposed a Self-Repair Algorithm for Ad Hoc On-demand Distance Vector (SRAODV) routing in mobile Ad Hoc networks. The proposed algorithm used to detect the link breakage with help of intermediate nodes as well as the backward pre-hop node. SRAODV improved the throughput and decreased the packet delivery delays which were justified with the help of simulated results.

Ruixin Niu et al. (2006) had proposed a Likelihood Ratio (LR) based fusion rule which was based only on the knowledge of channel statistics. With the help of simulated results under different channel Signal to Noise Ratio (SNR), it has been proved that the proposed method outperformed both the Equal Gain Combiner (EGC) as well as Chair-Varshney fusion rules.

Duarte-Melo E. J. and Mingyan Liu (2007) considered key organization in a Heterogeneous Sensor Network (H sensors) that embodies a bit number of fit top notch sensors like PDAs and innumerable end sensors (L sensors). The makers showed a capable key organization arrangement considering the asymmetric key pre-movement for H sensors. The exceptional H-sensors were utilized to give essential, capable and practical key set up arrangements for the L-sensors.

An execution and security examination showed that this arrangement could out and out decrease the sensor stockpiling essentials while achieving favored security over the couple of current sensor framework key organization arranges. Yang Xiao et al (2007) clarified the pair shrewd key foundation plan which is a standout amongst the most productive key foundation plans in remote sensor systems in light of the fact that it offers numerous extra elements contrasted with different plans, including hub to-hub confirmation and strength to hub replication. For a system of n hubs in the pair astute plan, the key pre-appropriation is finished by doling out every hub a remarkable pair savvy key with the various hubs in the system, i.e., n - 1 sets shrewd keys, which are held in every hub's memory so that every hub can correspond with every one of the hubs in its correspondence range. With every hub imparting an interesting key to each other hub in the system, this plan offers hub to-hub confirmation.

Each centre can affirm the identity of the centre point it is talking with. This arrangement in like manner offers extended quality to network, get as an exchanged off centre point does not reveal information about diverse centers that are not direct comparing with the got centre point. Du et al (2007) proposed a steering driven key administration plan, which just sets up shared keys for neighbor sensors that correspond with one another. The proposed method is used to further expand the effectiveness of the key administration plan. The execution assessment and security examination demonstrate that the directing driven key administration plan furnishes better security with noteworthy decreases on correspondence overhead, storage room and vitality utilization than a percentage of the current sensor key administration plans.

Razia Haider et al. (2007) have presented Energy Aware Greedy Routing (EAGR) protocol in which the routing is done based on the residual energy of the nodes. It is assumed that the geographic location and energy levels of all nodes including the destination nodes are known. The distances between the neighboring nodes to the destination node are computed and the average of all these distances is computed. The neighboring node with maximum energy is selected as the forwarding neighbor node, provided its distance is less than or equal to the average distance. The packet overhead factor is increased in this approach as all information of the neighbor needs to be maintained.

Tamilarasi et al. (2007) has displayed a versatile timeout component for on-interest directing conventions in MANETS. The quantity of control bundles is diminished by uprooting the stale courses in MANETS, in this manner the packet delivery ratio is expanded.

Denis C. Daly (2007) had proposed an energy efficient beneficial On Off Keying (OOK) transceiver for short range WSN. The transceiver was plotting with an envelope identifier based building configuration with a significantly versatile Radio Frequency (RF) front end to achieve predominant execution. The transceiver finished a fast beneficiary start up time, which helped for successful operation in low commitment cycle, energy starved circumstances.

EdithC.H.Ngai (2007) had proposed a novel algorithm for detecting the intruder in the sinkhole attack. In this method, first a list of suspected nodes was created by consistent data checking and then the intruders were nabbed by analyzing the network flow information. The proposed algorithm exhibited robust performance and dealt with multiple malicious nodes with reasonably low communication and computational overhead.

Luiz H.A et al. (2007) had proposed Transmission Power Control (TPC) technique which could dynamically adjust the transmission power. In this technique, the minimum power level required for successful message transmission known as ‘ideal transmission power’ has been calculated using a closed control loop method. It has been concluded that the proposed technique used to improve the reliability of the link, enhanced the network utilization and reduced the probability of hidden and exposed terminals.

A novel Decentralized Predictive Congestion Control (DPCC) protocol for WSN had been introduced by Maciej Zawodniok (2007) based on hop by hop feedback information. In this approach, the embedded channel estimator algorithm along with an adaptive flow control and scheduling methods were utilized to detect congestion and channel quality. This protocol ensured congestion control and Quality Of Service (QOS) with the help of fair scheduling algorithm even in fading channels.

Kui Ren et al. (2007) proposed a various leveled key administration plan named Hierarchical Key Management Protocol for Heterogeneous sensor systems, which is taking into account an irregular key pre-circulation. It means to develop a protected tree rather than a complete joined diagram as in the current plans.

Bahrololum.M (2008) had proposed Hierarchical Gaussian Mixture Model (HGMM), a novel type of Gaussian Mixture which was used to detect network based attacks as anomalies using statistical pre-processing classification. The proposed model learnt patterns of normal and intrusive activities using Gaussian probability distribution functions.

Zhen Cao et al. (2008) had proposed a Feedback Based Secure Routing (FBSR) convention in which input from the neighbouring hubs served as the dynamic data of the present system, with which sensor hubs settled on sending choices in a safe and the vitality mindful way.

In light of a progressive system model and bivariate polynomial key era instrument, Yun Wang et al. (2008) proposed a key administration plan. This plan ensures that two imparting gatherings can build up an extraordinary pairwise key between them. It gives adequate security, settled key stockpiling overhead, full system network and low correspondence overhead.

Noureddine et al. (2008) tended to the way key foundation issue when two neighbouring sensor hubs with an unreliable connection oblige their trusted neighbours to make a protected connection. The outcomes and investigations identify with situations for a WSN, having high and low integration. Specifically, the technique could find intermediaries with less trouble than those calculations obliging shorter ways and indicating great execution for scanty remote sensor systems. Likewise, it has been indicated, through reproductions, that the proposed calculation is more effective as far as the dynamic time to transmit the essential data to discover one intermediary and five intermediaries. The time to focus an intermediary is specifically identified with force utilization. A more extended time important to choose an intermediary will bring about an additional energy to be expended. In this way the proposed calculation is likewise control productive.

Najet Boughanmi and Ye Qiong Song (2008) portrayed two new directing measurements to examine the execution of AODV convention in real time application. Limited correspondence postponement and lingering vitality of the nodes are considered as parameters for directing to empower burden adjusting in the system

Bhalaji (2009) had proposed a new routing mechanism known as association based routing, in order to combat the common selective packet dropping attack. Associations between nodes were used to identify and isolate the malicious nodes from the active data forwarding and routing. The proposed method had been applied over the Dynamic Source Routing (DSR) protocol in order to enhance the security.

Chunguo Li et al. (2009) had proposed a joint Power Allocation (PA) scheme for a class of Multiple Input Multiple Output (MIMO) relay systems. In this method, two joint PA optimization problems were formulated based either on maximization of capacity or minimization of Mean Square Error (MSE).Finally a convex optimization problem has been formulated using the derived lower and upper bounds for the capacity and MSE respectively.

A direct rule encryption part of introducing systems security, in light of the Execute Only Memory (XOM) model known as the Latency Aware Simple Encryption for Embedded structures security (LASE) has been proposed by Kerry Courtright (2009).The proposed algorithm has been implemented using Field Programmable Gate Array (FPGA) and showed significant improvement on performance and power usage.

Oludele Awodele et al. (2009), had presented an Intelligent Intrusion Detection and Prevention System (IIDPS), which could have monitored a single host system from the following three different layers namely files analyser, system resource and connection layers. This system dealt with a multi – layered approach and made use of signature and anomaly detection approaches in order to achieve a better detection and prevention capabilities.

Saleem et al. (2009) had presented the model of self-optimized multipath routing algorithm for WSN based on ant colony optimization, which showed improved results during route discovery in terms of throughput and delay.

Since complex security measures entail a higher consumption of energy and computation power, keeping resource-starved sensor networks available is a challenging job. However, failure of the base station or CH will eventually threaten the entire WSN. Thus, availability is of primary importance for maintaining an operational network.

A two-hop neighborhood information-based routing protocol have been proposed for real-time WSN by Yanjun Li et al. (2009), where the steering choice was made taking into account the novel two-bounce speed coordinated with vitality adjusting instrument. The proposed plan has prompted lower bundle due date miss proportion and higher vitality proficiency.

Abedelaziz Mohaisen et al. (2009) have recognized four classes of security in processing frameworks. The real resources in registering frameworks are the equipment, the product and the information in sensor systems. The objective is to secure the system itself. The four classes of security are interference, capture attempt, alteration and creation which are clarified underneath.

Kiran Rachuri et al. (2009) have presented Increasing Ray Search (IRS), an energy efficient query resolution technique. The basic principle of IRS is to route the search packet along a set of trajectories called rays that maximize the likelihood of discovering the target information by consuming the least amount of energy.

Jinsu Kim (2010) had proposed a group based vitality productive key administration convention to battle with malignant node assaults by setting up a typical shared key through the key of key ring or through validation by a dependable establishment. Higher energy efficiency could be achieved by the proposed protocol when nodes were in mobile.

A force mindful, versatile, various levelled and chain based convention known as Clustered Chain based Power Aware Routing (CCPAR) had been proposed by Koushik Majumder (2010), that used the intermittent assignments of the Cluster Head (CH) part to distinctive nodes in view of the most astounding leftover battery limit for guaranteeing the even scattering of force by all nodes.

Rajagopal Kannan et al. (2010) had considered an obliged hugeness change called Minimum Energy Scheduling Problem (MESP) for WSN. A few assessment plans have been centred on for discovering the ideal transmission organize by discretizing force levels over the snag channel. A two-variable assessment game plan has been made for discovering the ideal changed transmission power estimation of every inside which could happen in the base essentials timetable and along these lines broadened the lifetime of WSN.

Mohammad et al (2010) had demonstrated a paper in which, the model of self-upgraded multi way managing estimation for WSN and its outcomes were exhibited. Certain parameters like vitality level, suspension and speed were measured. These estimations were utilized to locate the ideal and framed course for WSN.

Nike Gui (2010) had proposed a secure organizing and Aggregation Protocol with Low Energy (STAPLE) cost for WSN which used constrained hash chain and multi-way system to perform security for WSN, besides added to a structure creating model to control correspondence expense achieved by multi-way coordinating.

With a specific end goal to shield the system layer from noxious assaults, an answer of Umpiring System (US) that gives security to steering and information sending operations has been proposed by Ayyaswamy Kathirvel (2010). In the proposed framework, every node in the way from source to destination had double parts to perform: parcel sending and umpiring. US didn't matter any cryptographic procedures on the directing and bundle sending messages.

Amid the umpiring part, every node in the way firmly checked the conduct of its succeeding hub and if any misconduct has been seen then the umpiring hubs were instantly hailed off the liable hub. The proposed system has been actualized by adjusting Adhoc On interest Distance Vector (AODV) steering convention.

Ishwar Ramani (2010) had proposed programming based agreeable booking structure known as Covenant, which was given a rich arrangement of components for applications that obliged hubs to participate with one another to fulfil framework wide goals. The constant execution has been demonstrated in home passage situation which outlined the naturalness of Covenant practically speaking.

Vasanthi et al. (2010) made an execution examination of three sensor system directing conventions, to be specific, Rumour routing, Stream Enable Routing (SER) and Sensor Protocols for Information via Negotiation (SPIN) so as to discover a vitality proficient directing calculation which would be suitable to the natural qualities of WSN. These conventions were named information driven, progressive and area based separately. It has been inferred that Rumour routing was the ideal determination for little and medium size WSN.

Chin-Ling Chen and I-Hsien Lin (2010) proposed an area mindful element session-key administration convention for a framework based remote sensor system. This convention enhances the security of a mystery key which is powerfully overhauled. This dynamic upgrade can bring down the likelihood of the key being speculated effectively. Along these lines right now, known assaults can be shielded. By using the nearby data, this plan can likewise constrain the flooding district keeping in mind the end goal to decrease the vitality that is devoured in finding directing ways.

Anjana Devi (2011) had proposed cross layer interruption location, structural planning to find the noxious hubs and diverse sorts of Denial of administration (DOS) and sinkhole assaults by abusing the data accessible crosswise over distinctive layers of convention stack with a specific end goal to enhance the precision of discovery.

This methodology utilized an altered width grouping calculation for proficient recognition of the inconsistencies in the adhoc activity, furthermore for distinguishing more current assaults produced. In this approach, an adaptive association rule mining algorithm has been used to speed up the intrusion detection process when compared to the conventional techniques.

Debaditya Ghosh et al. (2011) had formulated a path in which WSN could accomplish boundless life and record-breaking preparation with general security to send data to the base station with least power dispersal utilizing multimode sensor nodes and sort classification of created data. They effectively changed their past rendition of Query Check Source (QCS) convention which came about into a superior execution of WSN.

In this approach, the bundle size was advanced and two bits of IP packet header have been adjusted with a specific end goal to decrease the handling unpredictability. Additionally this methodology encouraged the discovery of unpredictable circumstances at the network layer itself.

Tejomayee Nath (2011) proposed a multipath directing plan called Multipath On-interest Routing (MOR) keeping in mind the end goal to minimize the course break recuperation overhead. This plan gave various courses in the essential way to the destination alongside source hub. The essential way expected to be the first way gotten by the source node in the wake of starting the course disclosure, which was normally the briefest way. The proposed strategy decreased the course blunder transmitted amid course break recuperation and the overhead of extra course revelation endeavors.

Shobha (2012) had proposed a secure, energy efficient and dynamic routing protocol for WSN using a link state routing algorithm known as Open Shortest Path First (OSPF) which was designated as Interior Gateway Protocol (IGP) and proved that it was cost effective, secure and simpler to configure.

SeoHyunOh (2012) had introduced a malevolent and breaking down node recognition plan utilizing double weighted trust assessment in a various leveled sensor system. Vindictive nodes were adequately identified even in the vicinity of characteristic blames and clamor.

Saeed Ebadi et al. (2012) had proposed a Multi Hop Clustering (MHC) algorithm in which each CH was selected based on residual energy and node degree. The proposed method helped to reduce the overall energy consumption and improved the lifetime of WSN.

Vishal Parbat et al. (2012) proposed ZKP which facilitated the execution of recognizable proof, key trade and other essential cryptographic operations, without uncovering any mystery data amid the discussion and with littler computational necessities in examination to open key conventions.

Xuxun Liu (2012) had a detailed survey on various cluster-based routing protocols and suggested a novel cluster routing method using the simulated results.

Haixia Zhao et al. (2013) had presented a location pairwise key based Secure Geographical and Energy Aware Routing (SGEAR) protocol which could overcome bogus routing information, Sybil attack, and selective forwarding attack.

Hyeon Myeong Choi et al. (2013) proposed a protected steering system for recognizing false report infusions and wormhole assaults in WSN. The proposed strategy was in light of a Statistical En-course Filtering (SEF) plan for recognizing false reports and used affirmation messages for identifying wormholes. Low vitality utilization was a huge accomplishment of the proposed strategy.

Xiaoyong Li (2013) proposed a trust based protocol known as LDTS which facilitated to achieve system efficiency and energy saving. With minimum memory consumption, this protocol helped to achieve higher throughput even in the presence of malicious nodes.

Zair Hussain et al. (2013) analyzed the lifetime of different routing protocols for WSN based on various network parameters such as network lifetime, energy consumption of a node and hop distance. The influence of anchor node failure also had been briefly described.

A new range free localization scheme with dual mobile beacon for WSN had been proposed by Zhenbo Shi et al. (2013). The proposed scheme reduced the number of messages exchanged between unknown node and beacon node which resulted in reduced energy consumption. Improved localization accuracy and reduced channel noise interference were the added features of the proposed scheme.

Ankit Thakkar (2014) self-tended to the need to style a convention to support the system lifespan and gives information to the sink with a limited deferral. A steering guideline was anticipated by presenting Energy Delay Index for Trade-off (EDIT) to improve vitality and deferral. The EDIT is utilized to choose bunch heads and "next jump" by considering vitality and/or delay necessities of a given application. An exchange off in the middle of Energy and Delay are found by considering varying sorts of separations in the middle of CH and its member node alluded to as geometer separation and Hop-number. Explained talks connected with the impact of 'next jump' decision as to vitality and defers in an extremely multi bounce correspondence were consulted with the help of re-enacted results.

Degan Zhang et al. (2014) had proposed an Energy Balanced Routing Method in light of Forward Aware Factor (FAF-EBRM) in which the following bounce hub was chosen with attention to connection weight and forward vitality thickness. An unconstrained recreation instrument has additionally been imagined for neighborhood topology. The proposed strategy had heartiness, adaptation to non-critical failure and lessened the likelihood of progressive hub breakdown.

Duc Chinh Hoang et al. (2014) had actualized a unified group construct convention, which was based with respect to a music-based meta-heuristic advancement calculation known as Harmony Search Algorithm (HSA). The proposed convention minimized the intra bunch separations between the group individuals and their CHs and enhanced the vitality conveyance of the WSN. This system used the solid reckoning capacity of the Base Station (BS) to get executed in a sensible time of time for continuous operation. Broadened life time and quick union were the huge results of the proposed convention.

A few protocols explored in literatures (He et al. 2003, Chipara et al 2006, Parvaneh Rezayat et al. 2010) deliver the messages in the network based on their deadline. Though these protocols reduced the communications delay involved in message transfer by dynamically varying the transmission power, the power utility in the network can still be reduced.

# CHAPTER 3

# EFFICIENT AND SECURE ROUTING PROTOCOL FOR WSN

## 3.1 INTRODUCTION

In this chapter the proposed protocol ESRP, has been formulated and then analyzed with the help of simulated results. The energy efficiency has been achieved by the implementation of centralized cluster formation method as opposed to the conventional distributed approach. The effectiveness of this approach has been proved by comparing the simulated results with its distributed counterparts.

Due to the attributes of being a network and utilizing wireless communications, the security demands for WSN are unique. Security requirements in WSN to ensure trustworthy, secure connections as well as communications are a combination of the specifications for computer network and wireless communication security. The complete ESRP protocol has been formed by incorporating lightweight security features in order to combat with various real time security threats.

The influence of the security features has been studied so as to ensure that the performance of ESRP should not be degraded. The scalability of ESRP has been analysed before it could compared with other similar protocols. Finally the superiority of ESRP has been proved by comparing it with four similar protocols such as LDTS, SLEACH, SEDR and SERP.

## 3.2 ESRP WORK FLOW

Figure 3.1 shows the various phases of the proposed work. In the first phase, various advantages of the ESRP centralized cluster formation method have been explored by comparing it with the distributed approaches of LEACH protocol.

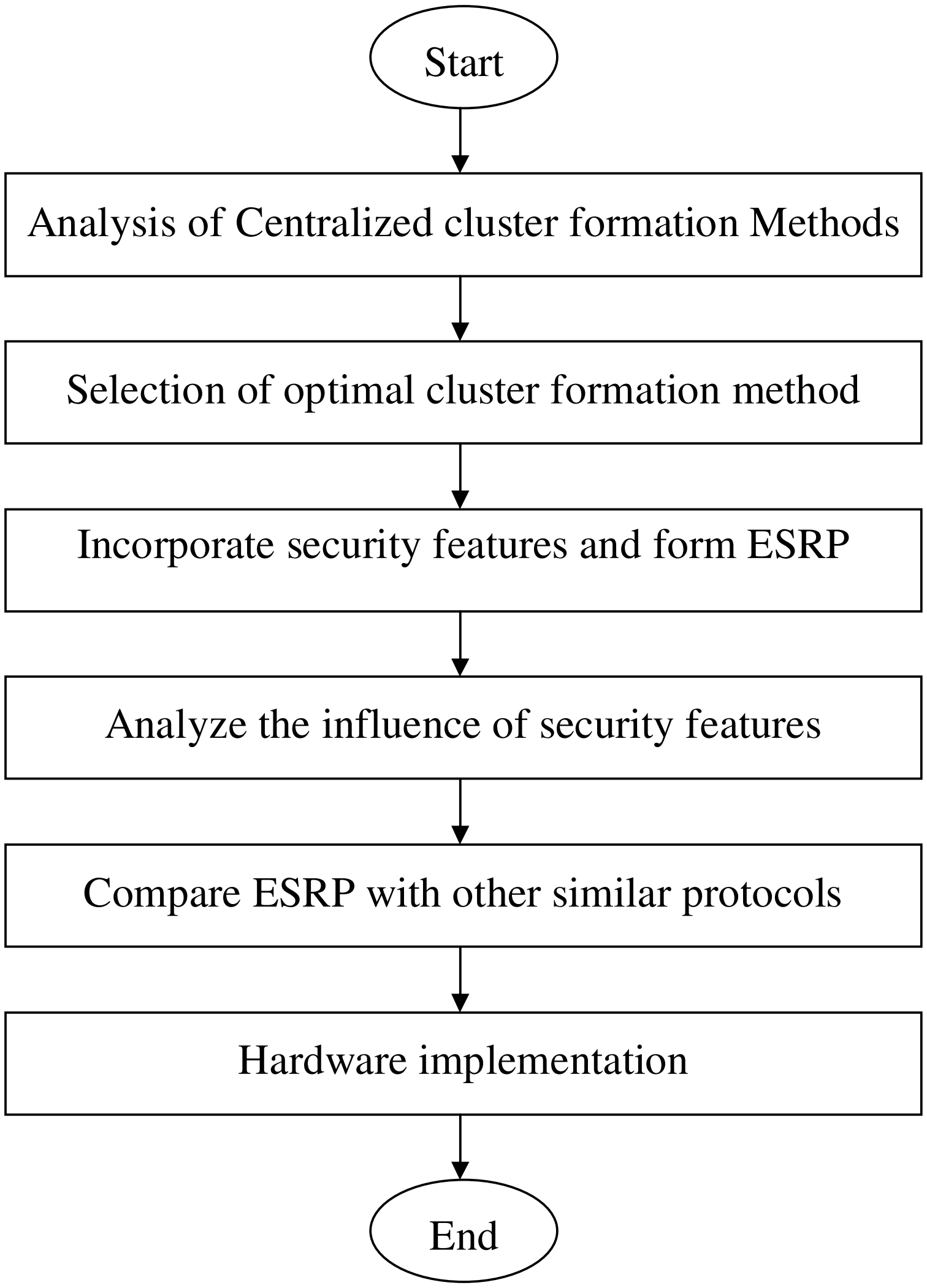

**Figure 3.1 Various Phases of ESRP**

Based on the optimal cluster formation method, the deployed wireless sensor nodes were grouped in to various clusters in the second phase. In order to ensure reliable data delivery, various security schemes have been incorporated in the third phase and their influence over the network performance has been analyzed during the fourth phase.The performance of the proposed protocol has been compared with other similar protocols with the help of simulated results during the fifth phase. Finally, the real time implementation of the proposed protocol has been carried out with the help of hardware wireless sensor nodes.

## 3.3 ESRP Packet structure

Figure 3.2 shows the structure of a signal packet, which is transmitted by the sink to each CH during the beginning of cluster formation phase as well as during each cluster reformation phase.

| **CH ID (1)** | **Public key(2)** | **Private key(3)** | **CM ID(4)** | **Neighbor CH ID(5)** | **R (6)** | **R (7)** | **R (8)** |
|---|---|---|---|---|---|---|---|
| 1 byte | 1 byte | 1 byte | 4 bytes | 1 byte | 1 byte | 1 byte | 1 byte |

**Figure 3.2 Signal packet structure**

The field numbers are indicated in brackets. The one byte long first field is used to identify the particular CH for which the packet is intended. The second and third fields are used to carry a one byte long public key and private key respectively. The public and private key values are changed by the sink during every cluster reformation phase.

The fourth field, which occupies 32 bits have been arranged in a 16 bit *16bit array is used to indicate CM ID for the corresponding CH. The one byte long, fifth field is used to indicated the neighbor CH ID. The next three fields are reserved, indicated by R.

The packet structure of a CH is shown in Figure 3.3.The field numbers were mentioned in the brackets. The first field which is marked in green color, occupies one bit in length, decides the type of routing structure that can be followed by a node.

| | | | | | |
|---|---|---|---|---|---|
| **Hierarchical / Flat Routing (1)** | **CH/CM (2)** | **Node ID (3)** | **Energy Value (4)** | **Next CH ID (5)** | **CM ID (6)** |
| 1 bit | 1 bit | 1 byte | 1 byte | 1 Byte | 1 Byte |
| **CM Energy Value (7)** | **CM Payload (8)** | **Secret key (9)** | **Public key (10)** | **Prover / Verifier (11)** | **CM Energy Value (12)** |
| 1 Byte | Variable | 1 Byte | 1 byte | 2 bits | 1 Byte |
| | **Adversary trapping enable bit (13)** | **Mine detection Enable bit (14)** | | **Promiscuous hearing enable bit (15)** | |
| | 1 bit | 1 bit | | 1 bit | |

**Figure 3.3 Packet structure of a CH**

If this bit is '0', then a node has to follow the flat routing structure, else it has to follow the hierarchical structure. All the remaining bits are validated for a hierarchical structure, only when this bit is ‘1’.

The one bit long, second field which is marked in purple color decides whether a node has to act as a CH, if it is ‘1’ or as a Cluster member (CM) if it is ‘0’.

The fields which are used to implement ESRP security features have been marked in blue color and all the remaining fields are marked in red color. A one byte logical Identification (ID) of each node is stored in the third field. The energy of a CH is indicated by a one byte long fourth field. The one byte long fifth field is used to store the logical ID of an upstream CH, which is required to transfer the aggregated data, during the inter cluster routing phase.

The next three fields are used to store the logical ID, energy value and the pay load of a CM respectively during intra cluster data collection phase. The one byte long fields nine and ten are used to store the sink specified ‘secret key’ and ‘public key’. As shown in the figure, the various security features of ESRP, have been enabled or disabled in a CH, by the sink dynamically using the fields eleven through fourteen, during the initial ‘cluster formation’ phase and subsequently followed during each ‘cluster reformation’ phase. The decision is based on the residual energy of each CH.

The outcome of each security method has been specified as a status in the fields fifteen to seventeen which are one byte long. Based on these status, a node can be excluded from the network structure, if found malicious or an intruder by the sink during the cluster reformation phase. Figure 3.4 shows the packet structure of a CM.

| Node ID | Energy value | CH ID | Payload |
|---|---|---|---|
| 1 byte | 1 byte | 1 byte | Variable |

**Figure 3.4 Packet structure of a CM**

As shown in the figure, each CM holds its logical ID, Energy value and the logical ID of their corresponding CH, each in one byte long fields.

## 3.4 THEROTICAL ENERGY CALCULATION

The theoretical value of energy in Joules (J) that could be consumed by a WSN has been calculated in this sub-division. As shown in Radio Model, there are three units contributing to the energy consumption: the Communication unit, Sensing unit and Computing unit.

The various energy values in J specified in the Figure 3.5 as wells as the other parameters used in the forthcoming energy consumption calculation have been defined bellow

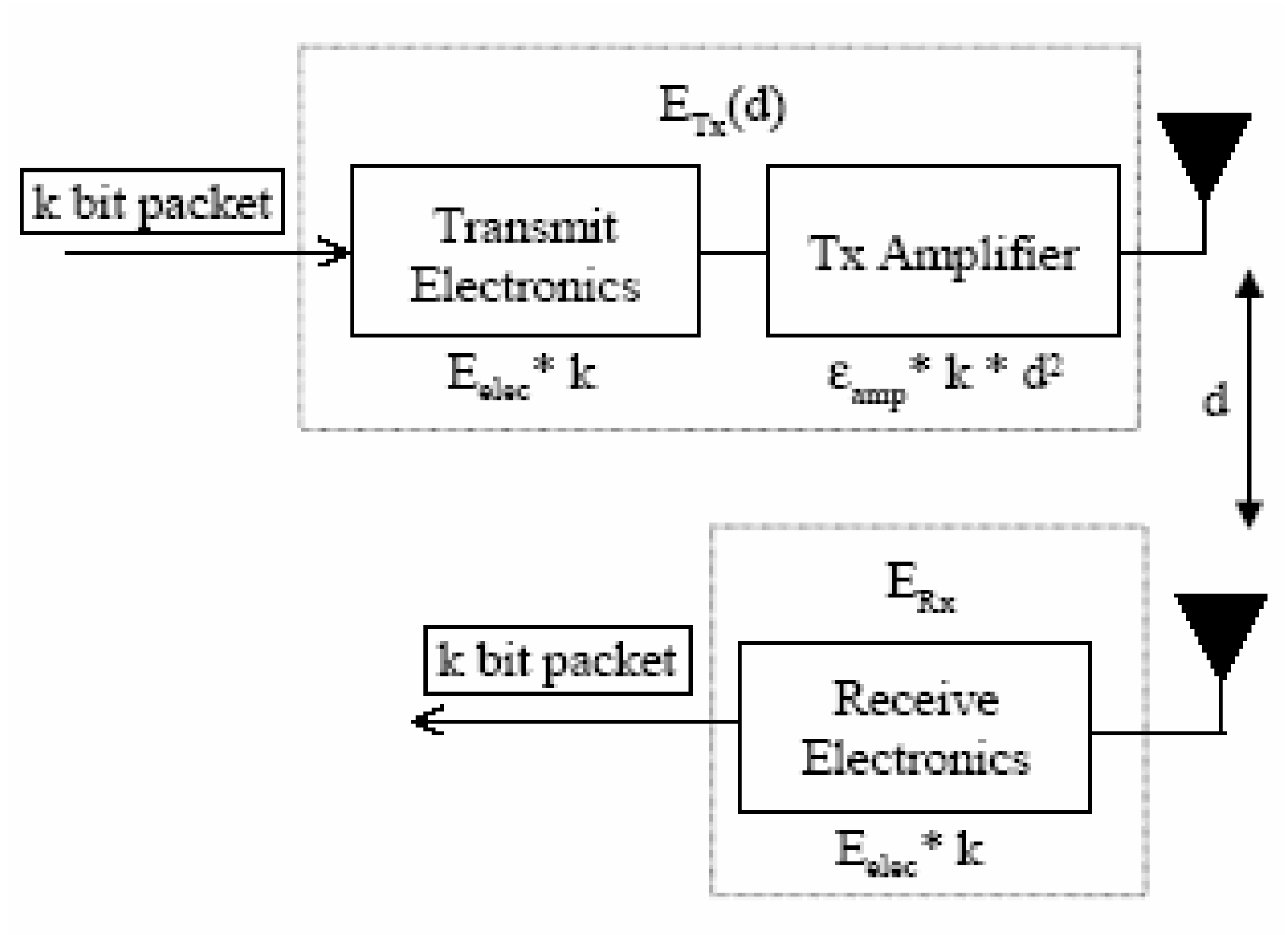

**Figure 3.5 Radio model of a wireless sensor node**

- $E_{elec}$ is the energy needed for modulating or demodulating one bit of the circuit.
- $E_{amp}$ is the energy for the amplifier circuit to transmit one bit to an area of radius d = 1 meter
- $E_{node}$ is the energy consumption of a complete sensor node.
- $E_{transceiver}$ is the total energy consumption during transmission and reception.
- $E_{CPU}$ is the total energy consumption of the processing unit.

- $E_{CPU\ data}$ is the total energy consumption of CPU board for processing data.
- $E_{CPU\ signal}$ is the total energy consumption of CPU board for processing signal.
- $E_{mem\ rd}$ is the total energy consumption of memory unit for reading a data.
- $E_{mem\ wr}$ is the total energy consumption of memory unit for storing a data.
- $E_{radio}$ is the total energy consumption of the radio board.
- $E_{sensor}$ is the energy consumption of the sensing unit which 2/3 of $E_{radio}$
- K is the number of bits per message which is 1000 for data messages and 64 for signal messages.
- L is the number of bits required to store one key entry which is eight bits for ESRP.

The functions various units shown in the Figure 3.5 were explained bellow.

**Communication unit**: The communication unit is responsible for wirelessly communication among nodes. The Transmit Electronics represents electronics circuit performing signal modulation. Tx

Amplifier is used to amplify the modulated signal and output it to the antenna. The Receive Electronics is used to decode the modulated signal

In a real device, the transmit module normally stays in sleep mode. It only wakes up when there is any bit that needs to be sent. The receiver module performs the reverse function. It needs to be ON when waiting to receive messages.

**Sensing unit**: This unit consists of Microcontroller Unit, shortly known as MCU, which is a combination of Central Processing Board (CPU) and Memory board as well as the Radio board. These boards work in two modes: full action and sleep. In the sleep mode, the energy dissipation is almost zero. The full action mode consumes energy depends upon the voltage and current ratings specified in their respective data sheets.

Generally, that the current of the MCU board in full operation is equal to that of the radio board in the receiving mode and the current of the sensor board in full operation is around 2/3 of the current of the radio board in receiving mode. The current of the memory board to write data is around 2 times of the current of the MCU board in full operation, and the current of the memory board to read data is ½ of the current of the MCU board in full operation.

The following assumptions have been made as the basis for the energy consumption calculation of ESRP during the simulation.

$E_{elec}$ = 50 nJ/ bit

$E_{amp} = 100$ pJ/ bit/ $m^2$ = 0.1 nJ / bit / $m^2$

Data packet size = 1000 bits = 125 bytes

Signal packet size = 64 bits = 8 bytes

Data rate = 1000 bits /sec

Optimized radius 'd' <= 50 m

$$E_{node} = E_{transceiver} + E_{CPU} + E_{sensor} \quad (3.1)$$

$$E_{transceiver} = E_{TX\ data} + E_{TX\ signal} + E_{RX\ data} + E_{RX\ signal} \quad (3.2)$$

$$E_{TX\ data} = E_{elec}*K + E_{amp}*K*d^2 \quad (3.3)$$

$$E_{TX\ signal} = E_{elec}*K + E_{amp}*K*d^2 \quad (3.4)$$

$$E_{RX\ data} = E_{elec}*K \quad (3.5)$$

$$E_{RX\ signal} = E_{elec}*K \quad (3.6)$$

$$E_{CPU} = E_{CPUdata} + E_{CPU\ signal} + E_{mem\ rd} + E_{mem\ wr} \quad (3.7)$$

$$E_{CPU\ data} = E_{elec}*K \quad (3.8)$$

$$E_{CPU\ signal} = E_{elec}*K \quad (3.9)$$

$$E_{mem\ rd} = 2*E_{elec}*L \quad (3.10)$$

$$E_{mem\ wr} = 0.5*E_{elec}*L \quad (3.11)$$

$$E_{radio} = E_{elec}*\text{data rate} \quad (3.12)$$

$$E_{sensor} = E_{radio}*2/3 \quad (3.13)$$

Total Energy consumed by 100 nodes = $E_{node}$ *100 (3.14)

Total Energy consumed by the WSN

for 3600 s = Energy consumption for 100 nodes/3600 (3.15)

The values of the parameters such as $E_{TX\ data}$, $E_{TX\ signal}$, $E_{RX\ data}$ and $E_{RX\ signal}$ have been calculated with the help of Equation (3.3) to Equation (3.6) respectively as shown below.

$E_{TX\ data} = 50*10^{-9}*1000+0.1*10^{-9}*1000*50^2 =$
300µJ/message=0.3 µJ/bit

$E_{TX\ signal} = 50*10^{-9}*64+0.1*10^{-9}*64*50^2 = 19.2$
µJ/message=0.3 µJ/bit

$E_{RX\ data} = 50*10^{-9}*1000 = 50$ µJ/message = 0.05 µJ/bit

$E_{RX\ signal} = 50*10^{-9}*64 = 3$ µJ/message = 0.05 µJ/bit

Using Equation (3.2), the transceiver energy consumption has been calculated as follows.

$E_{transceiver}$ = 300+19.2+50 + 3 = 372 µJ/message

The values of the parameters such as $E_{CPU\ data}$, $E_{CPU\ signal}$, $E_{mem\ rd}$ and $E_{mem\ wr}$ have been calculated with the help of Equation (3.8) to Equation (3.11) respectively as shown below.

$E_{CPU\ data}$ = 1000 bits/message * 50 nJ /bit = 50 µJ/ message

$E_{CPU\ signal}$ = 64 bits/message * 50 nJ /bit = 3 μJ/ message

$E_{mem\ rd}$ = 2 * 50 nJ /bit * 8 bits = 0.8 μJ

$E_{mem\ wr}$ = 0.5 * 50 nJ /bit * 8 bits = 0.2 μJ

Using Equation (3.7), the energy consumption of the CPU board has been calculated as follows.

$E_{CPU}$ = 50 +3 +0.8 + 0.2 = 54 μJ

The energy consumed by the radio and sensor board has been calculated respectively as follows, using Equation (3.12) and Equation (3.13).

$E_{radio}$ = 50 nJ /bit * 1000 bits /s = 50 μJ/s

$E_{sensor}$ = 50 μJ/s * 2/3 = 33 μJ/s

Using Equation (3.1), the energy consumption of a sensor node has been calculated as follows.

$E_{node}$ = 372 + 54 + 33 = 459 μJ = 0.0005 J/s

The ESRP have been simulated with a maximum of 100 nodes for 3600 seconds. The total energy consumed by the 100 nodes as well as by the WSN, assuming all nodes are fully functional throughout the simulation time has been calculated using Equation (3.14) and Equation (3.15) as follows.

Total Energy consumed by 100 nodes = 0.0005 J/s * 100 = 0.05 J/s

Total Energy consumed by the WSN for 3600 s =0.05 J/s*3600 s=180 J.

## 3.5 ESRP SECURITY METHODS

In this proposed protocol the following security methods have been incorporated to overcome four major security threats namely Node compromise attack, Selective forwarding attack, DoS attack and Self intruder attack.

### 3.5.1 Modified Zero Knowledge Protocol

This security feature has been used to identify '**Node compromise attack**'. In order to overcome this attack, a key based security feature which is an enhancement of ZKP has been proposed.

The method is as follows:

- Once the clusters have been formed, the sink dynamically selects few CH for identity verification of their downstream counter parts.
- The selected CH can play any one of the following roles during inter cluster routing. Prover, Verifier, Prover & Verifier, none

- A one byte long secret key ‘S’, have been provided by the sink to the CHs those who have been selected to play the role of ‘Prover’, or ‘Prover& Verifier’.

- A one byte long public key ‘N’ has been distributed to all selected CHs by the sink.

- During the Inter cluster routing, each ‘Verifier CH’ demands answer for a sink specified question from each ‘Prover CH’, which involves complex mathematical combinations of ‘S’ and ‘N’. The nature of the answer is such that, no Verifier can extract ‘S’ of the any Prover from that answer.

- The answer has been forwarded to the sink by the ‘Verifier CH’.

- The sink, which is already computed the answer since it was the initiator of the question, should compare the computed answer with the received one from the ‘Verifier CH’.

- The Verifier CH has been instructed by the sink to receive the data from the ‘Prover CH’, if both answers were matched at the sink and each ‘Verifier CH’ has to acknowledge each ‘Prover CH’.

- When there is mismatch, sink blocks all network activities of the ‘Prover CH’ and informs about its malicious status

to all CHs. The entries of malicious nodes have been removed from the data base of the sink, during the cluster reformation phase.

- The Public and Private keys have been assigned dynamically by the sink, during every cluster reformation phase.

Table 3.1 shows the comparison between ZKP and MZKP which is implemented in ESRP.

**Table 3.1 Comparison between ZKP and ESRP-MZKP**

| ZKP | ESRP-MZKP |
|---|---|
| Distributed approach | Centralized approach |
| Complex process | Simple process |
| Key distribution is Static | Key distribution is Dynamic |
| Residual energy of nodes have not been considered during the implementation | Residual energy of nodes have been considered during the implementation |
| Un reliable, since a Prover can over hear the answer sent to the Verifier | Reliable, since Prover has to answer first to the Verifier. Hence by overhearing, a Prover cannot change the answer that has been sent to the Verifier |

From the table, it can be concluded that MZKP is more energy efficient than the conventional ZKP, since the amount of information exchanged between a Verifier and the sink is higher in the former method than in the later one.

### 3.5.2 Promiscuous Hearing

This method is used to overcome the '**Selective forwarding attack**' especially for multi-hop networks, which are often based on the assumption that participating nodes may faithfully forward the received messages. This method can be implemented with the help of trusted nodes, which can overhear their neighbor's activities. The method is as follows:

- Once the identity has been proved, every 'Prover CH' has been placed in the promiscuous hearing mode and listen the acknowledgment for the forwarded packet by its upstream 'Verifier CH'.
- When there is no acknowledgement has been listened, the status of the 'Verifier CH' has been marked as malicious by the 'Prover CH'.
- The sink verifies the 'Promiscuous hearing' status field of every CH, during the cluster reformation phase and block the network activity of the corresponding 'Verifier CH' if found malicious.

### 3.5.3 Trapping of Adversaries

This method have been focused for combating two types of attacks: the **Node compromise attack** and the **DoS attack**. These two attacks are similar in the sense that they both generate black holes.

A black hole is an area within which the adversary can either passively intercept or actively block information delivery. The nodes near the base station in a WSN forward a significantly greater volume of packets than nodes further away from the base station. An adversary can analyze the traffic patterns in order to identify the location of the base station within the WSN topology. Since the base station is a central point of failure, once the location of the base station is discovered, an adversary disables or destroys the base station, which may result in the failure of the WSN.

One remedial solution to these attacks is to exploit the network routing functionality. Specifically, if the locations of the black holes formed by the compromised nodes are known a priori, then information can be delivered over paths that bypass these holes, whenever possible.To intercept different packets, the adversary has to compromise or jam all possible routes from the source to the destination, which is practically infeasible. The key contributions of this work are as follows:

- In this technique, several dummy (fake) packets with Time To Live (TTL) field, were created and propagated in the network to introduce more randomness in the communication pattern.

- The generation of dummy packets by a CH have been decided by the sink, based on its residual energy during cluster formation and reformation phases through an ‘Adversary trapping enable bit’

- Even if an adversary can track a packet using time-correlation, it could not track where the real packet is going.

### 3.5.4 Mine Detection

This method has been used to identify and isolate **Self intruders** during intra cluster data aggregation phase. The method is as follows

- Every CH has been instructed by the sink to broadcast a dummy packet and should wait for ‘dummy ack’ packet from every CM.

- If any of the CM failed to acknowledge, then those CM have been marked as an intruder and the sink have been informed with that report.

## 3.6 ESRP ALGORITHM

### 3.6.1 The Complete Protocol

Figure 3.6 shows the various phases of the complete ESRP protocol. During the Initial phase, the sensors nodes were deployed deterministically. The nodes were grouped in to various clusters by the sink, after knowing their position and initial energy values. Once the cluster have been formed, each CH starts collecting the data from their CM in the Intra cluster routing phase and then forward the aggregated data to their upstream neighbors in the Inter cluster routing phase.

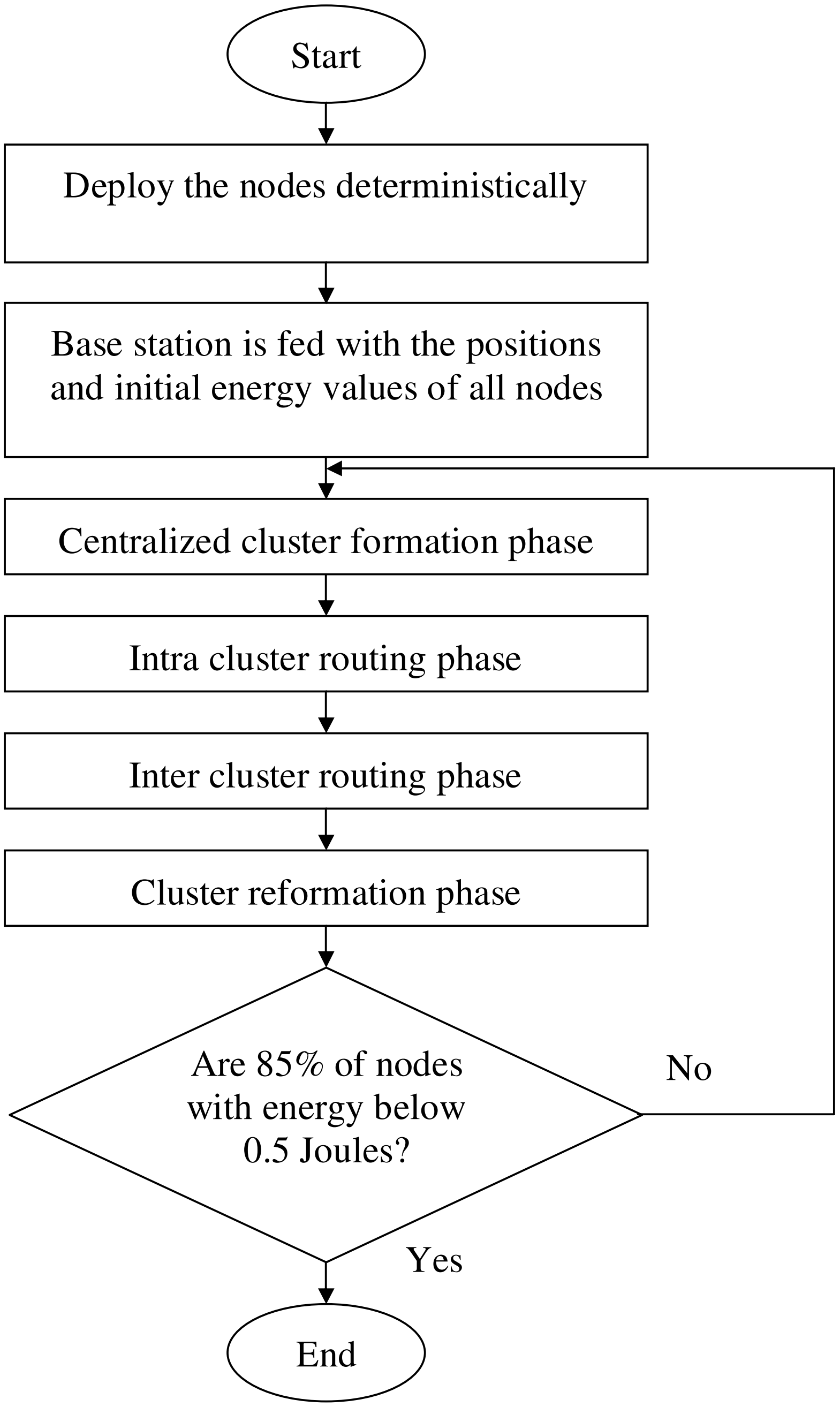

**Figure 3.6 Flow chart of the complete protocol**

After few iteration, the sink initiates the cluster reformation phase. Finally ESRP ceases to work when 85% of nodes were bellowing 0.5 Joules of energy.

### 3.6.2 Algorithm for Centralized Cluster Formation Phase

Figure 3.7 shows the flow chart of the Centralized cluster formation phase, which is explained bellow.

Step 1: The sink node is fed with the positions and initial energy value of all nodes. Logical Identities of nodes have been assigned by the sink node.

Step 2: Node with highest energy in the radio range of the sink have been deputed as CH 1.

Step 3: Among the neighbors of CH1, the node with farthest distance from CH1 and with highest energy have been selected as CH2.

Step 4: The CM were selected depends upon the radio coverage range of every CH.

Step 5: The memberships of common nodes have been decided by the sink, based on the number of CM in each cluster and also the distance to each CH.

Step 6: CH selection have been intimated to the member nodes as well as neighbor CH. Step 7: Security keys have been generated by the sink and have been distributed to all the CH.

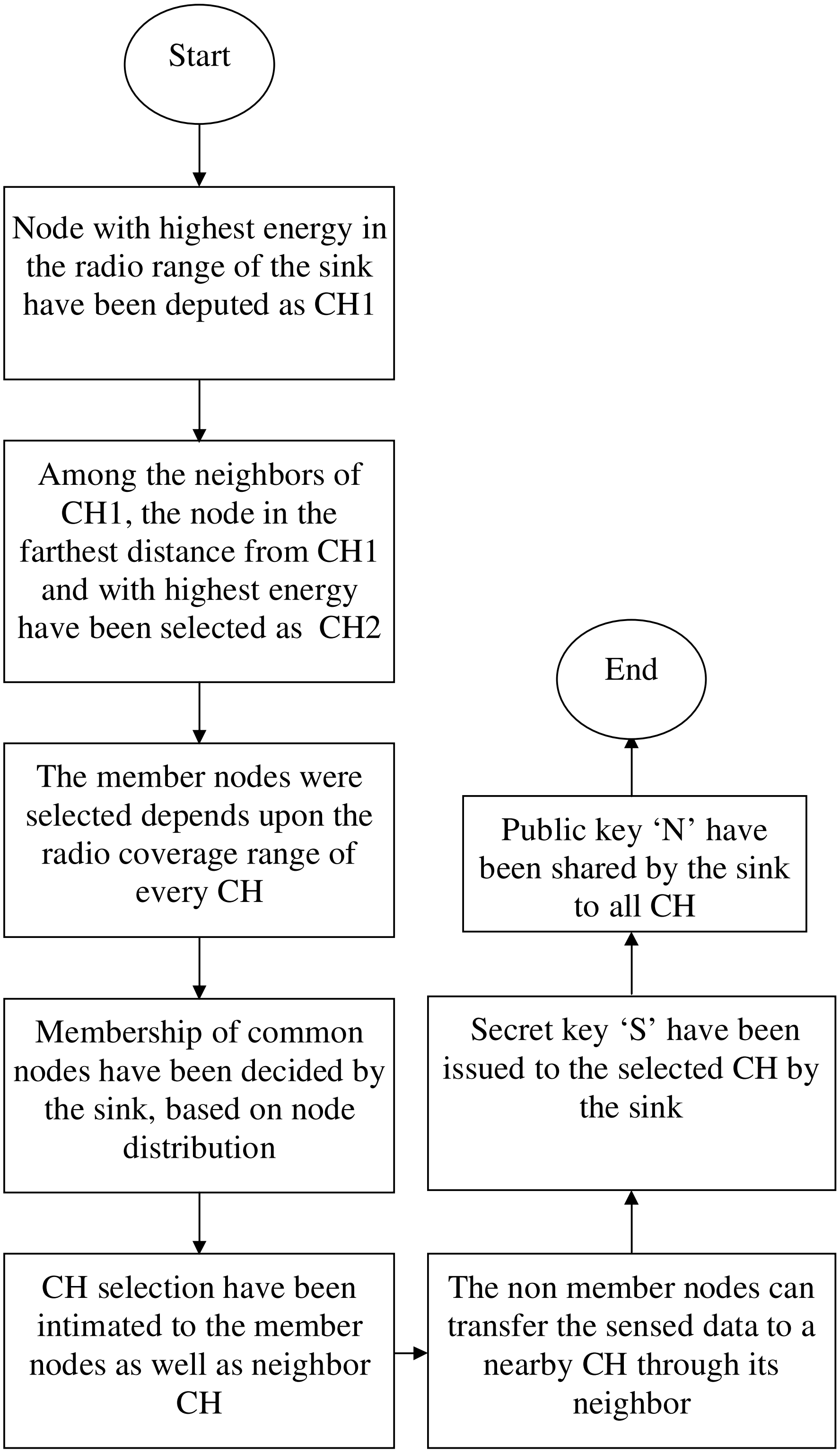

**Figure 3.7 Flow chart of Centralized cluster formation phase**

### 3.6.3 Algorithm for Intra Cluster Routing Phase

Figure 3.8 shows the flow chart of Intra cluster routing phase which is explained bellow.

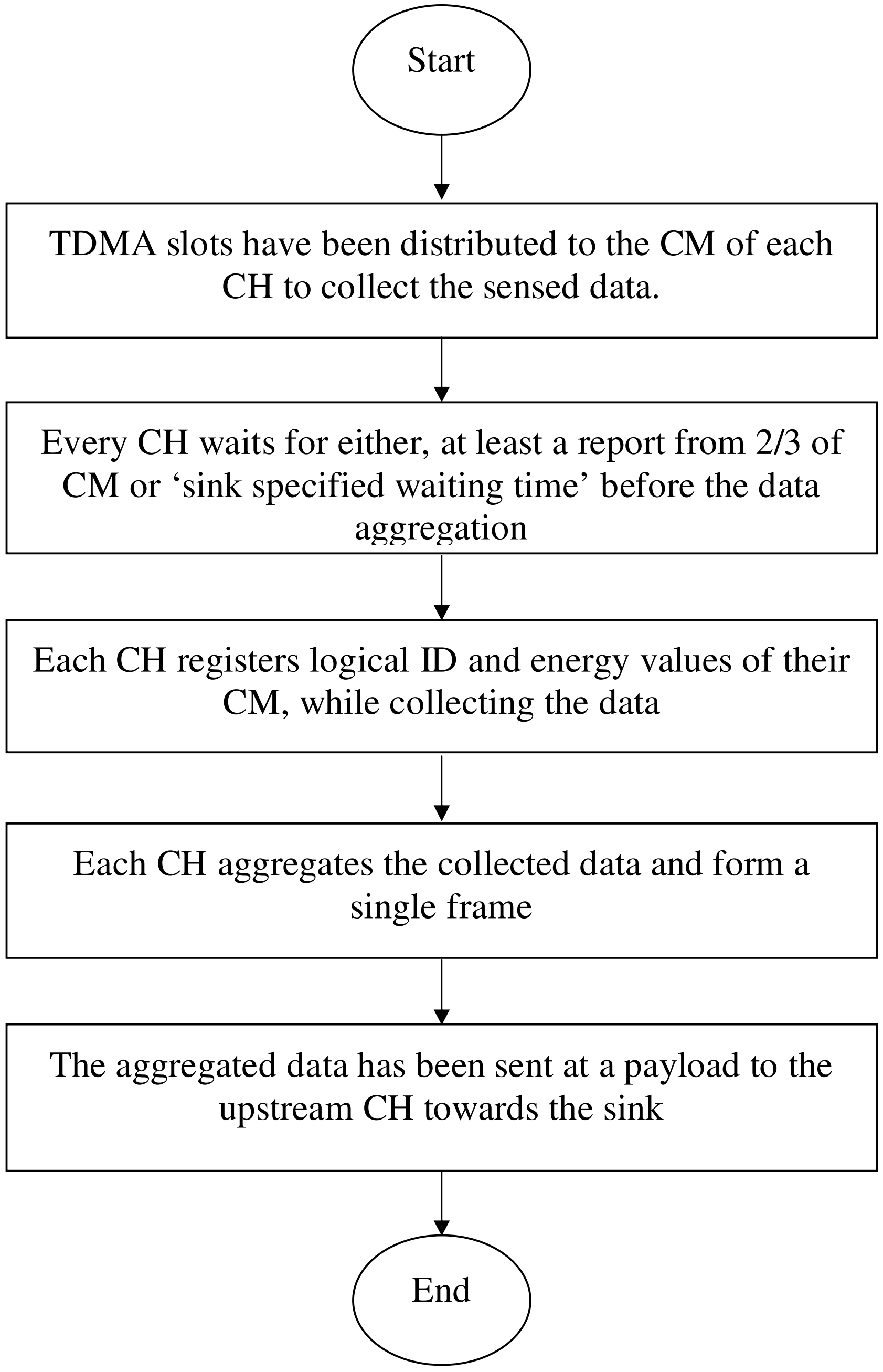

**Figure 3.8 Flow chart of Intra cluster routing phase**

Step 1: TDMA slots have been distributed to each member nodes by their respective CH in order to collect the sensed data.

Step 2: Every CH should wait for either, at least a report from 2/3 of their member nodes or 'sink specified waiting time' before aggregating the data.

Step3: The collected data have been aggregated by each CH and single frames have been formed.

Step 4: The frame have been sent as a payload to the upstream CH towards the sink.

Step 5: The node ID and the residual energy of every member nodes have been registered by their respective CH.

### 3.6.4 Algorithm for Inter Cluster Routing Phase

Figure 3.9 shows the flow chart of various security mechanisms incorporated during the Inter cluster routing phase, which is explained bellow.

Step 1: The 'Prover – Verifier' mechanism of Zero Knowledge Protocol (ZKP) have been employed, whenever a CH wants to transmit a data to its upstream CH, which is present along the route towards the sink.

Step 2: Once the identity have been proven, the 'Prover CH' have been placed in the promiscuous hearing mode in order to verify the acknowledgement of the forwarded packet from the 'Verifier CH'

Step 3: Fewer CHs have been instructed to create dummy packets and transmitted them in a direction opposite to that of original packets in order trap the adversaries.

Step 4: The aggregated data reaches the sink.

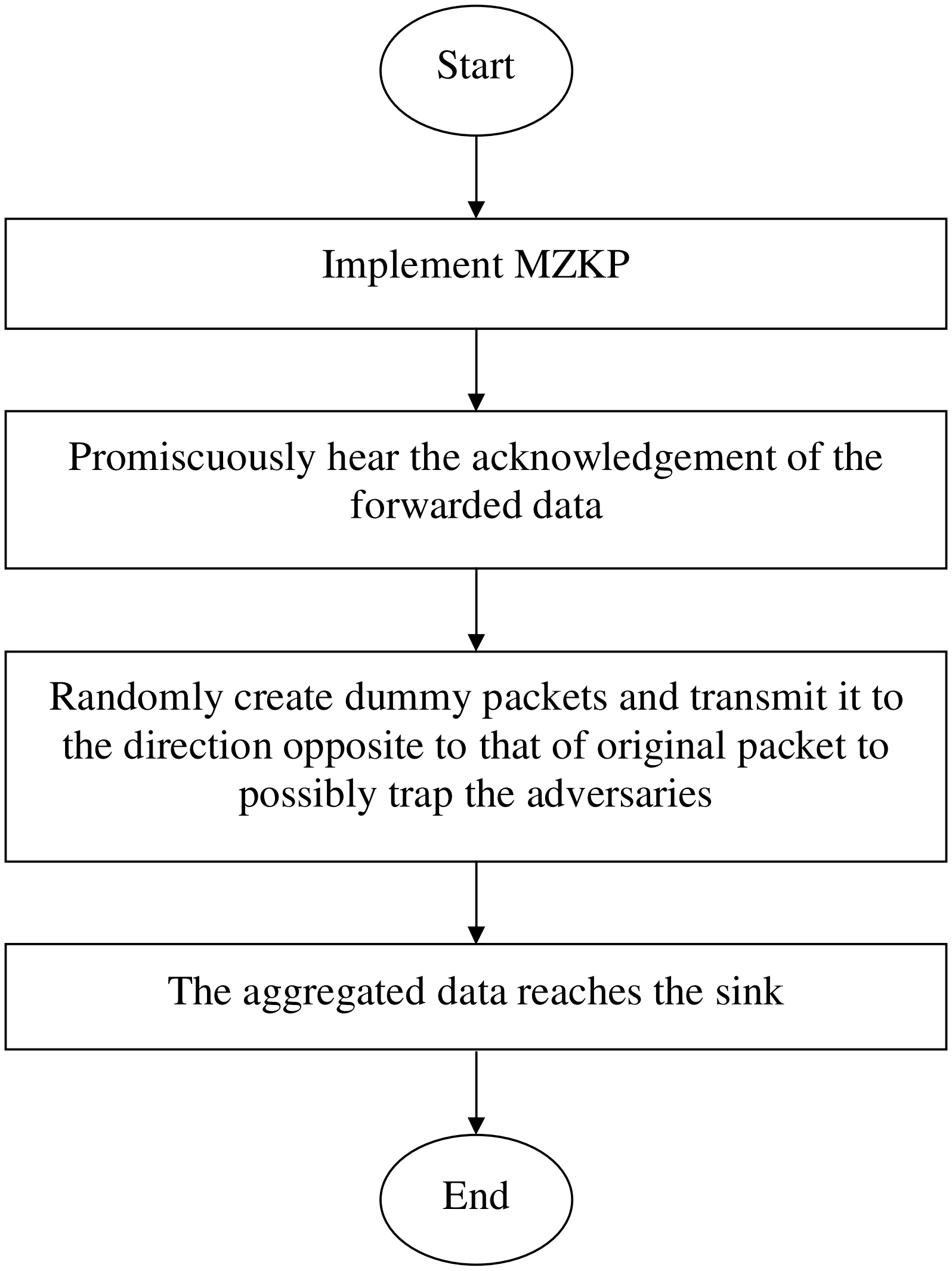

**Figure 3.9 Flow chart of various security mechanisms incorporated during the Inter cluster routing phase**

### 3.6.5 Algorithm for CH Reformation Phase

Figure 3.10 shows the flow chart of Cluster reformation phase, which is explained bellow.

Step 1: Mine Detection have been performed by the sink and based on the acknowledgement for the dummy packets, the self-intruders have been booked by every CH by marking on their corresponding status bits.

Step 2: After being collected the data from every CH, the sink can prevent any CH or a CM from all network activities based on their malicious or intruder status. The security related status bits were used for this purpose. The logical ID of the malicious nodes has been removed from the sink database.

Step 3: Based on their residual energy, the nodes were instructed to follow flat routing in a particular region, where 2/3 of nodes residual energy were bellowing 0.5 J.

Step 4: Sink forms new clusters in other regions, after each iteration, with the nodes those who are having considerable residual energies still. New CH's were also appointed and new CM's have been informed about their corresponding CH. Also sink informs the upstream and downstream counterpart for every newly selected CH's.

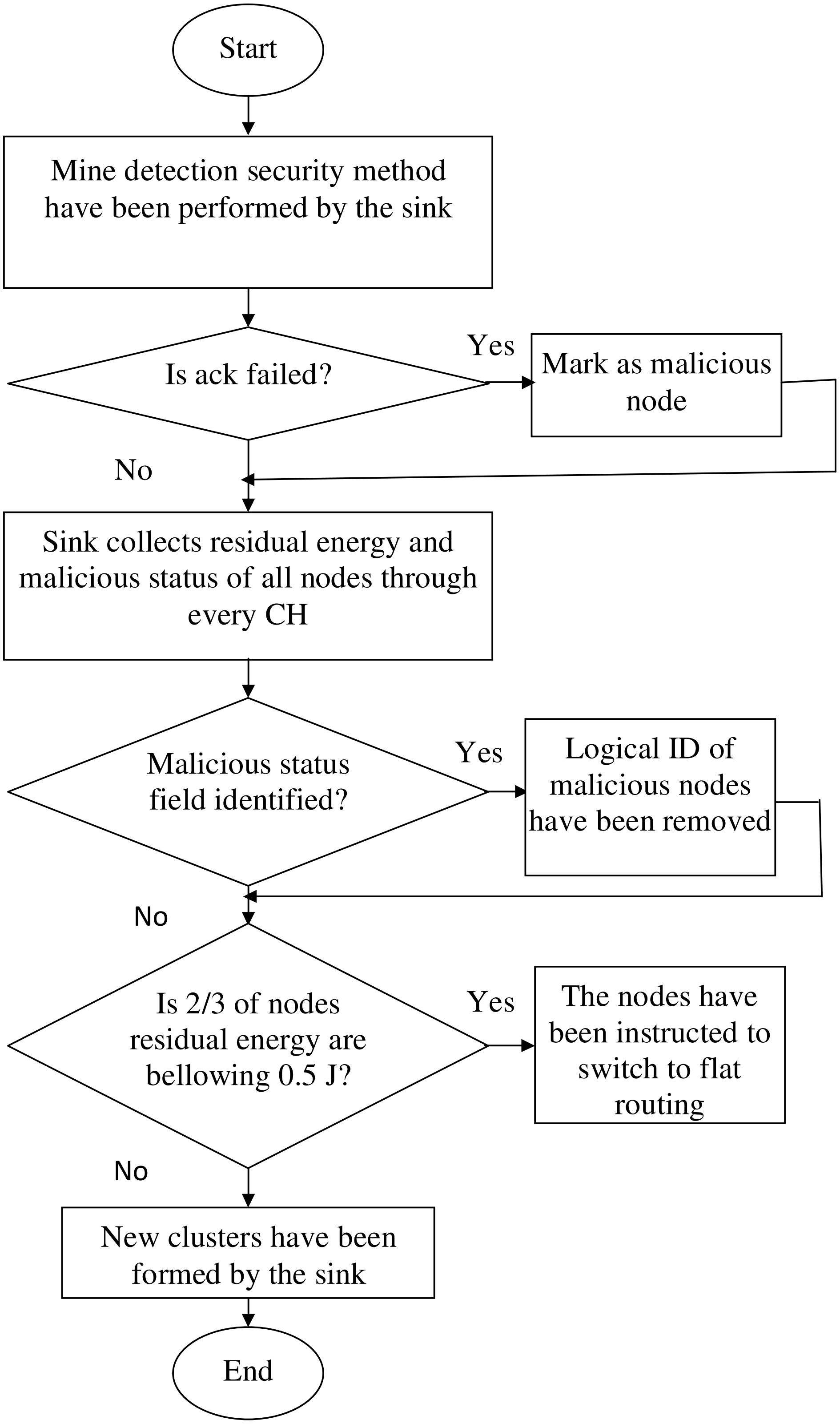

**Figure 3.10 Flow chart of Cluster reformation phase**

## 3.7 SIMULATION RESULTS AND DISCUSSIONS

In order to evaluate the performance of ESRP the following scenario as shown in Table 3.2, have been set in Network Similator-2 (NS-2). Initially the nodes were placed randomly in the specified area of 1900m*1100m. The initial energy of each node is taken as 2 J. Hence the initial energy of the WSN with 100 nodes have been assumed to be 200 J, which closely approximates the theoretically calculated value of 180 J in sub division 3.2.The Data Rate is set to 2 Mbps. The distributed coordination function of IEEE 802.11 is used as the MAC layer protocol.

During the simulation, 25 intruder nodes have been randomly introduced in order to evaluate the reliability of ESRP. The simulation has been carried for 3600 seconds. The values of parameters such as ‘Energy consumption’, ‘Message overhead’, ‘End to end delay’, ‘Network life time’ and ‘Number of nodes alive', were cumulatively calculated over the simulation period. Hence the difference in values of those parameters exhibits the performance of the corresponding protocol.

Among the parameters, ‘Network lifetime’ has been the optimization objective for most of the communication protocols for WSN. The positions of nodes significantly impact the network lifetime. For example, variations in node density throughout the area can eventually lead to unbalanced traffic load that causes the rapid drain of the energy reserve of some sensors.

**Table 3.2 Simulation setup**

| | |
|---|---|
| Area of Sensing field | 1900*1100 m |
| Number of sensor nodes | 100 |
| Simulation time | 3600 s |
| Frequency | 2.4GHz |
| Data rate | 2 Mbps |
| Traffic type | Constant Bit rate |
| Propagation limit | -111.0 dBm |
| Path loss model | Two ray |
| Number of clusters | 5 |
| Size of each cluster | 19 |
| Number of intruders | 25 |
| Initial energy of nodes | 2J |
| Initial energy of WSN | 200J |
| Routing protocol | ESRP |
| Maximum number overhead bytes | 650 |

The percentage variations in 'Number of nodes alive', End to end delay', 'Energy consumption', 'Network lifetime' and 'Message overhead' as shown in the comparison graphs were calculated using Equation (3.16) to Equation (3.20) respectively. In Equation (3.16) to Equation (3.20), the values of the parameters present in the numerator have been taken from the respective simulation parameters summary tables and the values of the parameters present in the denominator for Equation (3.16) to Equation (3.20) have been taken from Table 3.2.

Percentage decrease in Number of nodes alive = (Number of alive nodes at the end of simulation period/ Total number of alive nodes at the beginning of the simulation period) * 100 (3.16)

Percentage increase in End to end delay = (Total amount of time taken to complete five iterations/ Total simulation time) * 100 (3.17)

Percentage increase in Energy consumption = (Total amount of energy spent over the simulation period /Maximum energy considered over the simulation period) * 100 (3.18)

Percentage increase in Network lifetime = (Total number of clusters retained at the end of the simulation period /Total number of clusters available at the beginning of the simulation period) * 100 (3.19)

Percentage increase in Message overhead = (Total number of overhead bytes spent at the end of the simulation period / Maximum number of overhead bytes) * 100 (3.20)

In addition, improper node distribution may lead to the depletion of energy of nodes that are close to the base station at a higher rate than other nodes and thus shorten the network lifetime. ESRP has focused on prolonging the network lifetime rather than area coverage. The implicit assumption in the proposed protocol is that there are a sufficient number of nodes or that the sensing range is large enough such that no holes in coverage may result.

### 3.7.1 Comparison of ESRP Centralized Cluster Formation Method with its Distributed Approaches

In the figures shown below, ESRP Clustering method 1 (ESRPC1) indicates the centralized cluster formation approach, ESRP Clustering method 2 (ESRPC2) is based on the distributed cluster formation method of LEACH protocol, where every node has to involve in the formation of clusters and CH. The third method namely ESRP Clustering method 3 (ESRPC3) is based on the modified LEACH protocol, where any CH can select their successors dynamically without waiting for the cluster reformation cycle.

The selection is based on comparing the residual energy between current CH and its CM. The CM with highest residual energy has been selected as new CH.

Figure 3.11 shows the comparison between ESRP centralized cluster formation method with two of its distributed approaches in terms of 'Number of nodes alive'. As shown in the figure, it has been observed that ESRPC1 can retain 36 sensor nodes alive at the end of simulation, compared to 20 alive nodes in ESRPC2, and 8 alive nodes in ESRPC3.

Since the number of nodes alive at the end of the simulation decides the life time of a WSN, it has been observed that a WSN with ESRPC1 have longer life time than the other distributed cluster formation methods such as ESRPC2 and ESRPC3 used in the simulation.

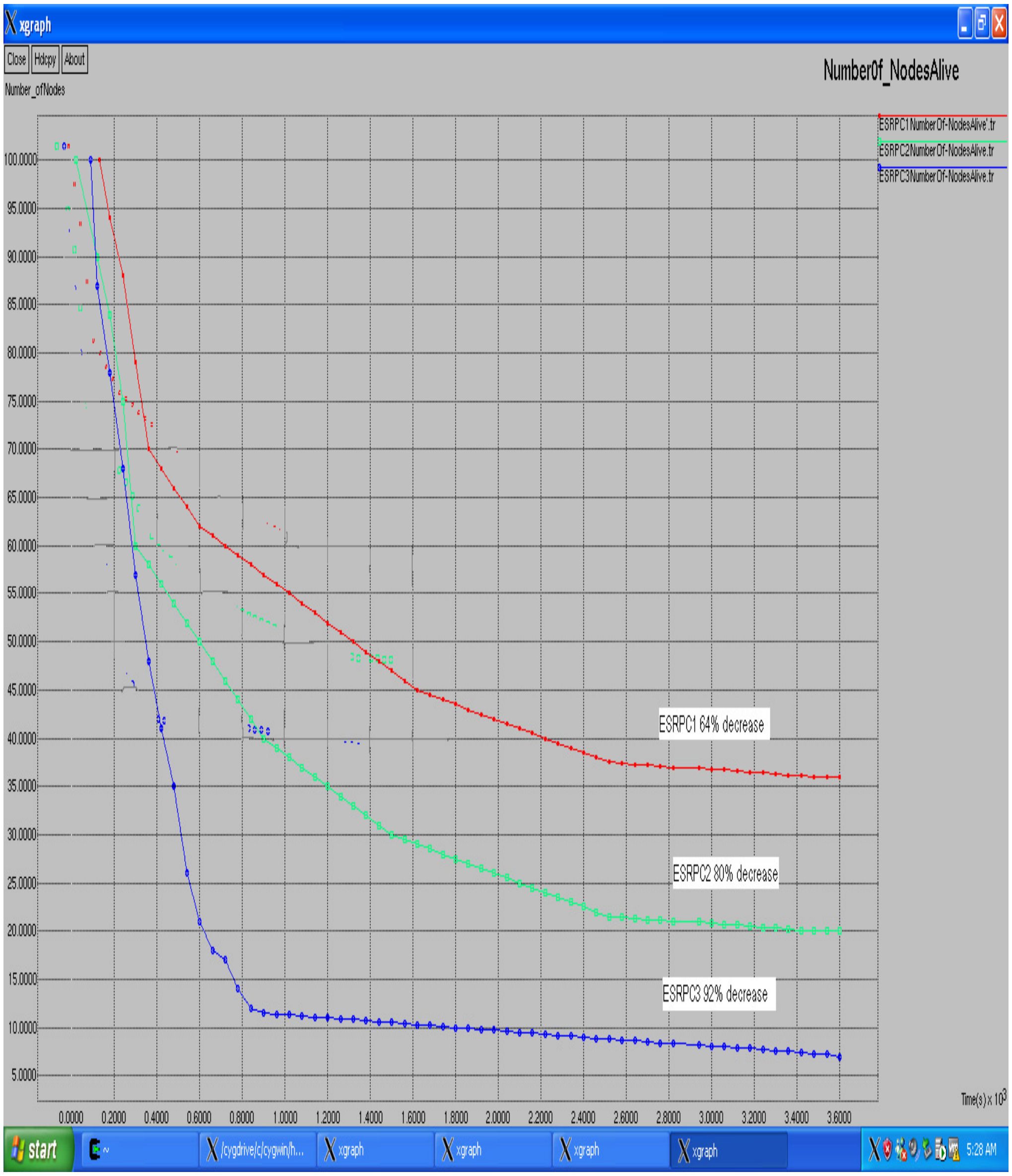

**Figure 3.11 Comparison of clustering methods in terms of 'Number of Nodes alive'**

Figure 3.12 shows the comparison between ESRP centralized cluster formation method with two of its distributed approaches in terms of 'End to end delay'. It can be observed that ESRPC1 took 1800 ms as apposed to 2800 ms by ESRPC2 and 3600 ms by ESRPC3 to complete the simulated scenario.

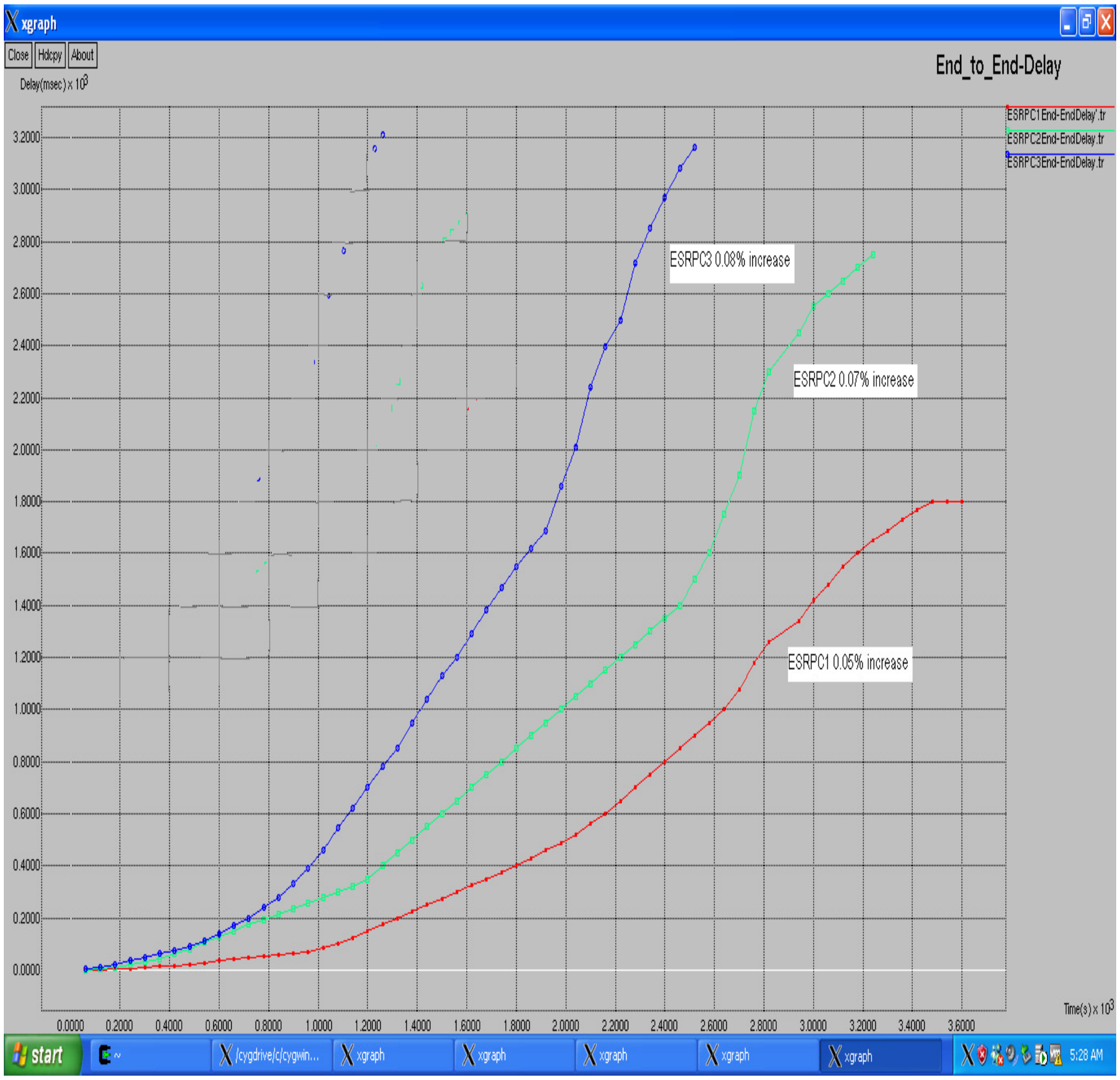

**Figure 3.12 Comparison of clustering methods in terms of 'End to end delay'**

As far as 'Energy consumption' is concerned, at the end of simulation period, as shown in Figure 3.13, the WSN with ESRPC1 consumed nearly 75J, which is 37.5% of the total 200J of energy available initially, whereas ESRPC2 consumed 112J, which is 56 %, and ESRPC3 consumed 120J which is 60% respectively. Hence it has been observed that the WSN with ESRPC1 consumed much lesser energy, whereas ESRPC2 and ESRPC3 have consumed much higher energy.

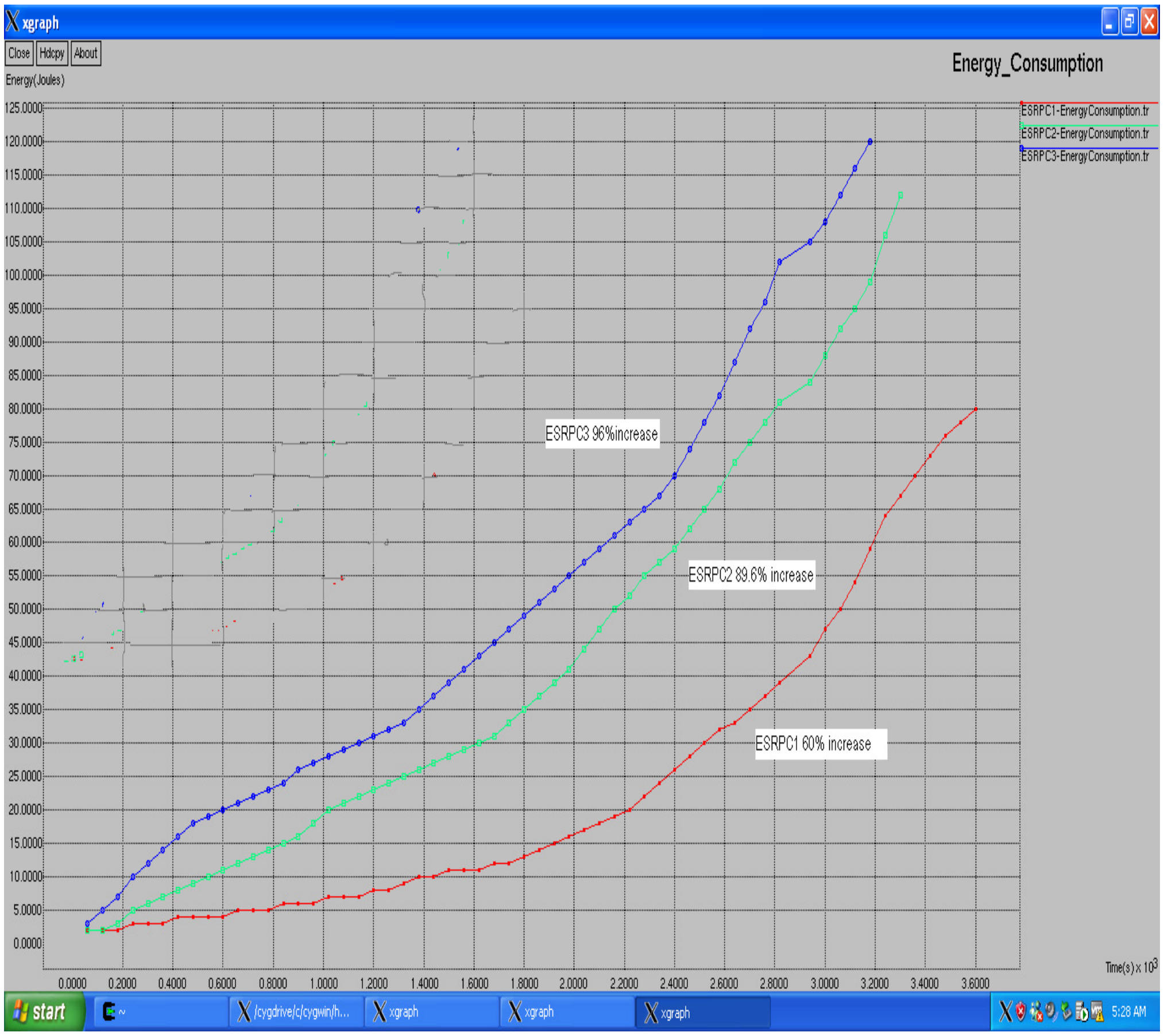

**Figure 3.13 Comparison of clustering methods in terms of 'Energy consumption'**

As shown in Figure 3.14, ESRPC1 have completed the cluster formation phase with less number of message overhead which is 250 bytes as compared to 450 bytes and 550 bytes encountered by ESRPC2 and ESRPC3 respectively.

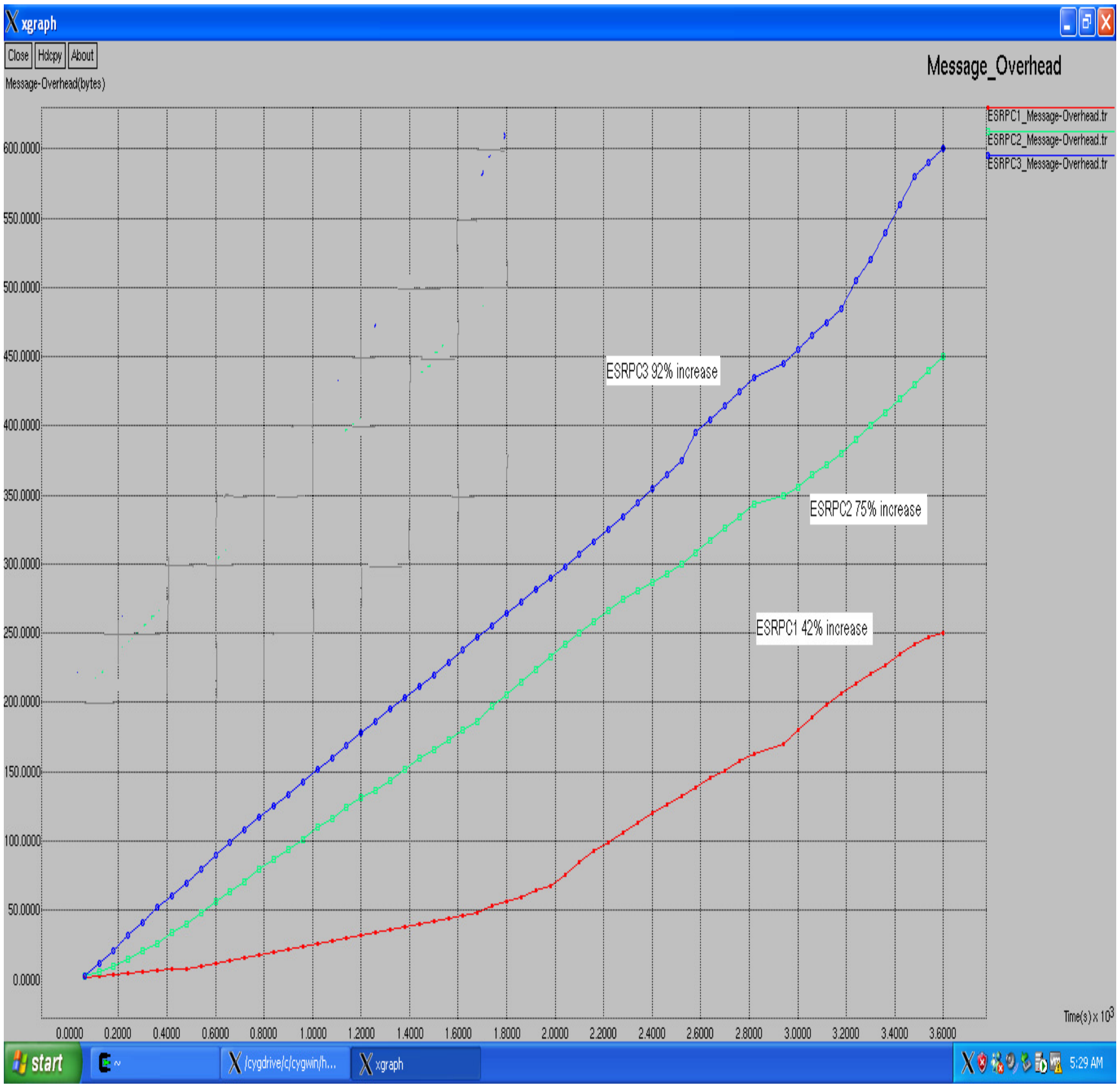

**Figure 3.14 Comparison of clustering methods in terms of 'Message overhead'**

By retaining more number of clusters at the end of simulation, as shown in Figure 3.15, ESRPC1 showed improved network life time compared to ESRPC2 and ESRPC3.Hence the centralized cluster formation approach have been justified and used in the proposed protocol while forming the clusters.

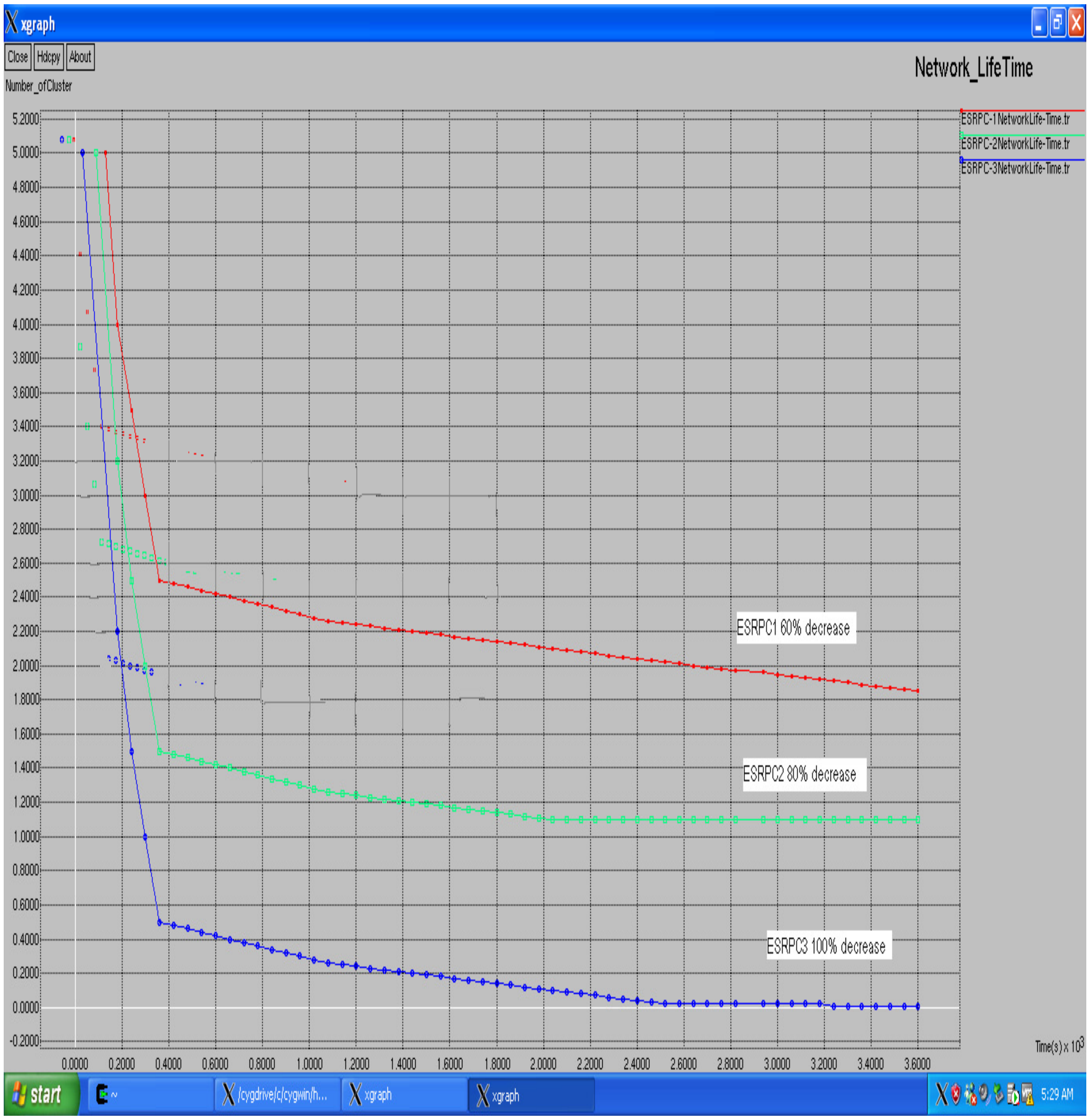

**Figure 3.15 Comparison of clustering methods in terms of 'Network life time'**

Table 3.3 summarizes the simulation parameters presented in the above mentioned figures.

**Table 3.3 Comparison of ESRP Centralized Cluster Formation with its Distributed Approaches**

| ESRP clustering method | Number of nodes alive | Energy consumption over simulation period (J) | Percentage Energy consumption | End to end delay (ms) | Message overhead (bytes) | Network life time (Number of clusters retained) |
|---|---|---|---|---|---|---|
| **ESRP C1** | 36 | 75 | 37.5 | 1800 | 250 | 2 |
| **ESRP C2** | 20 | 112 | 56 | 2800 | 450 | 1 |
| **ESRP C3** | 08 | 120 | 60 | 3200 | 550 | 0 |

Table 3.4 shows the Percentage variations realized in various simulation parameters during each cluster formation phase by the WSN.

**Table 3.4 Percentage variations realized in various simulation parameters during cluster formation phase**

| Cluster formation method | Percentage decrease in Number of nodes alive | Percentage increase in End to end delay | Percentage increase in Energy consumption | Percentage decrease Network life time | Percentage increase in Message overhead |
|---|---|---|---|---|---|
| **ESRP C1** | 64 | 0.05 | 60 | 60 | 42 |
| **ESRP C2** | 80 | 0.07 | 89.6 | 80 | 75 |
| **ESRP C3** | 92 | 0.08 | 96 | 100 | 92 |

As shown in Table 3.4 ESRPC1 exhibits better performance compared to ESRPC2 as well as ESRPC3.

### 3.7.2 Analyze the influence of ESRP security features

In the figures shown below, ESRP Clustering method (ESRPC) indicates the centralized cluster formation approach and ESRP indicates the protocol with the added security features.

Figure 3.16 shows the comparison between ESRP centralized cluster formation method and ESRP security features in terms of 'Number of nodes alive'.

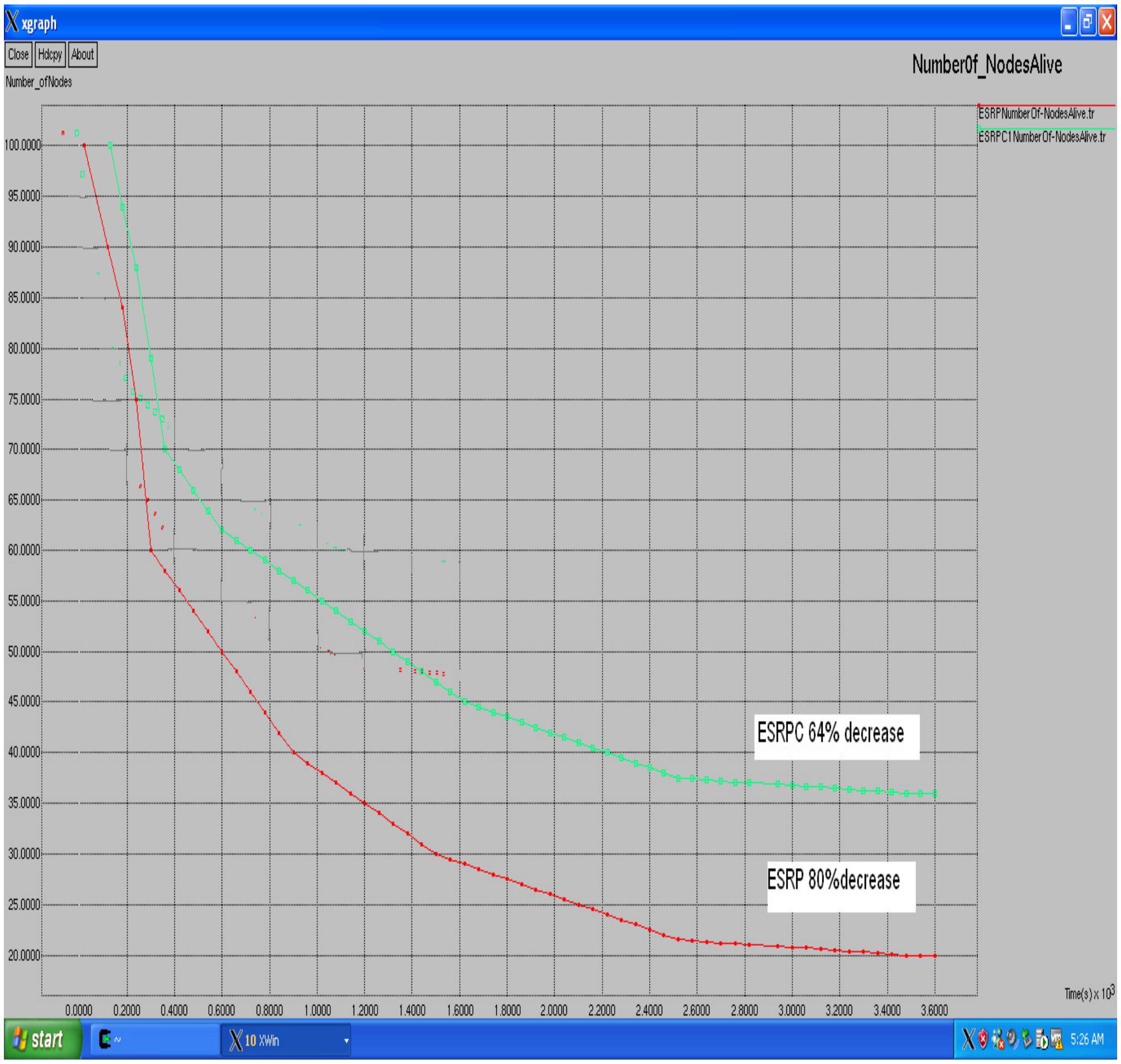

**Figure 3.16 Analyze the influence of ESRP security features in terms of 'Number of nodes alive'**

As shown in Figure 3.16, it has been observed that ESRPC can retain 35 sensor nodes alive at the end of simulation, compared to 20 alive nodes in ESRP. Since the number of nodes alive at the end of the simulation decides the life time of a WSN, it has been observed that a WSN with ESRPC have longer life time than ESRP.

As shown in Figure 3.17, it can be observed that ESRP could complete the simulated scenario in 2800 ms as apposed to 3200 ms by ESRPC.

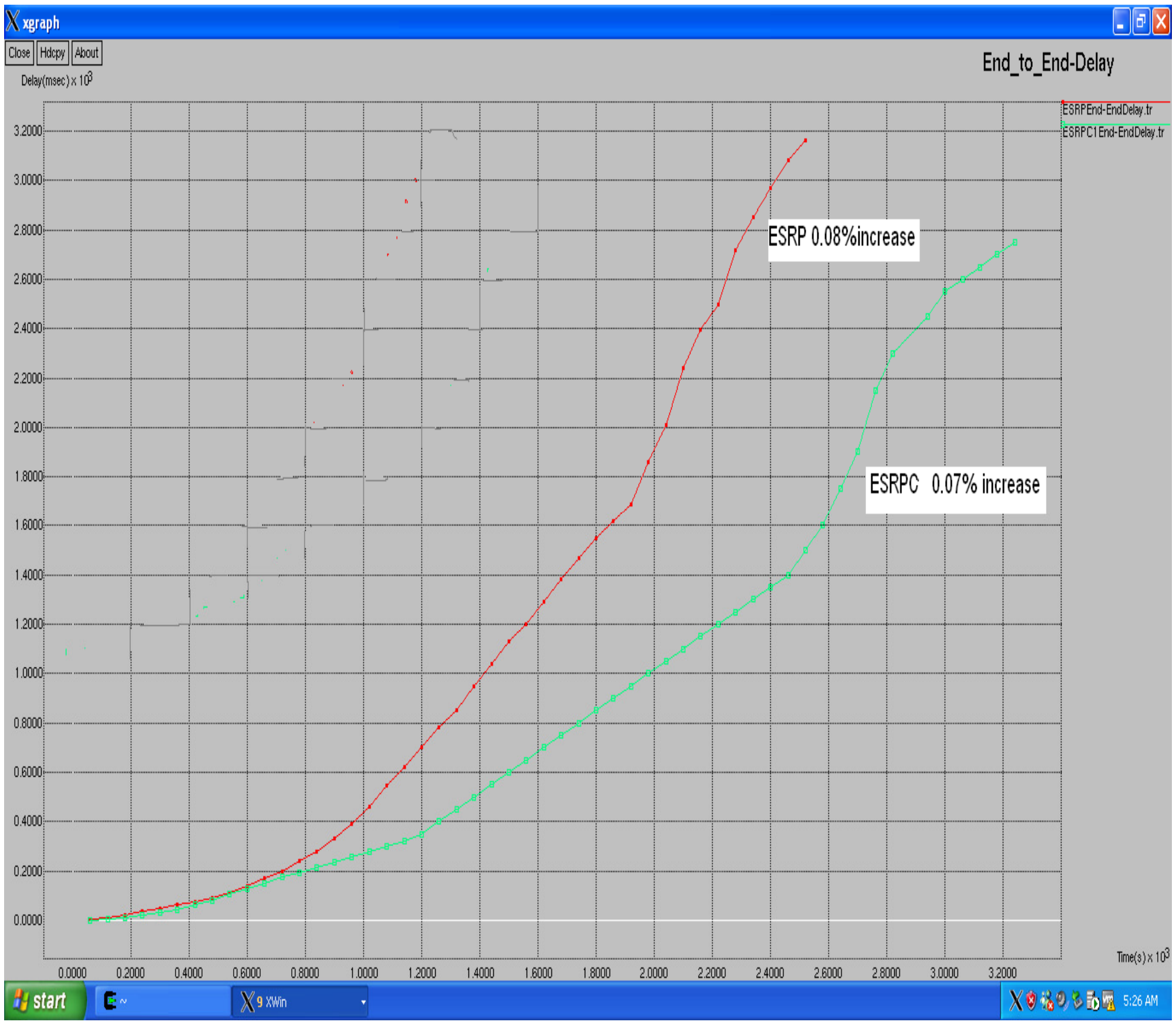

**Figure 3.17 Analyze the influence of ESRP security features in terms of 'End to end delay'**

As far as ‘Energy consumption’ is concerned, at the end of simulation period, as shown in Figure 3.18, the WSN with ESRPC consumed nearly 110J, which is 55% of the total 200J of energy available initially, whereas ESRP consumed 120J, which is 60 % of the total 200J of energy available initially.

Hence it has been observed that the WSN with ESRP consumed little higher energy, compared to ESRPC to deliver the data reliably.

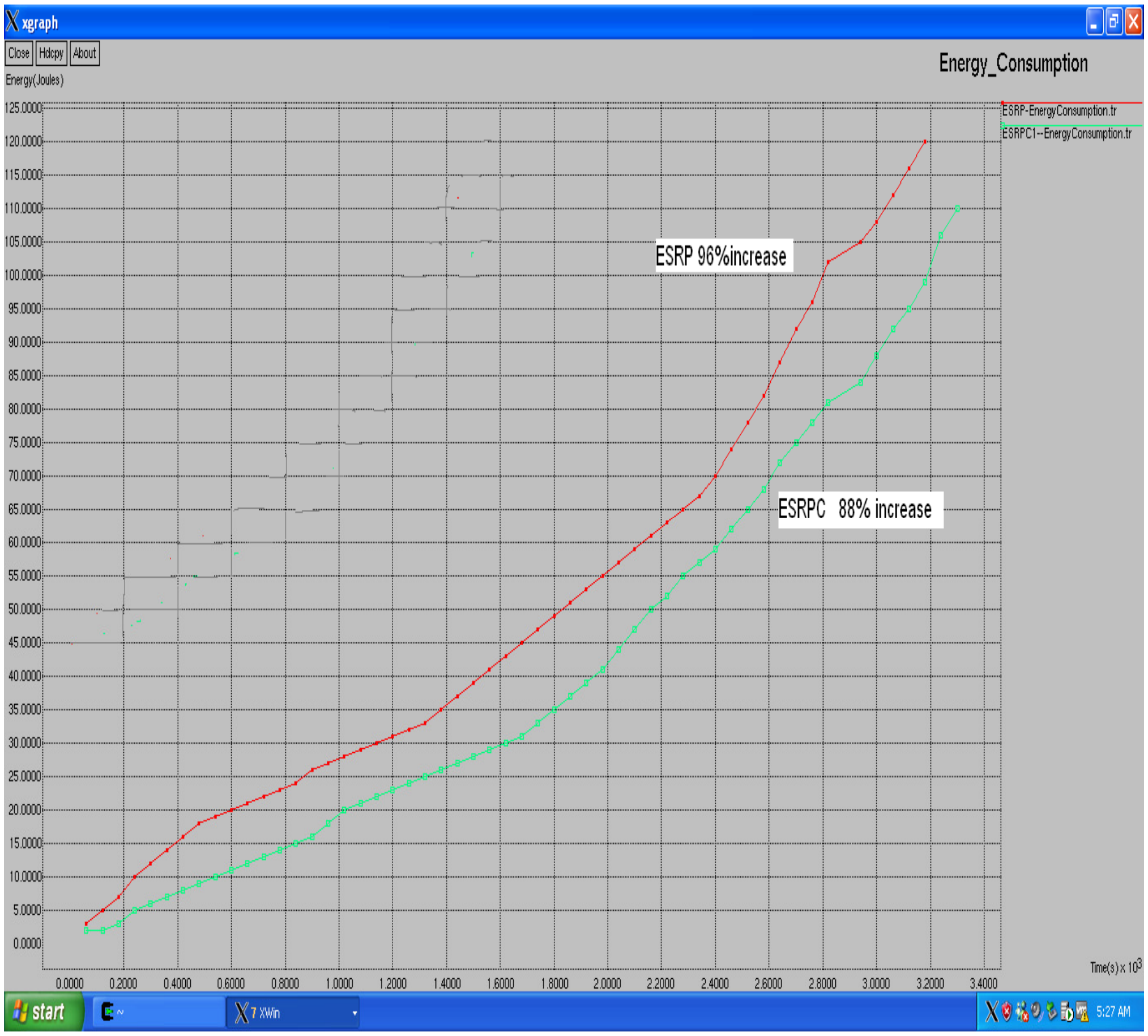

**Figure 3.18 Analyze the influence of ESRP security features in terms of ‘Energy consumption’**

Moreover with an additional 100 bytes of message overhead ESRP have managed to incorporate the light weight security features compared with ESRPC where no security methods have been incorporated as shown in Figure 3.19.

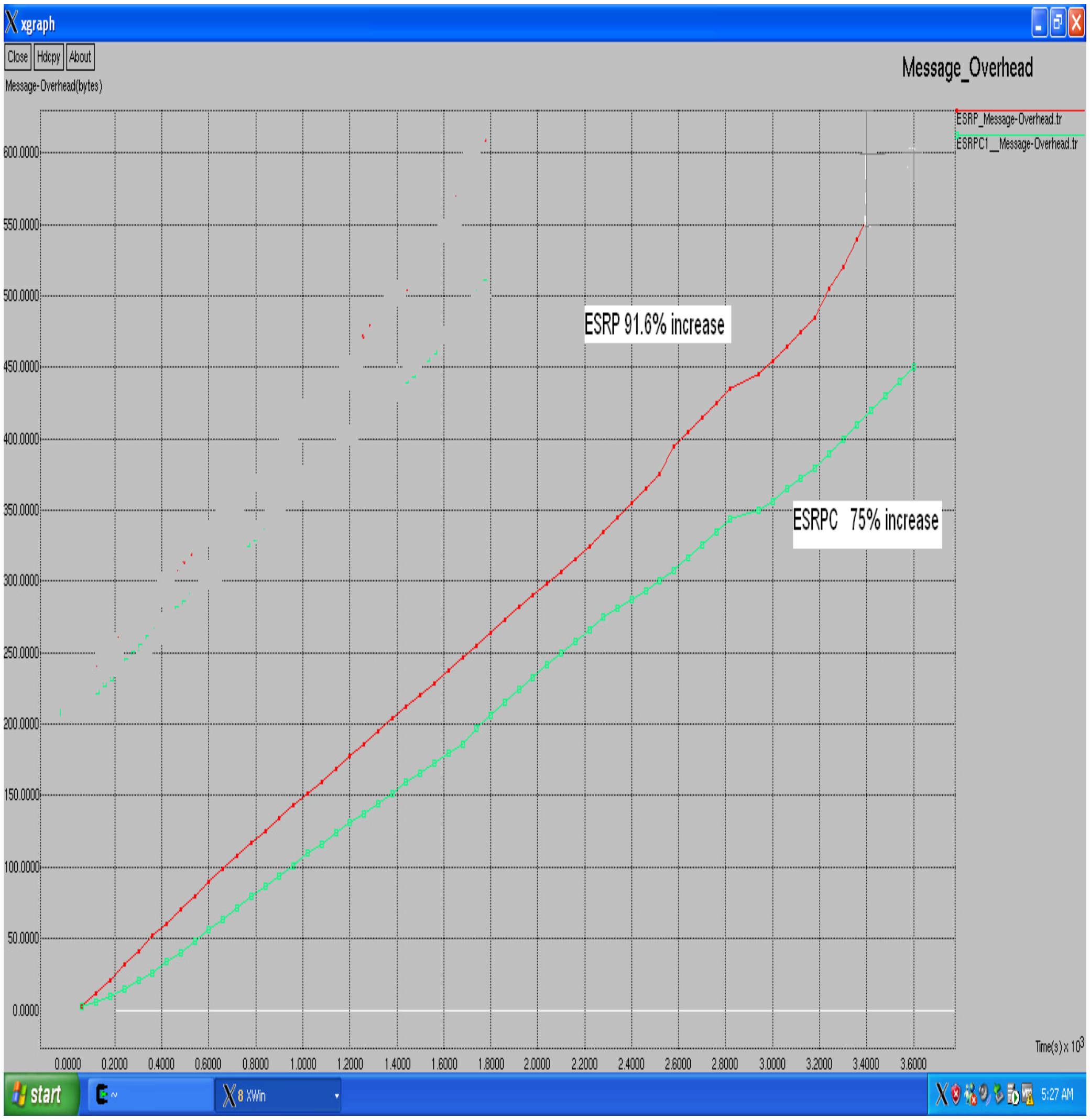

**Figure 3.19 Analyze the influence of ESRP security features in terms of 'Message overhead'**

Also it can be observed that there is no much difference between ESRP and ESRPC in terms of network life time as observed from Figure 3.20. Hence it can be concluded that the performance of ESRP have not been degraded by its light weight security features and they guaranteed reliable date delivery.

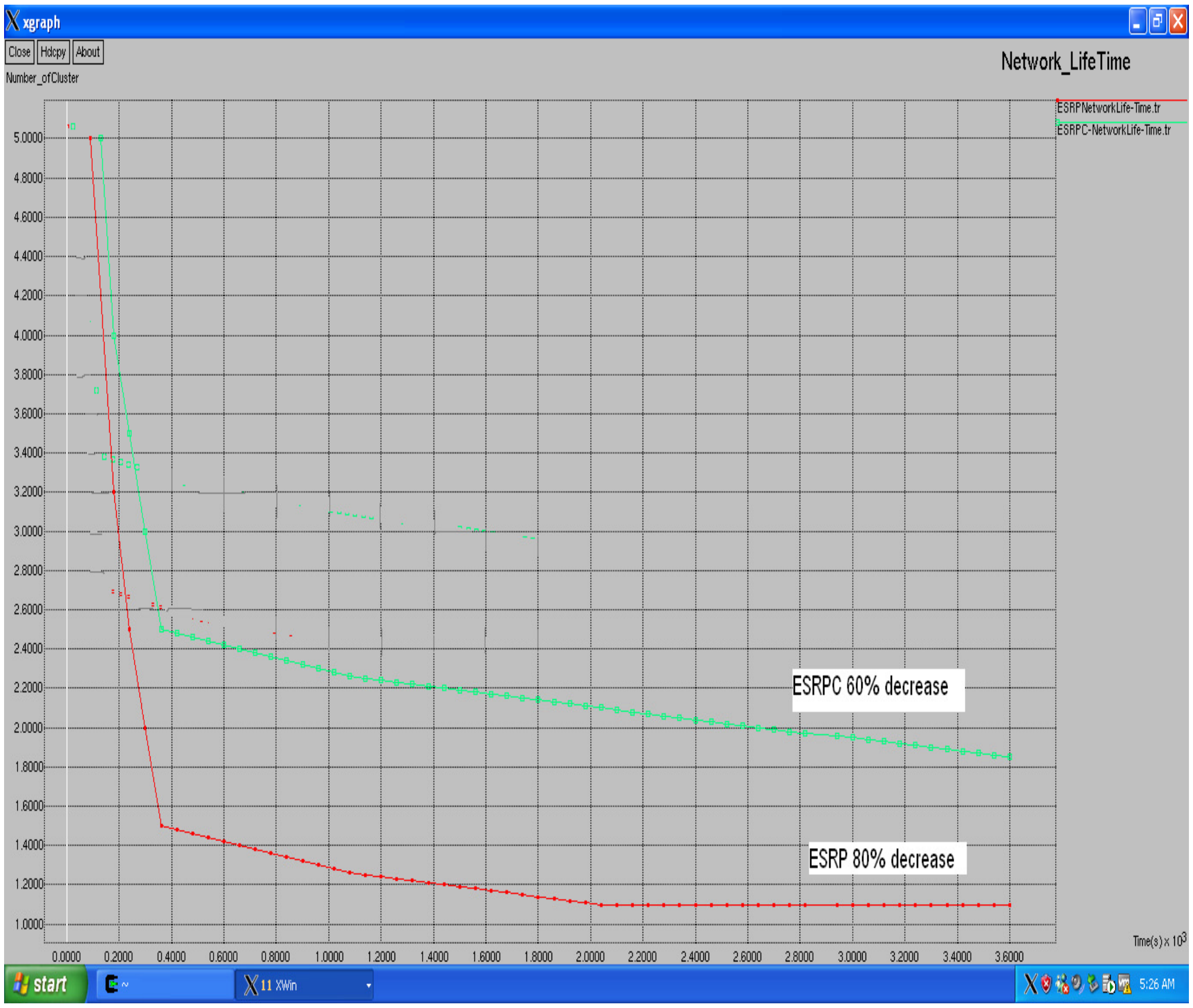

**Figure 3.20 Analyze the influence of ESRP security features in terms of 'Network lifetime'**

Table 3.5 summarizes the simulation parameters presented in the above mentioned figures.

**Table 3.5 Analyze the influence of ESRP security features**

| Metrics | Number of nodes alive | Energy consumption over simulation period (J) | Percentage Energy consumption | End to end delay (ms) | Message overhead (bytes) | Network life time (Number of clusters retained) |
|---|---|---|---|---|---|---|
| **ESRPC** | 35 | 110 | 55 | 2800 | 450 | 2 |
| **ESRP** | 20 | 120 | 60 | 3200 | 550 | 1 |

Table 3.6 shows the Percentage variations realized in various simulation parameters due to the influence of ESRP security features.

**Table 3.6 Percentage variations realized in various simulation parameters due to the influence of ESRP security features**

| Cluster formation method | Percentage decrease in Number of nodes alive | Percentage increase in End to end delay | Percentage increase in Energy consumption | Percentage decrease Network life time | Percentage increase in Message overhead |
|---|---|---|---|---|---|
| **ESRP C** | 60 | 0.07 | 88 | 60 | 75 |
| **ESRP** | 80 | 0.08 | 96 | 80 | 91.6 |

### 3.7.3 Comparison of ESRP with other Similar Protocols

In the figures shown ESRP have been marked with red color line and other protocols have been marked in green color line. Figure 3.21 shows the comparison between ESRP and LDTS in terms of ‘Number of nodes alive’. It has been observed that ESRP can retain 37 sensor nodes alive at the end of simulation, compared to 20 alive nodes in LDTS.

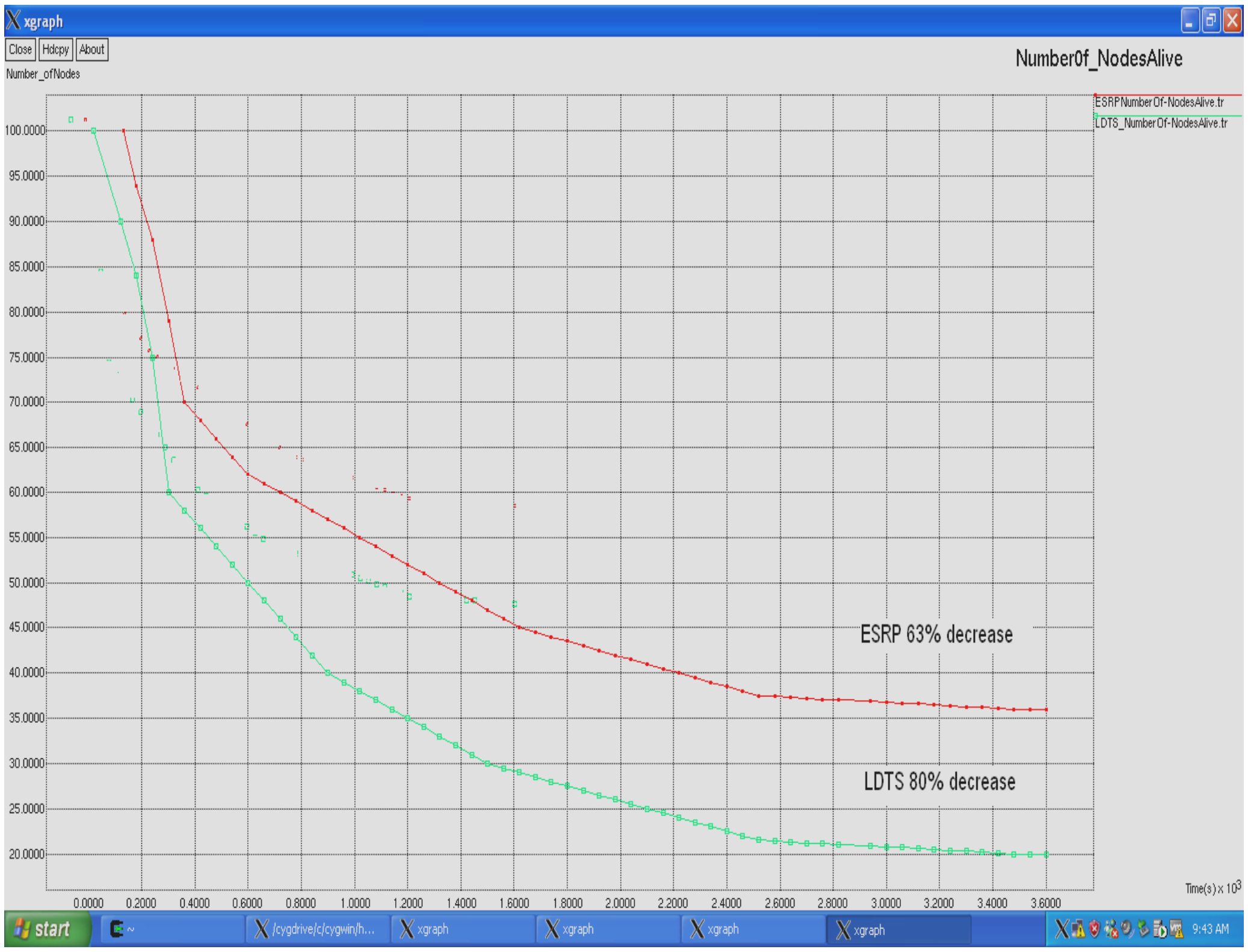

**Figure 3.21 Comparison of ESRP and LDTS in terms of 'Number of nodes alive'**

Figure 3.22 shows the comparison between ESRP and SLEACH in terms of 'Number of nodes alive'. It has been observed that ESRP can retain 37 sensor nodes alive at the end of simulation, compared to 30 alive nodes in SLEACH.

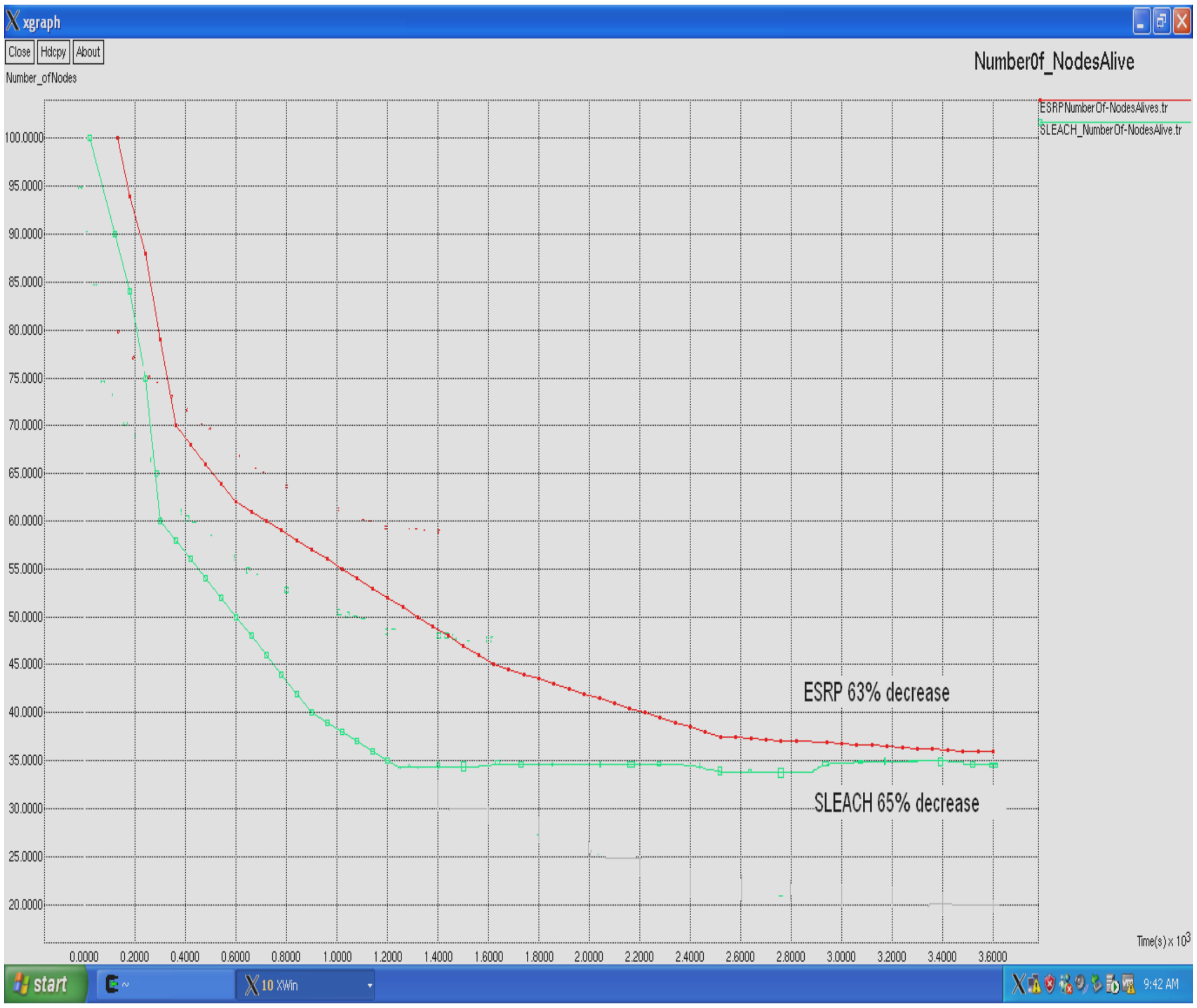

**Figure 3.22 Comparison of ESRP and SLEACH in terms of 'Number of nodes alive'**

Figure 3.23 shows the comparison between ESRP and SEDR in terms of 'Number of nodes alive'. It has been observed that ESRP can retain 37 sensor nodes alive at the end of simulation, compared to 26 alive nodes in SEDR.

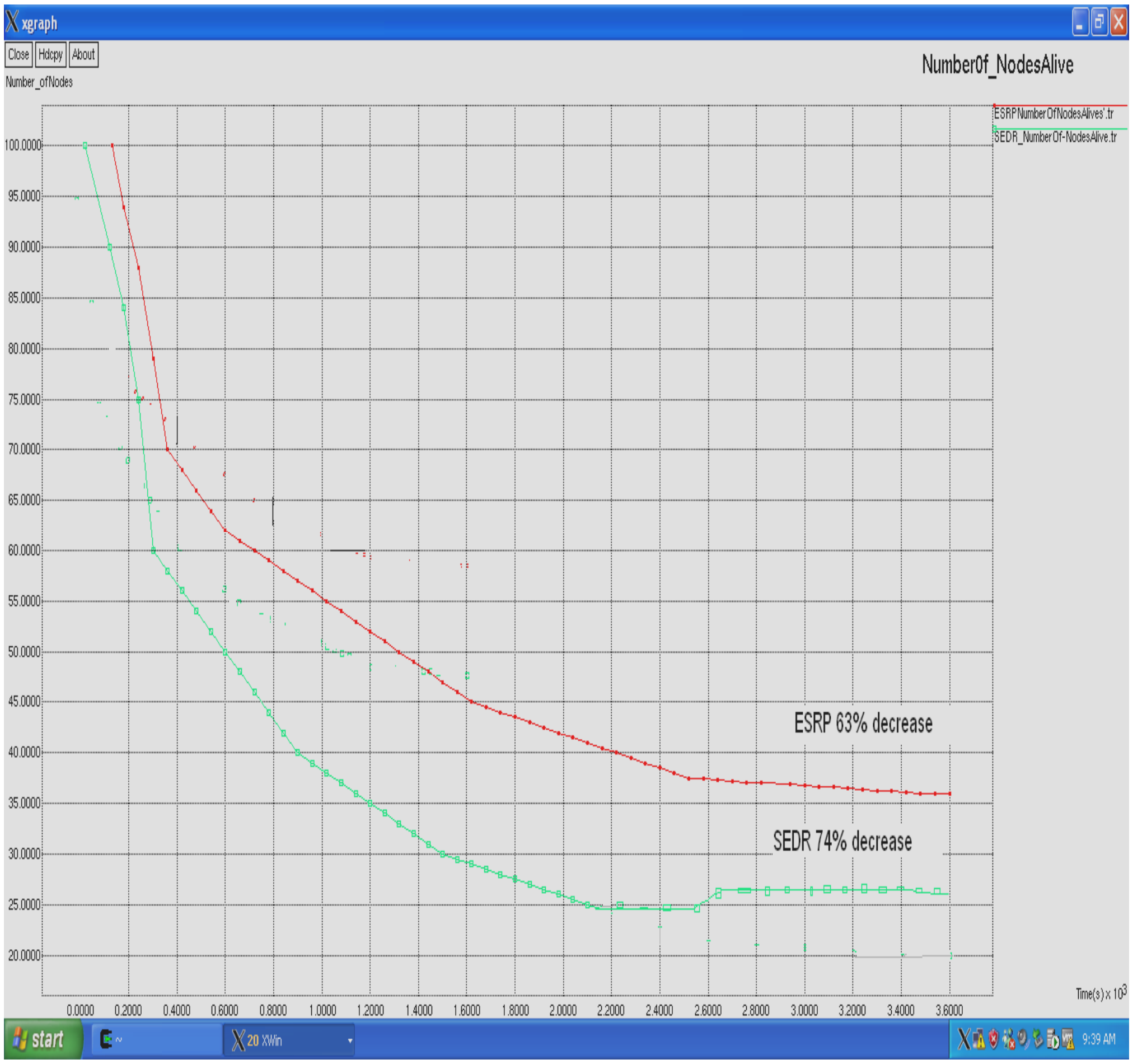

**Figure 3.23 Comparison of ESRP and SEDR in terms of 'Number of nodes alive'**

Figure 3.24 shows the comparison between ESRP and SERP in terms of 'Number of nodes alive'. SERP could retain only 10 alive nodes at the end of the simulation scenario where as ESRP could retain 37 sensor nodes alive at the end of simulation.

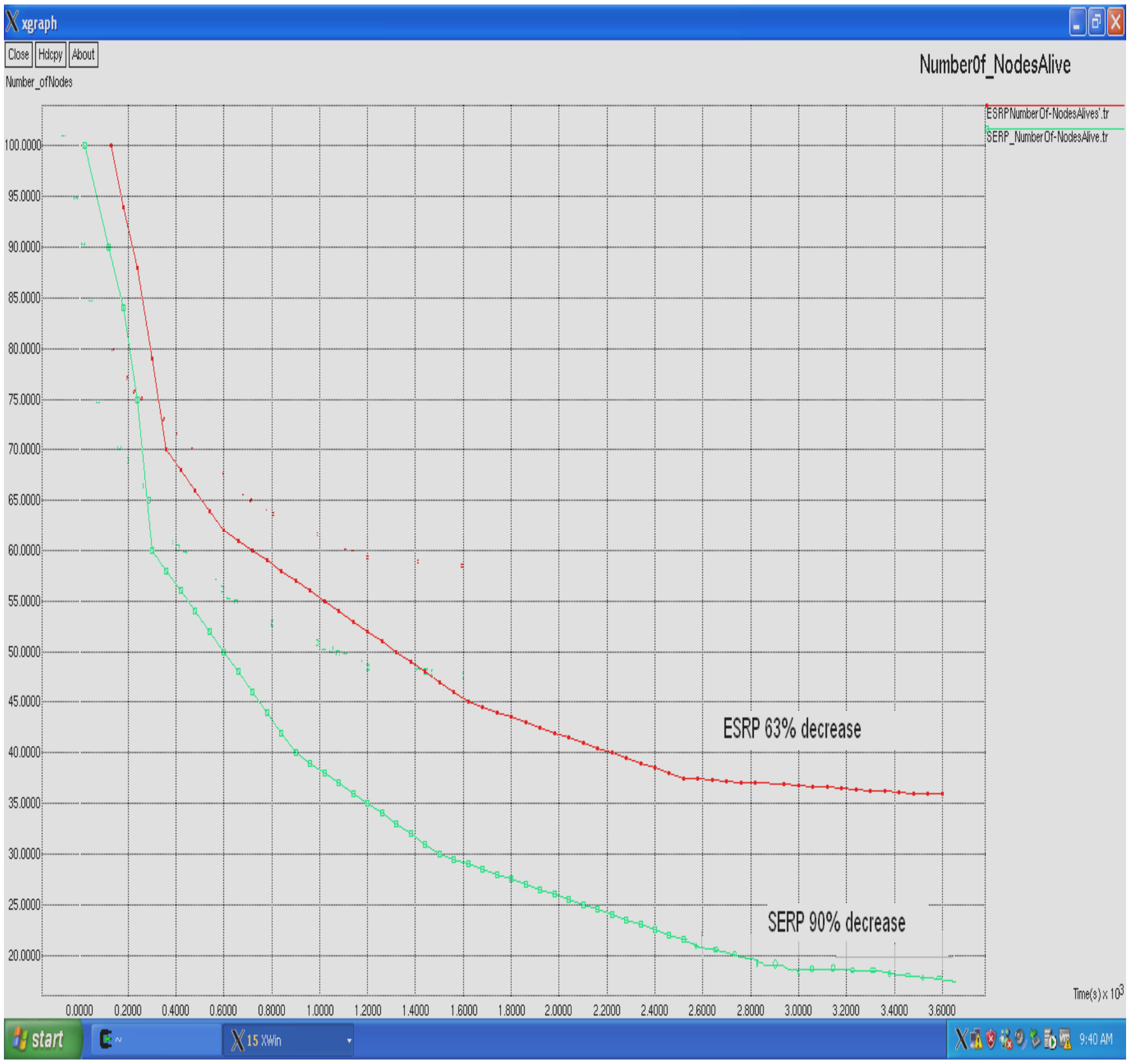

**Figure 3.24 Comparison of ESRP and SERP in terms of 'Number of nodes alive'**

Since the number of nodes alive at the end of the simulation decides the life time of a WSN, it has been observed that a WSN with ESRP have longer life time than the other protocols used in the simulation. Also it has been observed that SERP exhibits poor performance among the protocols compared, due to its poor multi path structure.

Figure 3.25 shows the comparison between ESRP and LDTS in terms of ‘Energy consumption’ in Joules. The WSN with ESRP consumed nearly 72J, which is 36% whereas LDTS consumed 192J, which is 96 % of the total 200J of energy available initially.

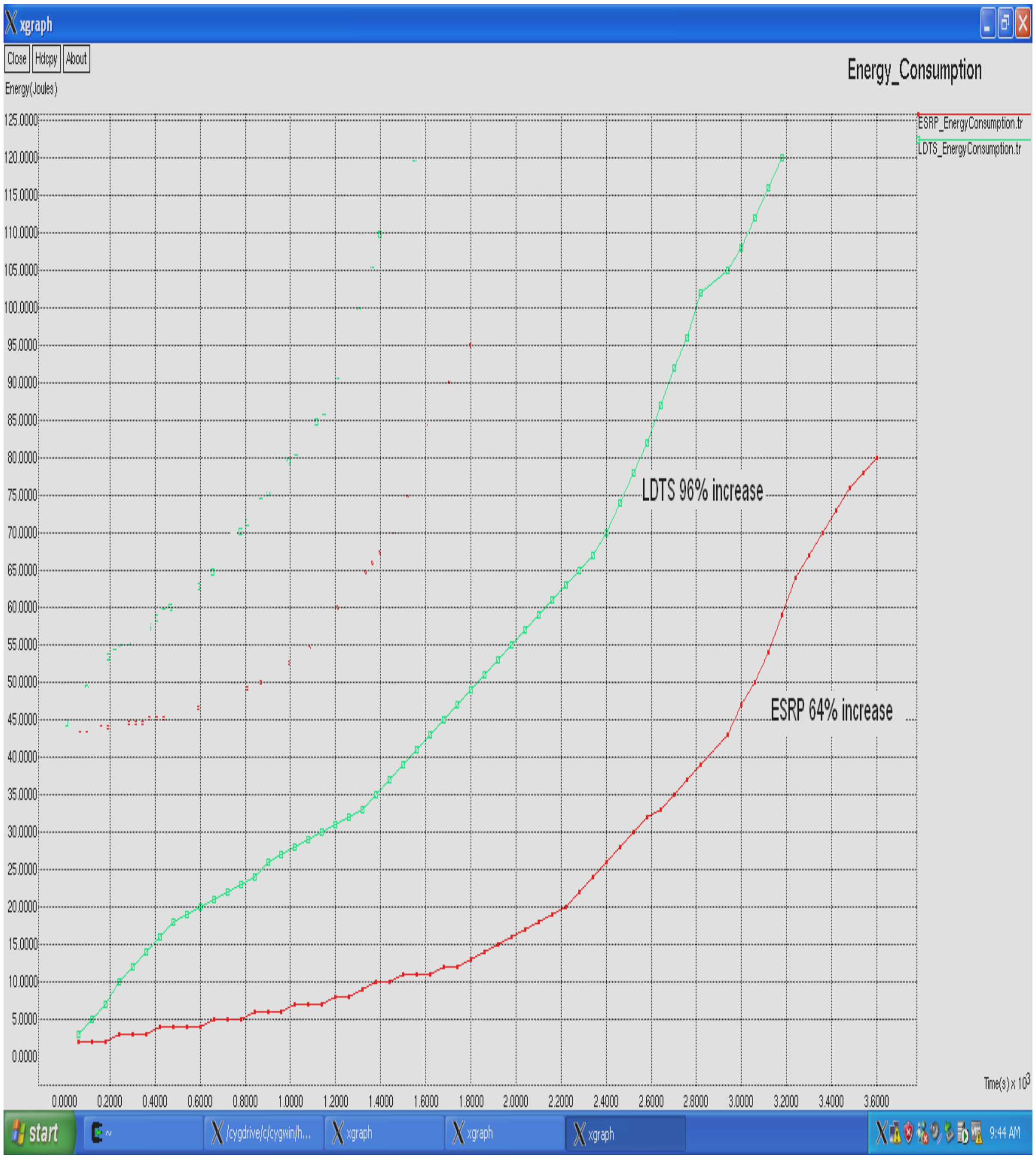

**Figure 3.25 Comparison of ESRP and LDTS in terms of ‘Energy consumption’**

Figure 3.26 shows the comparison between ESRP and SLEACH in terms of ‘Energy consumption’ in Joules. The WSN with ESRP consumed nearly 72J, which is 36% whereas SLEACH consumed 144J which is 72 % of the total 200J of energy available initially.

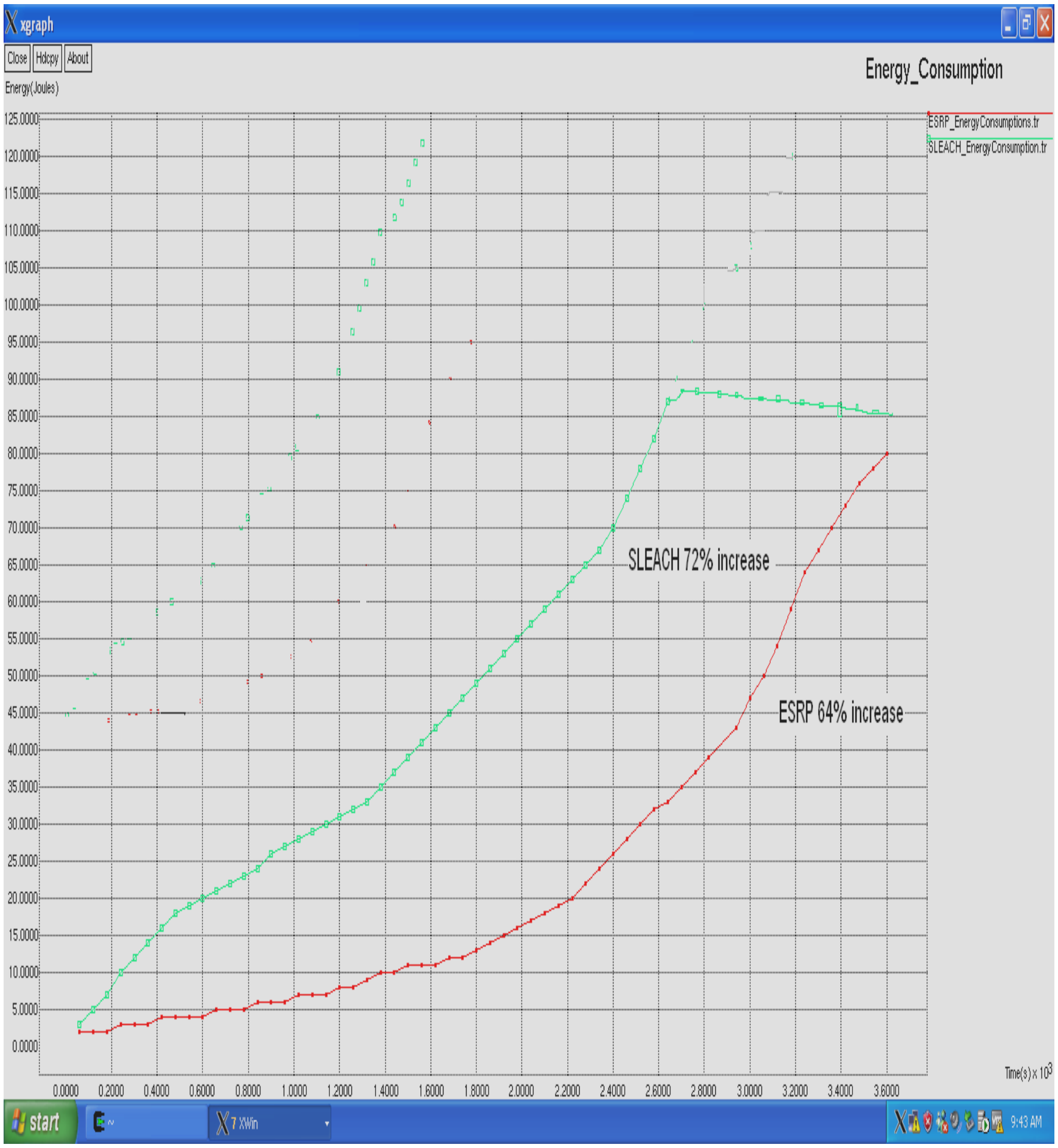

**Figure 3.26 Comparison of ESRP and SLEACH in terms of ‘Energy consumption’**

Figure 3.27 shows the comparison between ESRP and SEDR in terms of 'Energy consumption' in Joules. The WSN with ESRP consumed nearly 72J, which is 36% whereas SEDR consumed 164J which is 82% of the total 200J of energy available initially.

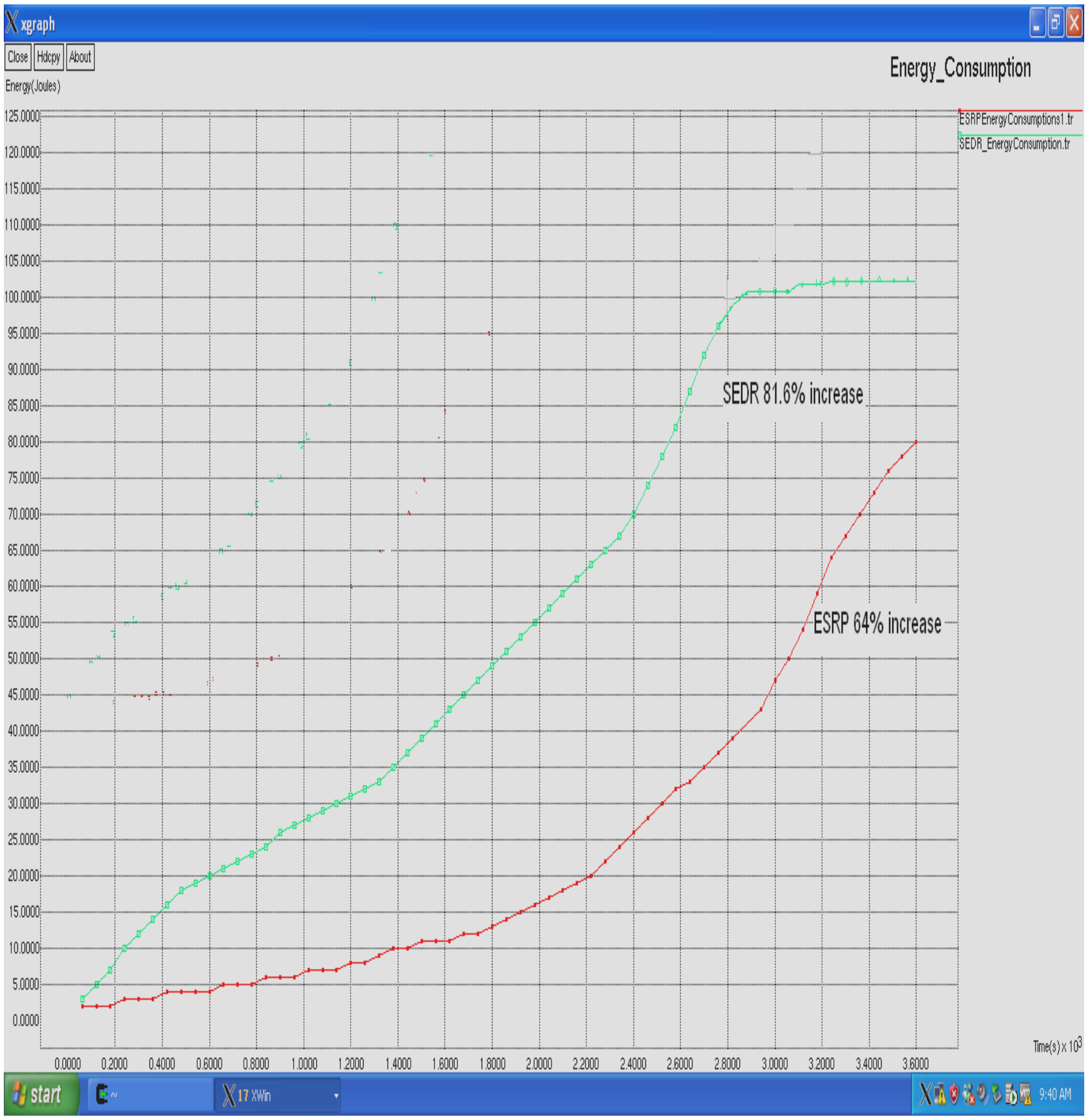

**Figure 3.27 Comparison of ESRP and SEDR in terms of 'Energy consumption'**

Figure 3.28 shows the comparison between ESRP and SERP in terms of 'Energy consumption' in Joules. The WSN with ESRP consumed nearly 72J, which is 36% whereas SERP consumed 184J which is 92% of the total 200J of energy available initially.

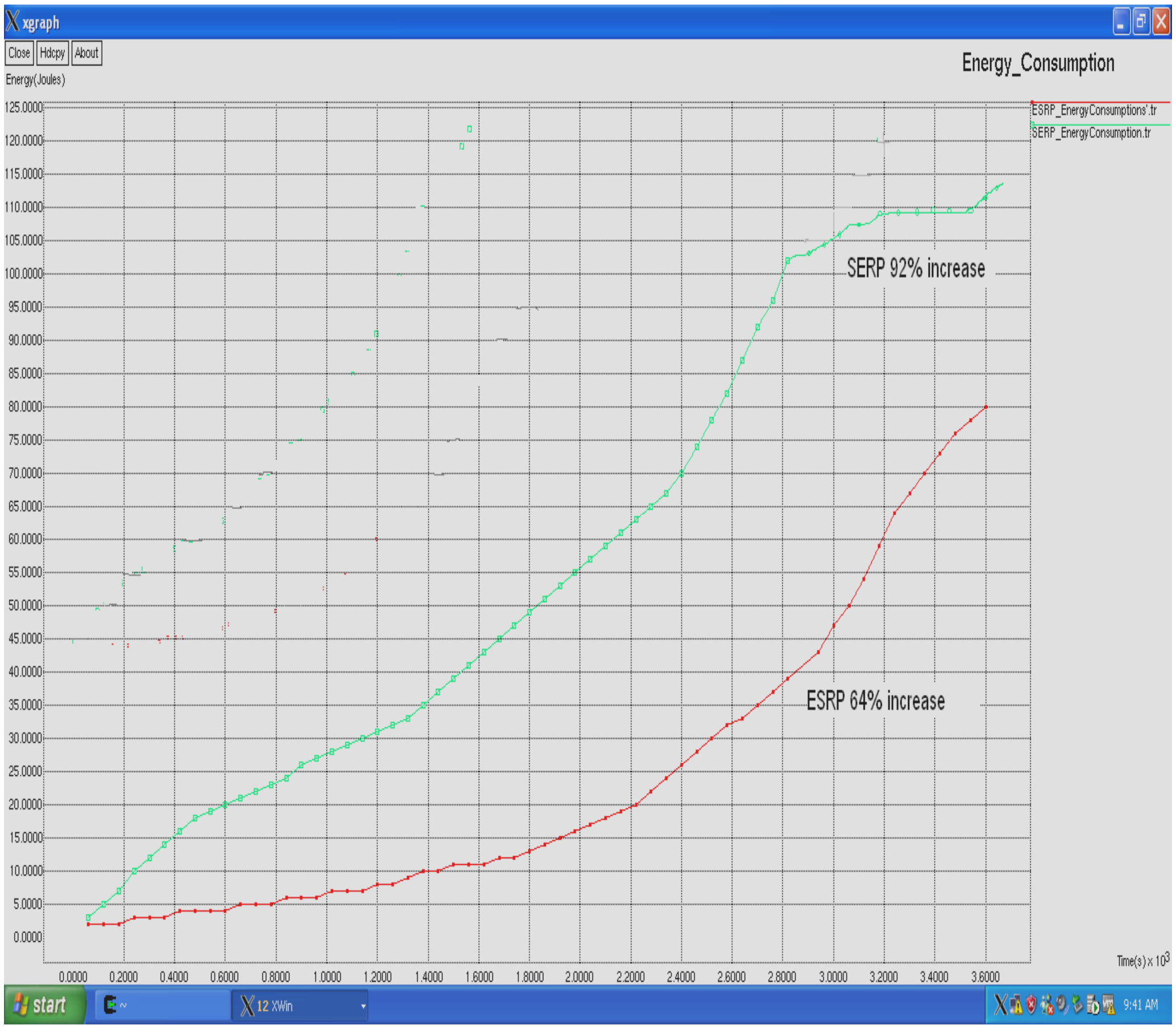

**Figure 3.28 Comparison of ESRP and SERP in terms of 'Energy consumption'**

Hence it has been observed that the WSN with ESRP consumed much less energy, whereas LDTS and SERP consumed much higher energy among the protocols compared.

The simulation is carried out for 3600 seconds. As far as delay is concerned, out of the five iterations carried out during the simulation period, ESRP have taken 2800 ms whereas LDTS have taken 3200 ms to complete the data transfer as shown in Figure 3.29.

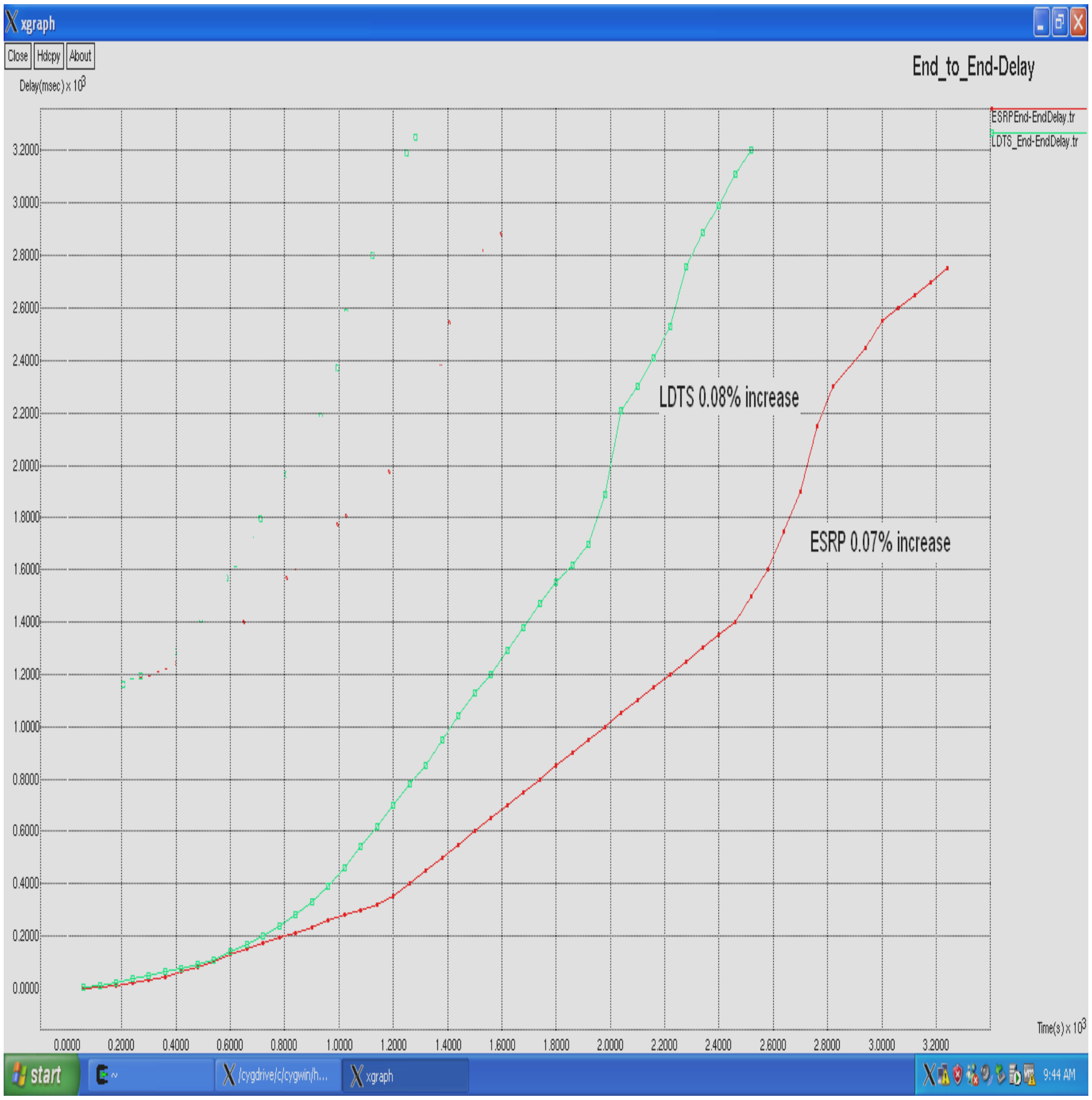

**Figure 3.29 Comparison of ESRP and LDTS in terms of 'End to end delay'**

As shown in Figure 3.30 in comparison with ESRP the SLEACH protocol have taken 2400 ms to complete the data transfer.

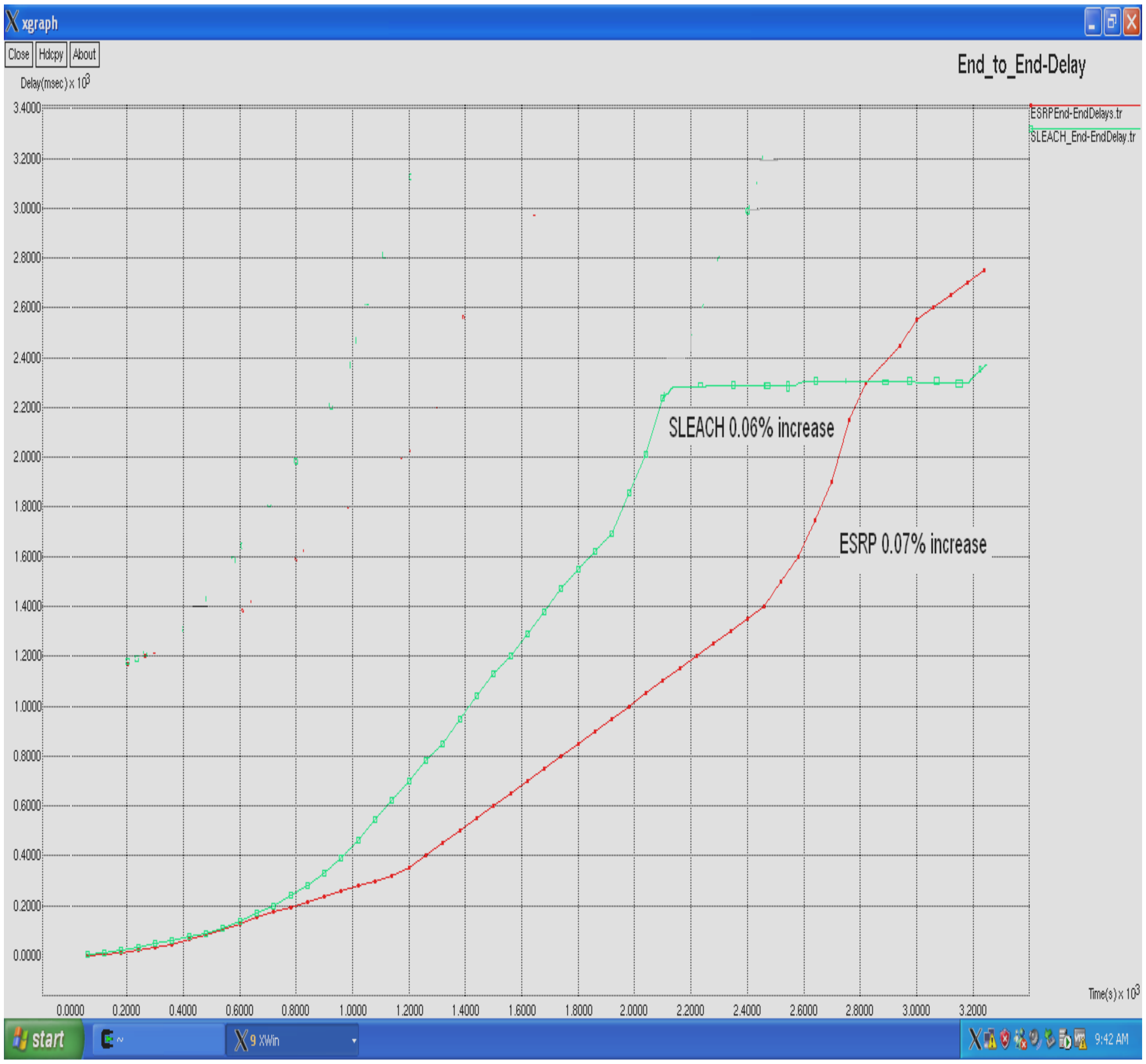

**Figure 3.30 Comparison of ESRP and SLEACH in terms of 'End to end delay'**

As shown in Figure 3.31 in comparison with ESRP the SEDR protocol have taken 3000 ms to complete the data transfer.

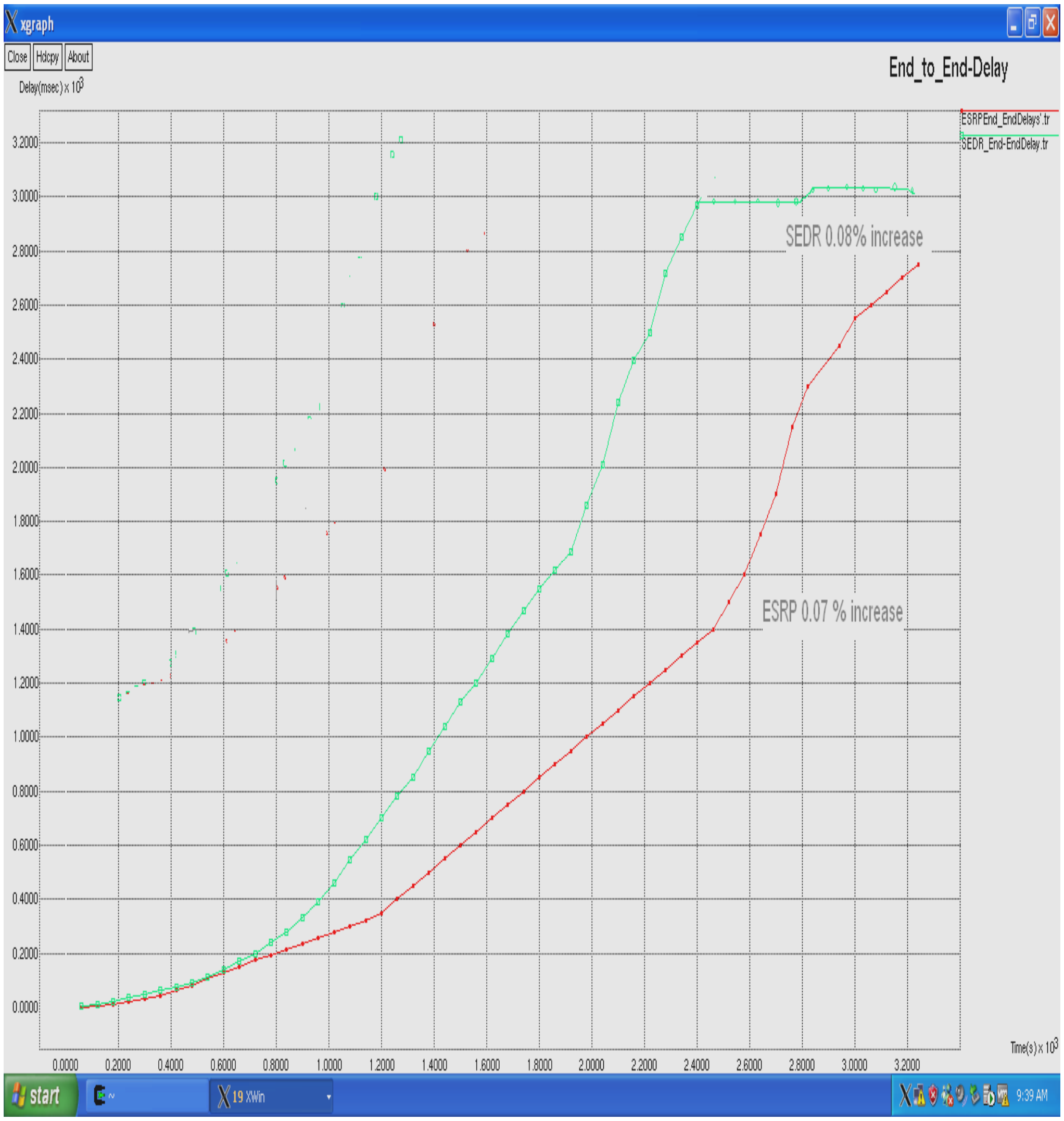

**Figure 3.31 Comparison of ESRP and SEDR in terms of 'End to end delay'**

As shown in Figure 3.32 in comparison with ESRP the SERP protocol have taken 3400 ms to complete the data transfer. The total percentage of time taken is 0.07%, 0.08%, 0.06%, 0.08% and 0.09% out of the total simulation time by ESRP, LTDS, SLEACH, SEDR, and SERP respectively. Hence ESRP exhibited better performance in terms end to end delay.

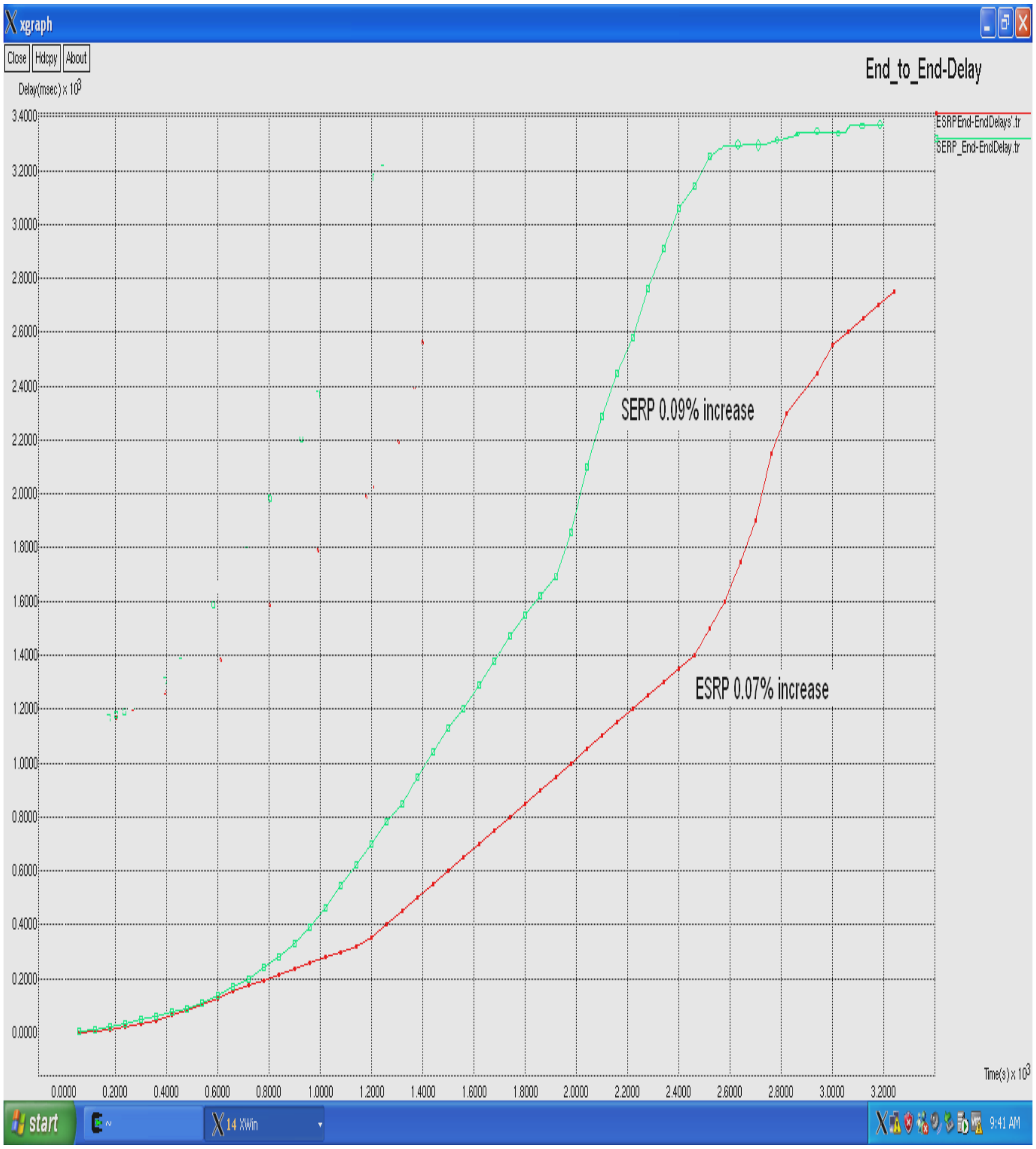

**Figure 3.32 Comparison of ESRP and SERP in terms of 'End to end delay'**

As far as message overhead is concerned ESRP had encountered much less overhead which is 450 bytes as compared to 600 bytes for LDTS as shown in Figure 3.33.

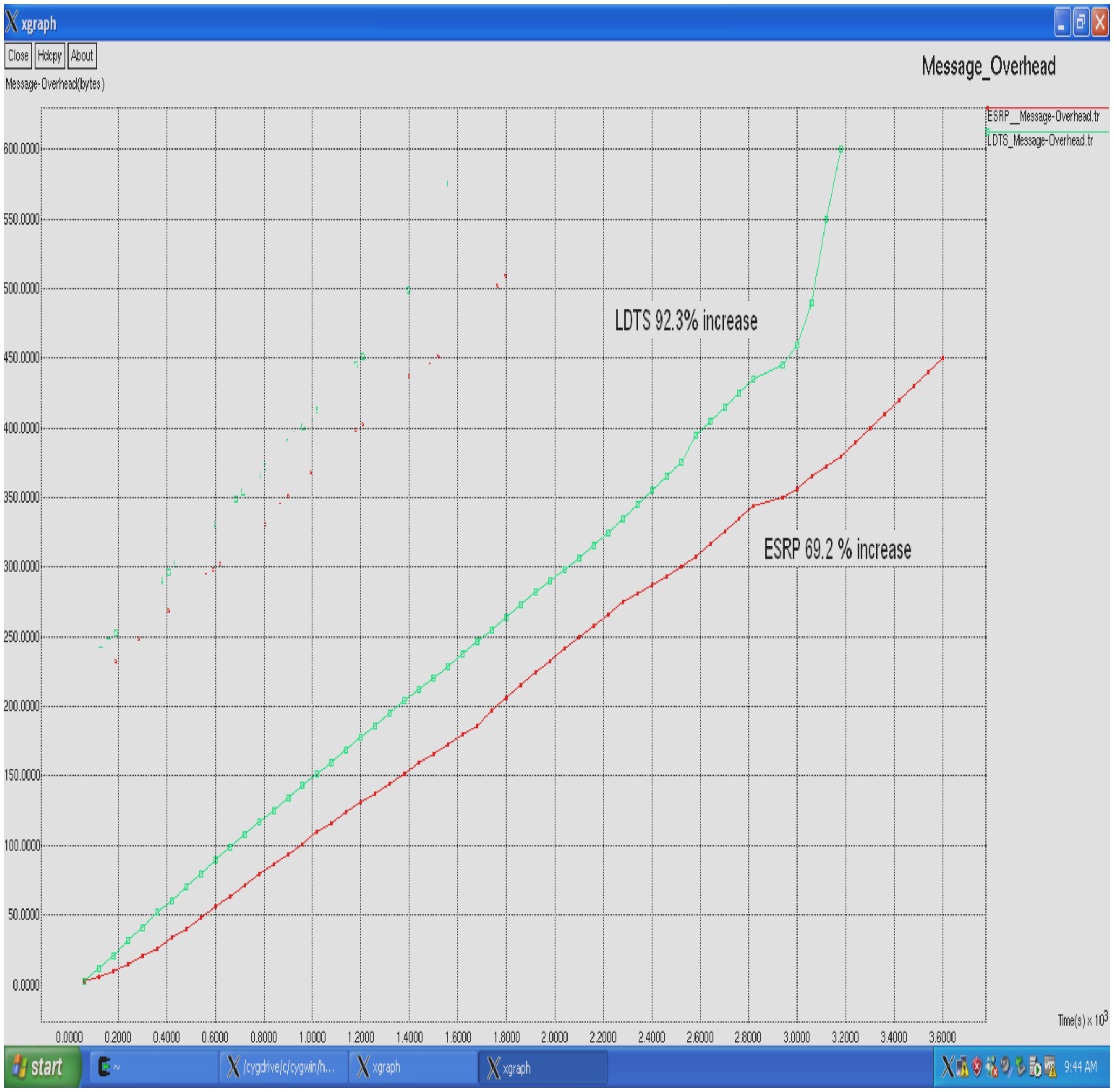

**Figure 3.33 Comparison of ESRP and LDTS in terms of 'Message overhead'**

The superiority of ESRP over LDTS can be shown in terms communication overhead which has been calculated as follows.

Let 'm' is the number of clusters and 'n' is the cluster size including the CH. It has been assumed that each CH needs to collect feedback reports from its CM, and the sink has to collect feedback reports from its CH.

During intra cluster data gathering phase, ‘n-1’ requests has been sent and ‘n-1’ responses has been received, whenever a CH wants to collect data from its ‘n-1’ CM. Let $C_{inraESRP}$ is the total number overhead packets encountered during the intra cluster routing phase of ESRP. It has been calculated as follows.

$$C_{intraESRP} = 2 * n\text{-}1 \text{ packets} \tag{3.21}$$

A single feedback request has been sent, during the inter cluster communication phase, whenever a CH wants to interact with another CH. Thus, the communication overhead is two packets. Also it has been noticed that, the sink has sent one request packet to each CH and received with one response packet from each CH, during the data collection. Hence for ‘m’ clusters, the total communication overhead is 2m packets.

Let $C_{interESRP}$ is the total number overhead packets encountered during the inter cluster routing phase of ESRP. It has been calculated as follows.

$$C_{interESRP} = 2 + 2 * m \text{ packets} \tag{3.22}$$

Let $C_{totalESRP}$ is the total number overhead packets encountered during both intra as well as inter cluster routing phases of ESRP.It has been calculated as follows

$$C_{totalESRP} = m * C_{intraESRP} + C_{interESRP} \tag{3.23}$$

With reference to the Table 3.2 as well as the uniform distribution assumption about the distribution of CM, as specified in the ESRP algorithm, it has been identified that the value of ‘m’ is 5 and the value ‘n’ is 20 for ESRP.

Hence the values of $C_{intraESRP}$ ,$C_{interESRP}$ and $C_{totalESRP}$ have been calculated using Equation (3.21) to Equation (3.23) respectively as follows

$C_{intraESRP}$ = 38 packets

$C_{interESRP}$ = 12 packets

$C_{totalESRP}$ = 202 packets

Let $C_{intraLDTS}$ is the total number overhead packets encountered during the intra cluster routing phase, $C_{interLDTS}$ is the total number overhead packets encountered during the inter cluster routing phase and $C_{totalLDTS}$ is the total number overhead packets encountered during both intra as well as inter cluster routing phases of LDTS.

These parameters have been calculated based on the literature as follows

$$C_{intraLDTS} = 2 * (n-2) * (n-1) + 2n \text{ packets} \quad (3.24)$$

$$C_{inter\ LDTS} = 2 * (m-1)^2 + 2m \text{ packets} \quad (3.25)$$

$$C_{totalLDTS} = m * C_{intraLDTS} + C_{interLDTS} \quad (3.26)$$

Using the same simulation set up as shown in Table 3.2, the values of $C_{intraLDTS}$ ,$C_{interLDTS}$ and $C_{totalLDTS}$ have been calculated using Equation (3.24) to Equation (3.26) respectively as follows

$C_{intraLDTS}$ = 233968 packets

$C_{interLDTS}$ = 42 packets

$C_{totalLDTS}$ = 11,698,82 packets

Based on the above calculations, it has been concluded that ESRP is highly suitable for small scale WSN, with small number of clusters and large size of clusters thus outperforming LDTS.

As shown in Figure 3.34, the total overhead of 480 bytes in SLEACH, resulted in an excess of 30 bytes of overhead in comparison with ESRP.

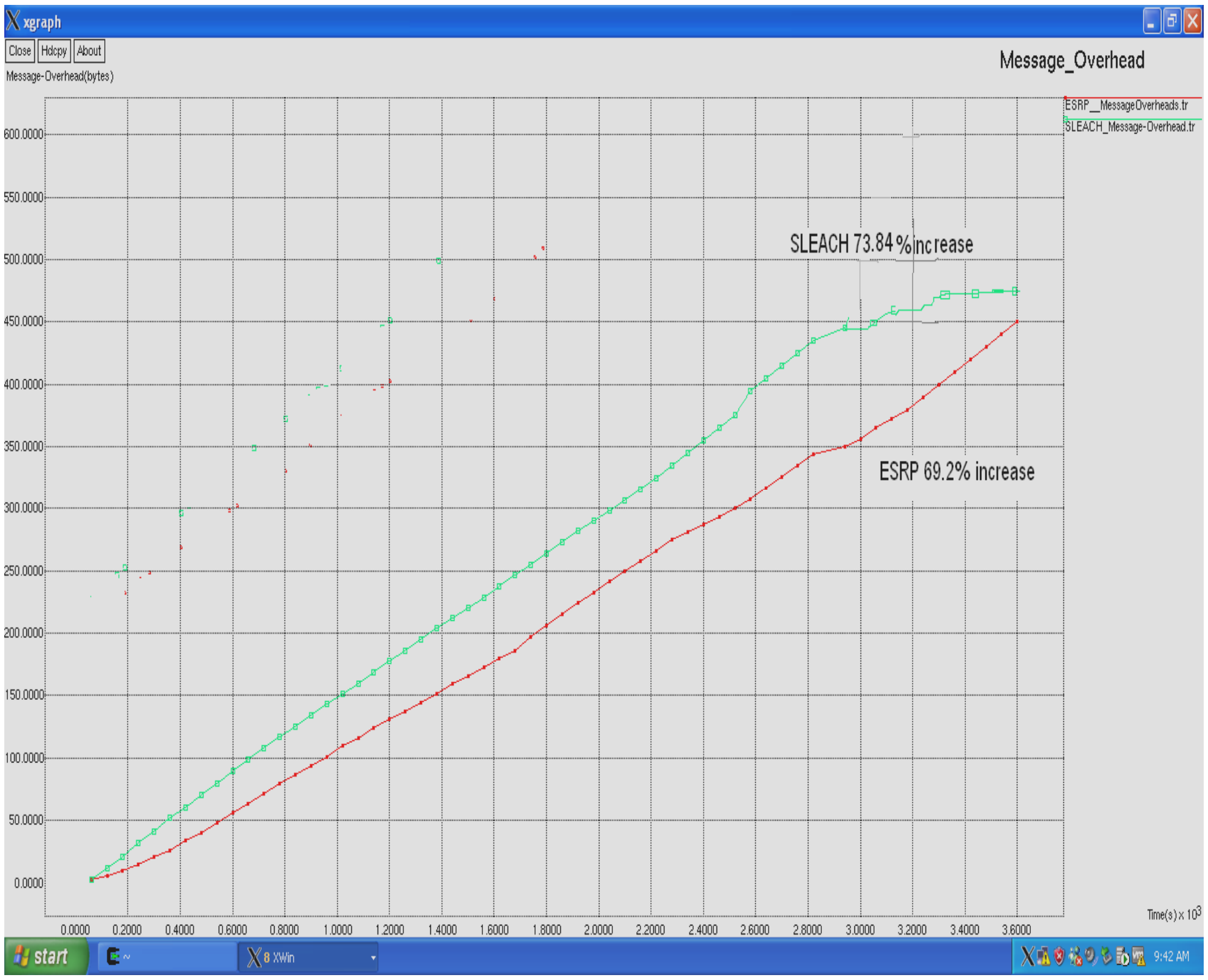

**Figure 3.34 Comparison of ESRP and SLEACH in terms of 'Message overhead'**

As shown in Figure 3.35, SEDR protocol spent nearly 525 bytes which is an additional 75 bytes of overhead in comparison with ESRP, to transfer the same amount of data during the simulation time.

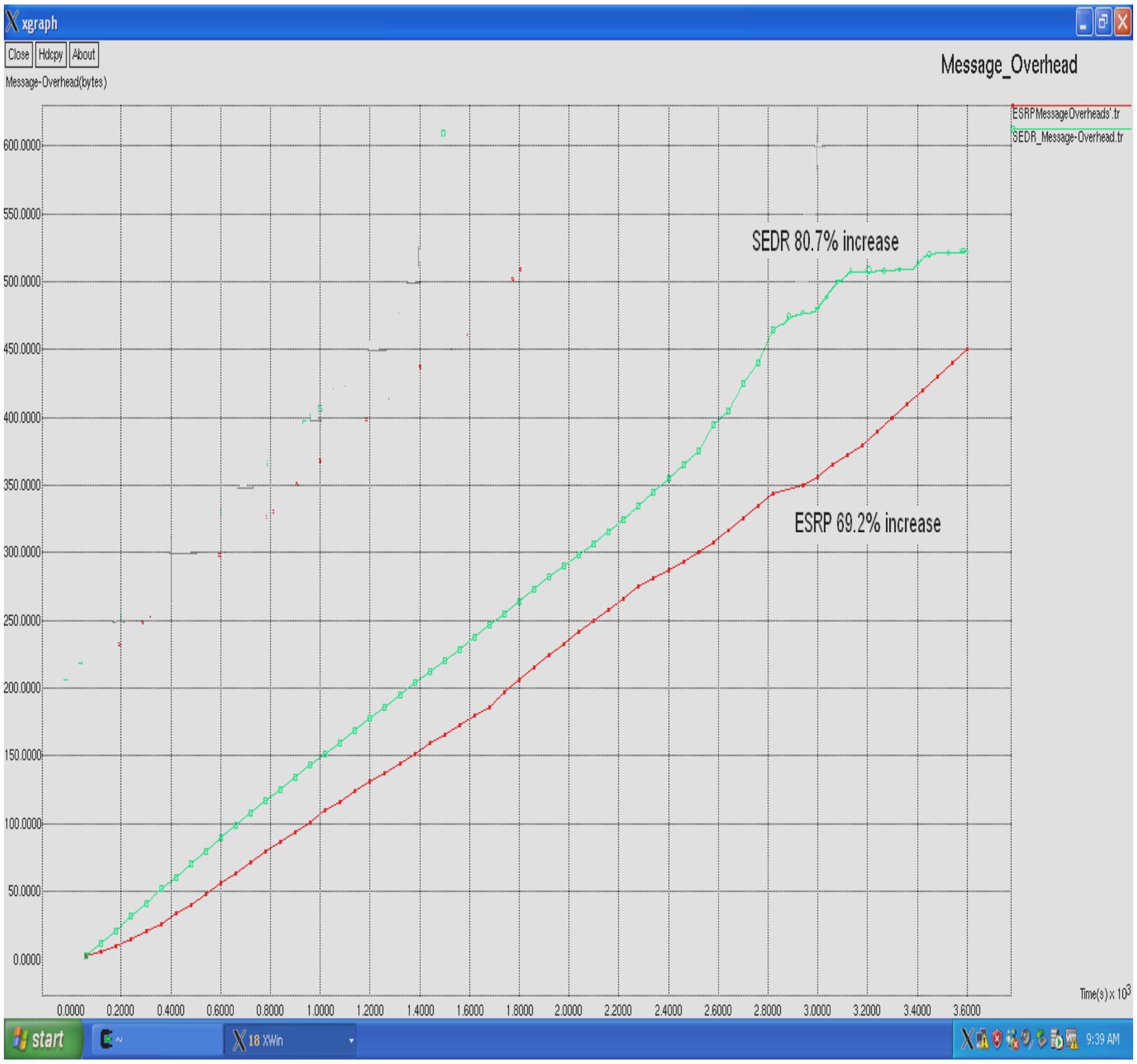

**Figure 3.35 Comparison of ESRP and SEDR in terms of 'Message overhead'**

Figure 3.36 shows the comparison of ESRP and SERP in terms of message overhead. A shown in the figure, SERP spent 630 bytes of overhead which is nearly 96.9% of the maximum 650 bytes during the simulation period.

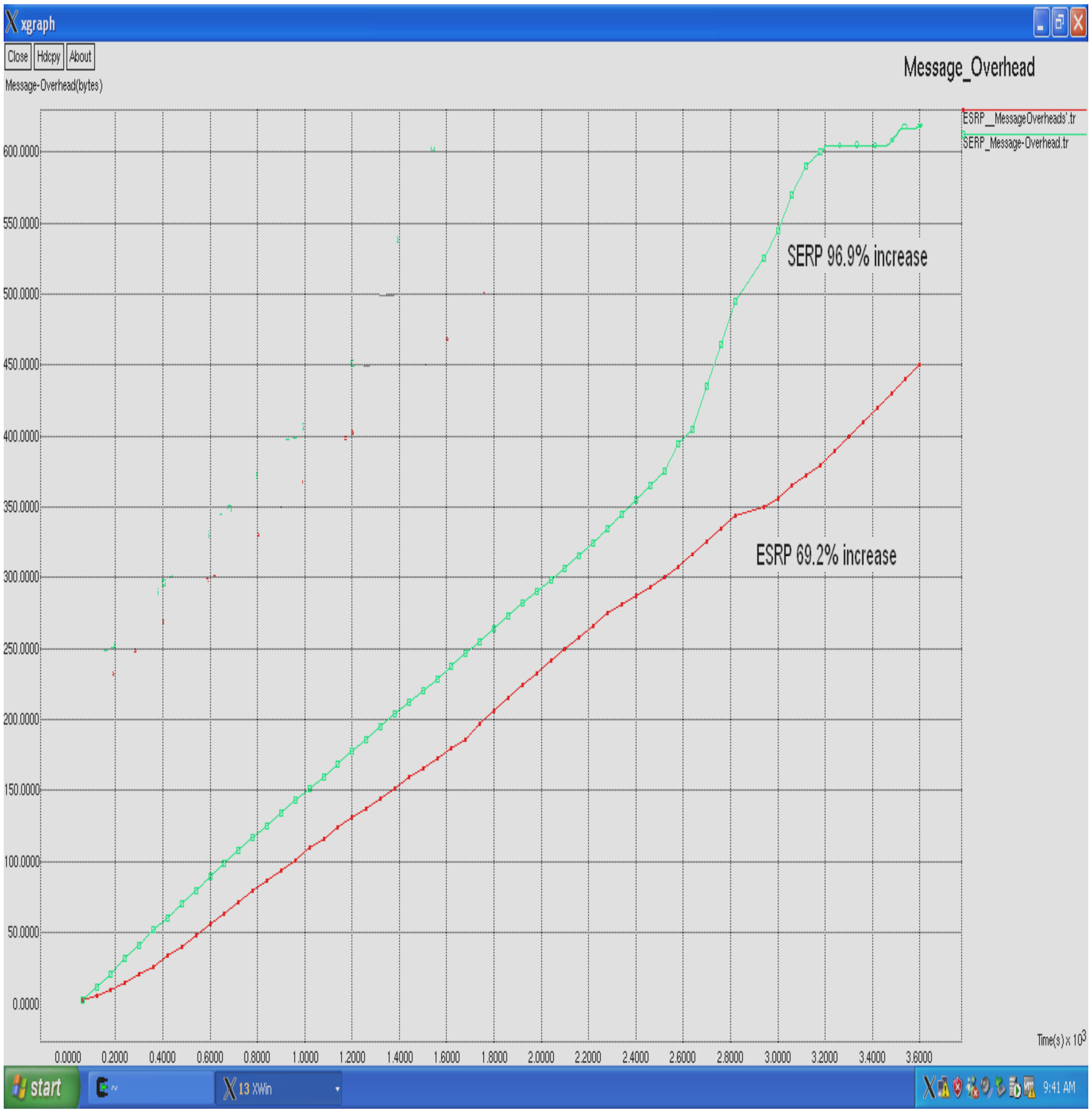

**Figure 3.36 Comparison of ESRP and SERP in terms of 'Message overhead'**

The results were summarized in Table 3.7. Except SLEACH, other protocols exhibits inferior performance in terms of message overhead compared to ESRP.

**Table 3.7 Comparison of ESRP with LDTS, SLEACH, SEDR and SERP**

| Metrics | Number of nodes alive over simulation period | Energy consumption over simulation period (J) | Percentage Energy consumption over simulation period | End to end delay (ms) | Message overhead (bytes) |
|---|---|---|---|---|---|
| **ESRP** | 37 | 72 | 36% | 2800 | 450 |
| **LDTS** | 20 | 192 | 96% | 3200 | 600 |
| **SLEACH** | 30 | 144 | 72% | 2400 | 480 |
| **SEDR** | 26 | 164 | 82% | 3000 | 525 |
| **SERP** | 10 | 184 | 92% | 3400 | 630 |

Table 3.8 shows the Percentage variations realized in various simulation parameters of ESRP in comparison with LDTS, SLEACH,SEDR and SERP.

**Table 3.8 Percentage variations realized in various simulation Parameters of ESRP in comparison with LDTS, SLEACH, SEDR and SERP.**

| Cluster formation method | Percentage decrease in Number of nodes alive | Percentage increase in End to end delay | Percentage increase in Energy consumption | Percentage increase in Message overhead |
|---|---|---|---|---|
| **ESRP** | 63 | 0.07 | 64 | 69.2 |
| **LDTS** | 80 | 0.08 | 96 | 92.3 |
| **SLEACH** | 65 | 0.06 | 72 | 73.84 |
| **SEDR** | 74 | 0.08 | 81.6 | 80.7 |
| **SERP** | 90 | 0.09 | 92 | 96.9 |

# CHAPTER 4

# HARDWARE IMPLEMENTATION

## 4.1 HARDWARE SPECIFICATIONS

The proposed protocols have been implemented in hardware with 28 wireless sensor nodes and 1 sink node which is connected to a personal Computer (PC). The embedded code for Peripheral Interface Controller (PIC) microcontroller have been written in 'Embedded C' and implemented with the help of Microchip MPLAB which is an Integrated Development Environment (IDE).

Texas Instruments CC2520 which is a 2.4 GHz IEEE 802.15.4 Zigbee® RF module have been used as the transceiver in every sensor node as well as for the sink node.

As mentioned in the simulation, the nodes were placed deterministically. Their location and initial energy have been informed to the sink.

## 4.2 HARDWARE RESULTS AND DISCUSSIONS

All 28 nodes were assigned to an initial energy of 2J, as mentioned in the simulation. The 28 nodes were placed in a 100m x 100m area.

As mentioned in the energy calculation model in CHAPTER 3, the communication range of each node was set as 50m.

Figure 4.1 shows the nodes in a grid structure, where each CH has been indicated by enlarged black color points. Figure 4.2 shows them in a hierarchical structure after clusters have been formed.

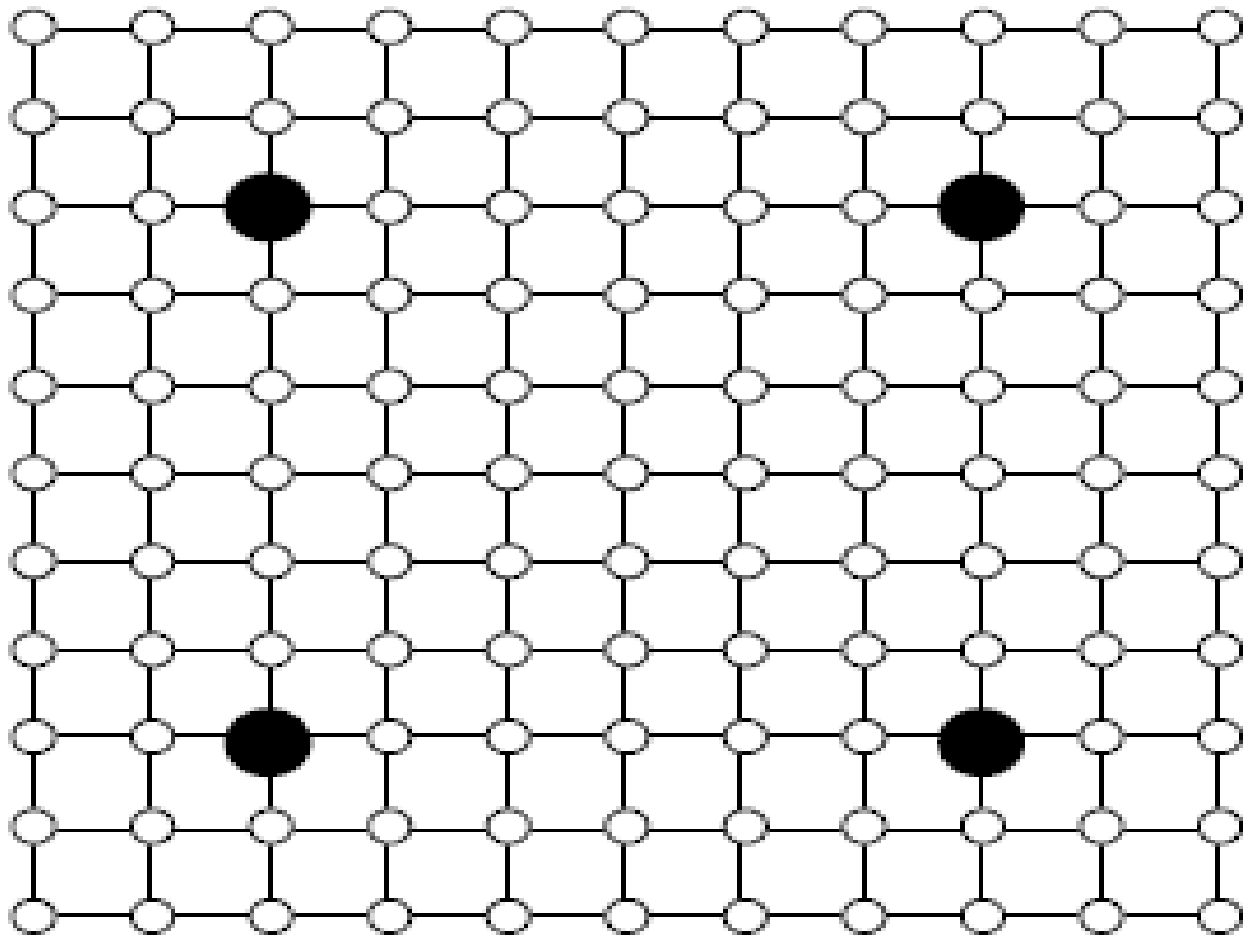

**Figure 4.1 Grid structure of sensor nodes**

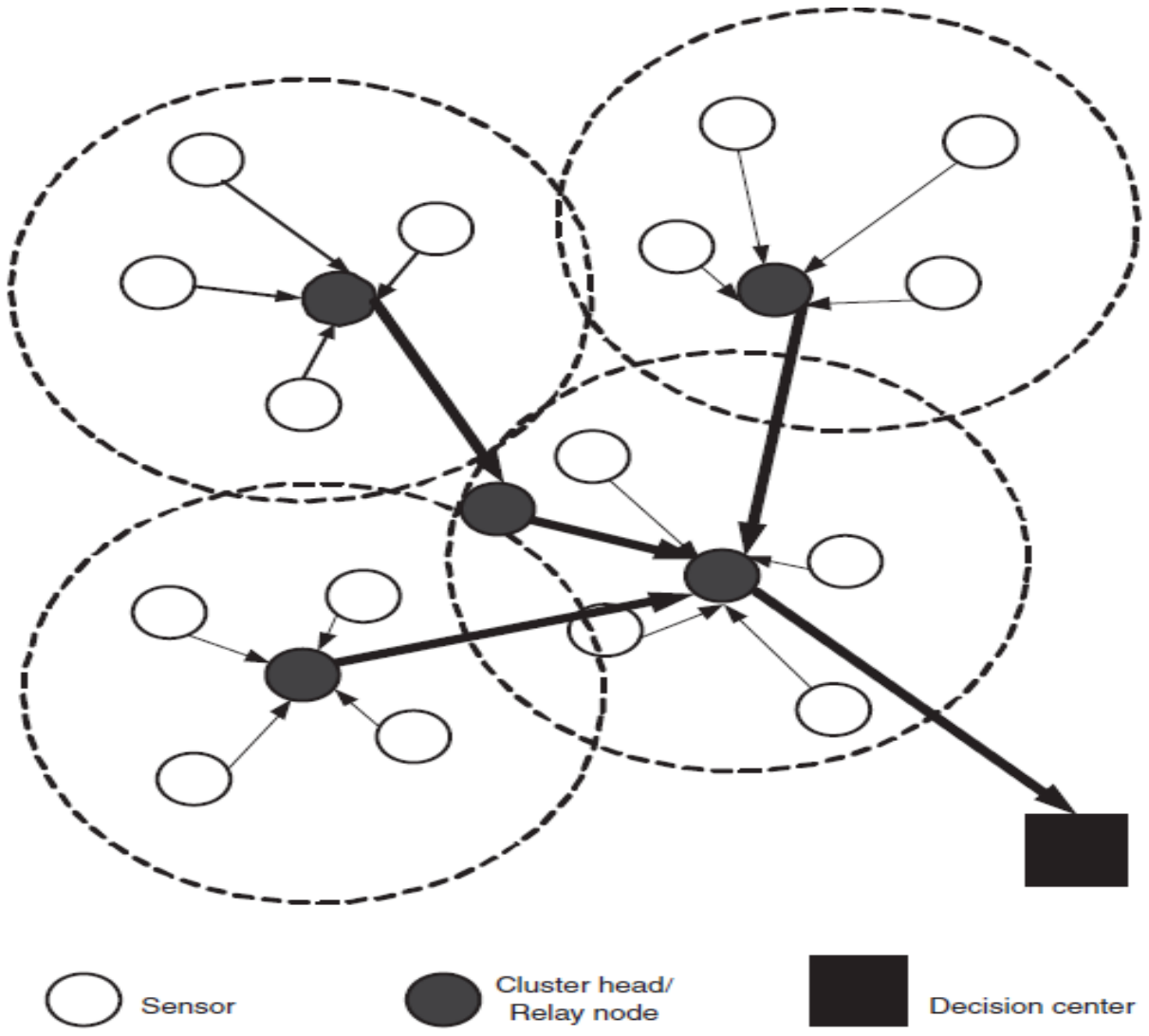

**Figure 4.2 Hierarchical structure of sensor nodes**

Figure 4.3 shows the initial energy values of each node, after been grouped into various clusters. In the figure, nodes 10 to 13 were assigned to upper case alphabets 'A to D' in the left hand side and to the lower case alphabets 'a to d' in the right hand side respectively, where as nodes 14 to 27 were assigned to upper case alphabets 'E to R' on both sides respectively, during their energy representation in milli Joules.

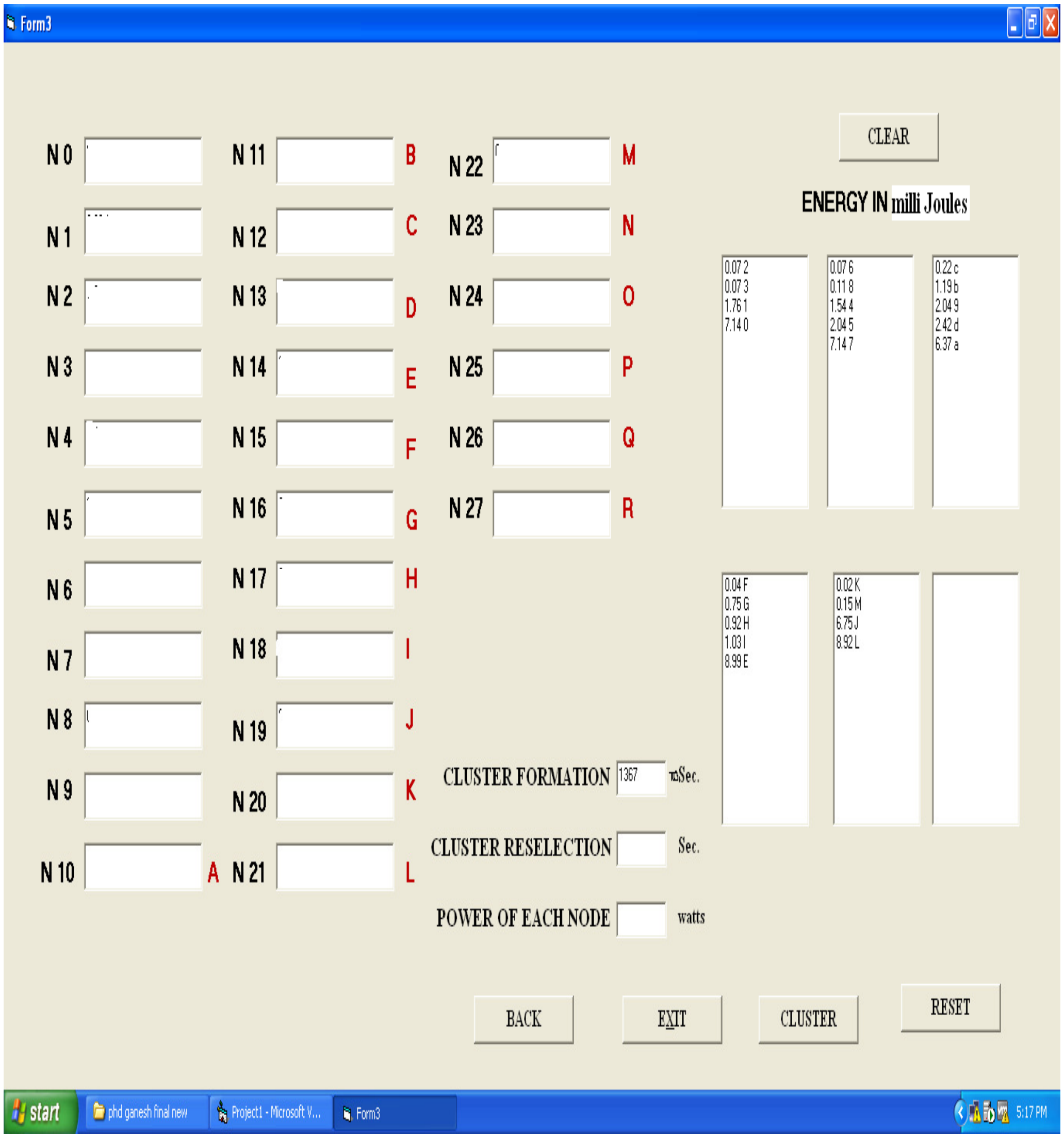

**Figure 4.3 Initial energy values of sensor nodes**

As shown in Table 4.1, the nodes were grouped in five clusters. In order to explore the hybrid routing feature of ESRP, nodes

'N to R' have been programmed to follow flat routing method. The variation in nodes energy is the result of the nature of environment selected for the deployment, where lots of obstacles were present.

The higher energy nodes in each cluster were marked in red color in Table 4.1.

**Table 4.1 Initial energy values of sensor nodes**

| Cluster number | List of nodes | Energy consumption in mJ |
|---|---|---|
| 1 | 2,3,1,**0** | 0.07,0.07,1.76,**7.14** |
| 2 | 6,8,4,5,7 | 0.07,0.11,1.54,2.04,**7.14** |
| 3 | C,b,9,d,**a** | 0.22,1.19,2.04,2.42,**6.37** |
| 4 | F,G,H,I,**E** | 0.04,0.75,0.92,1.03,**8.99** |
| 5 | K,M,J,**L** | 0.02,0.15,6.75,**8.92** |

Figure 4.4 shows the initial cluster formation phase. As shown, the centralized cluster formation of ESRP, took only 136 ms to complete the cluster formation task along with the selection of CH for each cluster.

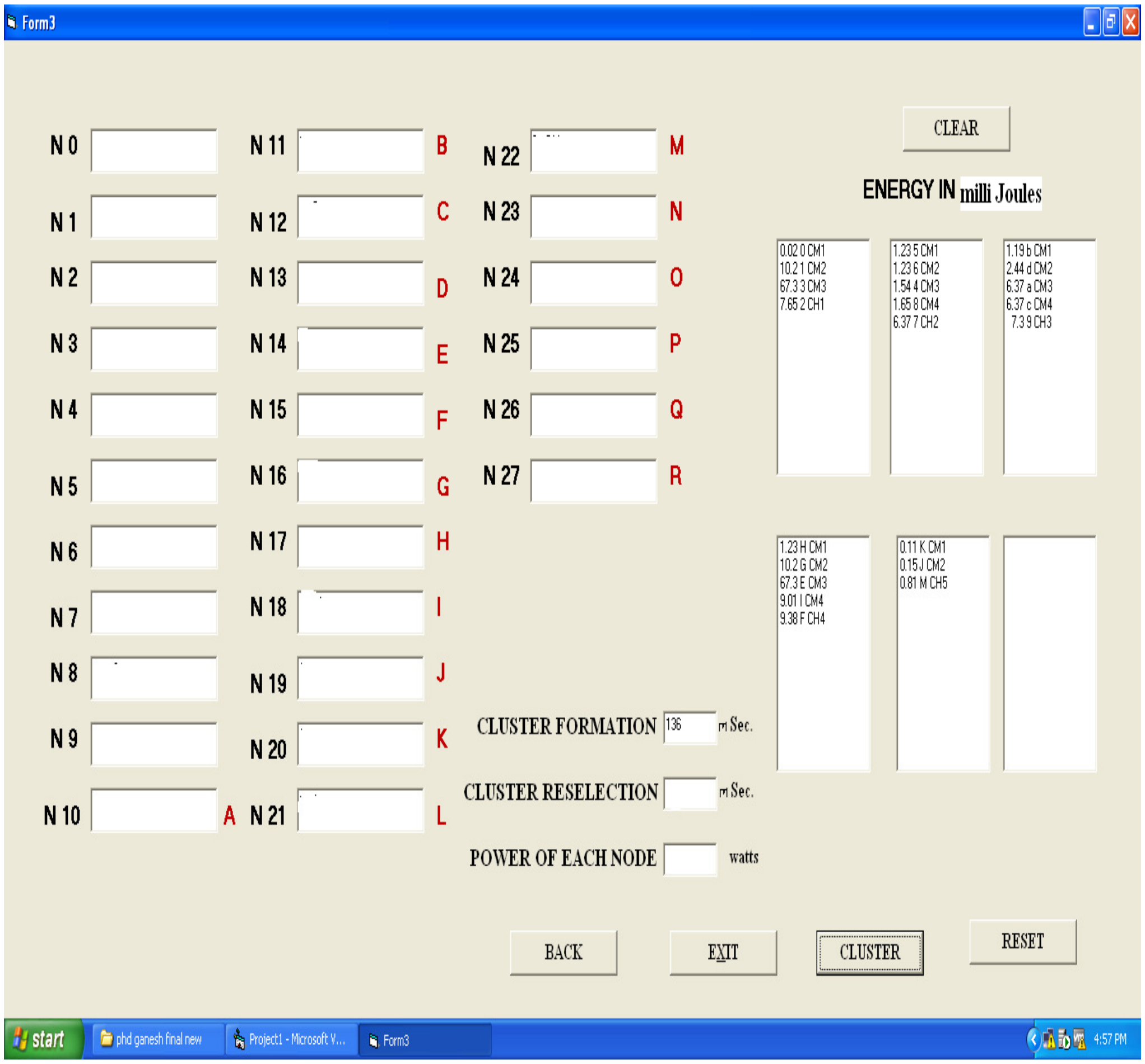

**Figure 4.4 Initial cluster formation phase**

The energy values of each node have been listed in Table 4.2. The node with highest energy, in each cluster has been marked with green color. Except for clusters 1 and 4, the node with highest energy in each cluster has been selected as CH by the ESRP. For clusters 1 and 4 which are marked in red color, lower energy nodes were selected by the ESRP as CH, even though higher energy nodes are available. This can be attributed to the weakness of ESRP algorithm when implemented in a real time, obstacle filled environment.

**Table 4.2 Initial cluster formation phase**

| Cluster number | Selected CH | Energy consumption of CH in mJ | List of CM nodes | Energy consumption in CM nodes in mJ |
|---|---|---|---|---|
| 1 | 2 | 7.65 | 0,1,**3** | 0.02,10.21,**67.33** |
| 2 | **7** | **6.37** | 5,6,4,8 | 1.23,1.23,1.54,1.65 |
| 3 | **9** | **7.30** | b,d,a,c | 1.19,2.44,6.37,6.37 |
| 4 | F | 9.38 | H,G,**E**,I | 1.23,10.2,**67.3**,9.01 |
| 5 | **M** | **0.81** | K,J | 0.11,0.15 |

Also, it can be noted that from Table 4.1 and Table 4.2 that node L, have been removed from the previous list due to its temporary mall function.

Figure 4.5 shows the data gathering from node c. Data have been collected from various member nodes like node 6, node 8, node 1, node I, node J and node M and are tabulated in Table 4.3. A constant overhead of 2 bytes have been assumed during the data gathering operation from all the above mentioned nodes.

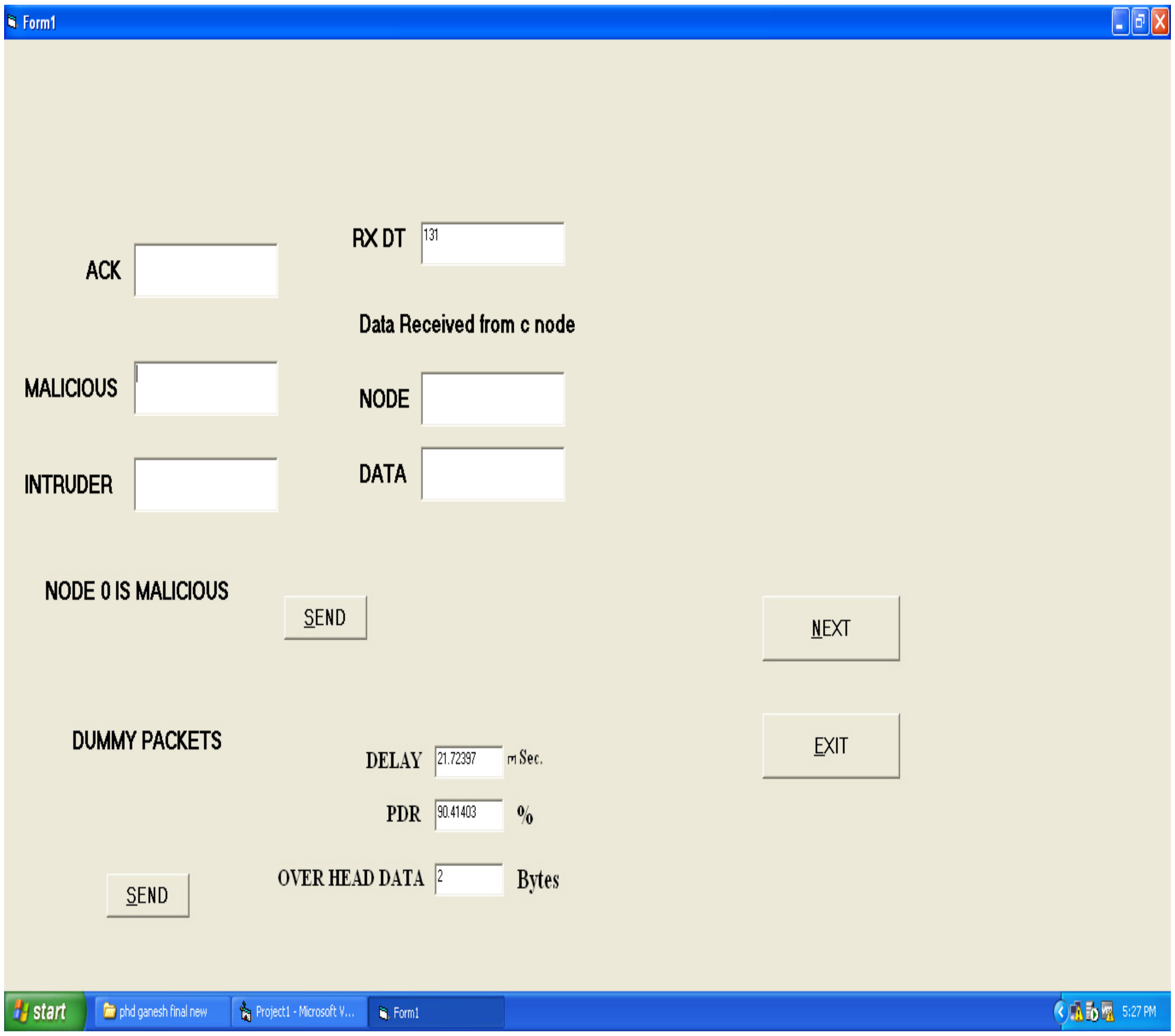

**Figure 4.5 Data gathering from node c**

The Packet Delivery Ratio (PDR) in percentage, which is defined as the ratio of ‘Number of data packets successfully delivered’ to the sink node to those generated by the source node, has been used as a parameter for the identification of abnormal node behavior in the implementation. Every node has been programmed to advertise a 100 % PDR during their abnormal behavior. A hardware switch has been used to change the status of a node from normal to intruder. Mine detection security method which is explained in CHAPTER 3 has been incorporated to identify the malicious nodes.

Table 4.3 shows the corresponding %PDR and the associated delay in ms. Among the listed nodes, node M itself a CH which is shown in red color. Nearly 90% PDR can be achieved in all communication. As stated earlier, different delays have been experienced due to the obstacle filled environment.

**Table 4.3 Data gathering from various nodes**

| Node number | Cluster number | CH | %PDR | Delay in ms |
|---|---|---|---|---|
| C | 3 | 9 | 90.41 | 21.72 |
| 6 | 2 | 7 | 90.70 | 213.39 |
| 8 | 2 | 7 | 90.77 | 86.65 |
| 1 | 1 | 2 | 90.94 | 6.92 |
| I | 4 | F | 90.05 | 0.87 |
| J | 5 | M | 90.99 | 31.69 |
| M | 5 | - | 90.76 | 41.03 |

Nodes 0 and 1 which are the current CM of cluster 1and node 8, which is a current CM of cluster 2, were identified as malicious nodes using the mine detection feature as shown in Figure 4.6, Figure 4.7 and Figure 4.8.

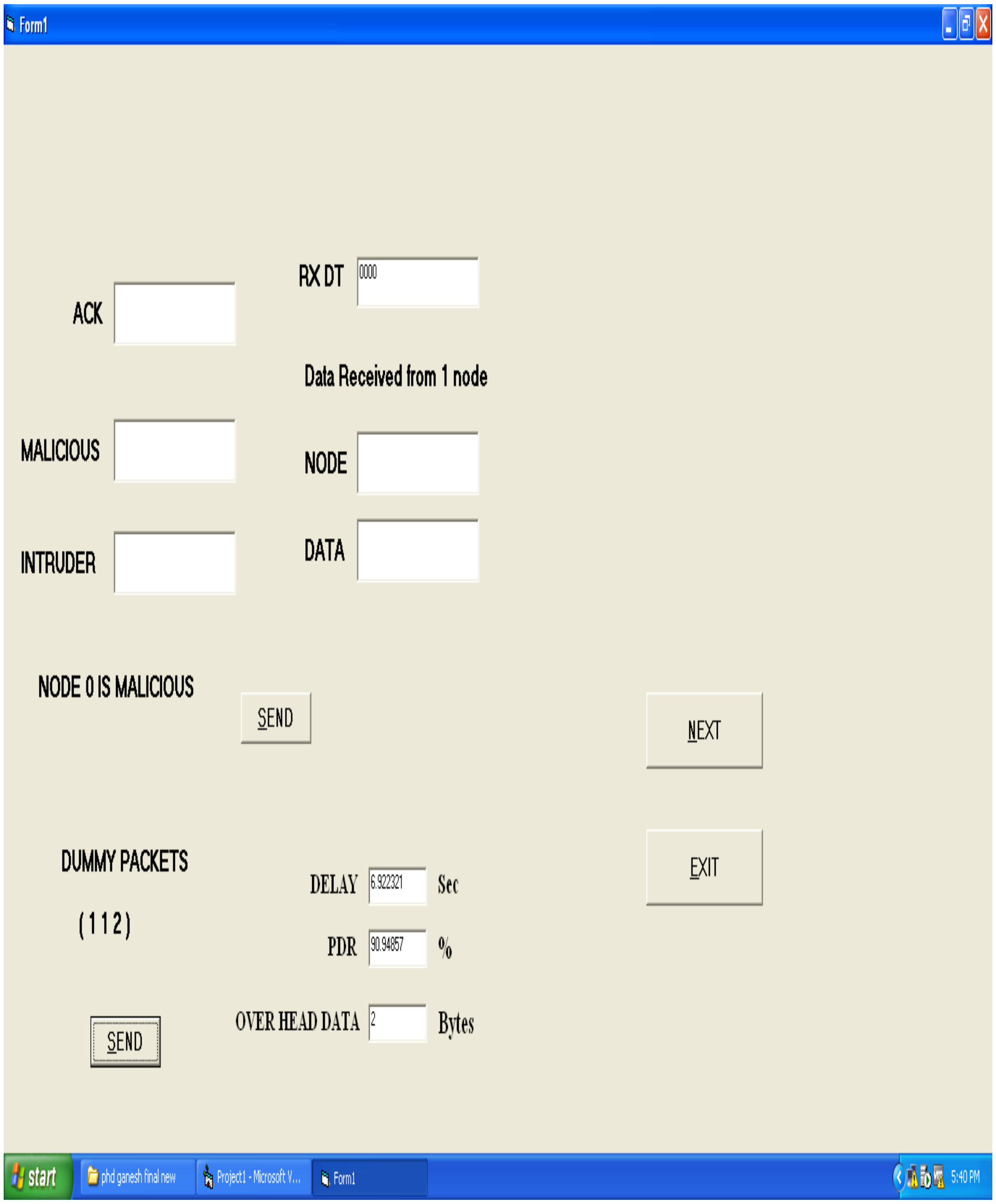

**Figure 4.6 Malicious status identification of node 0**

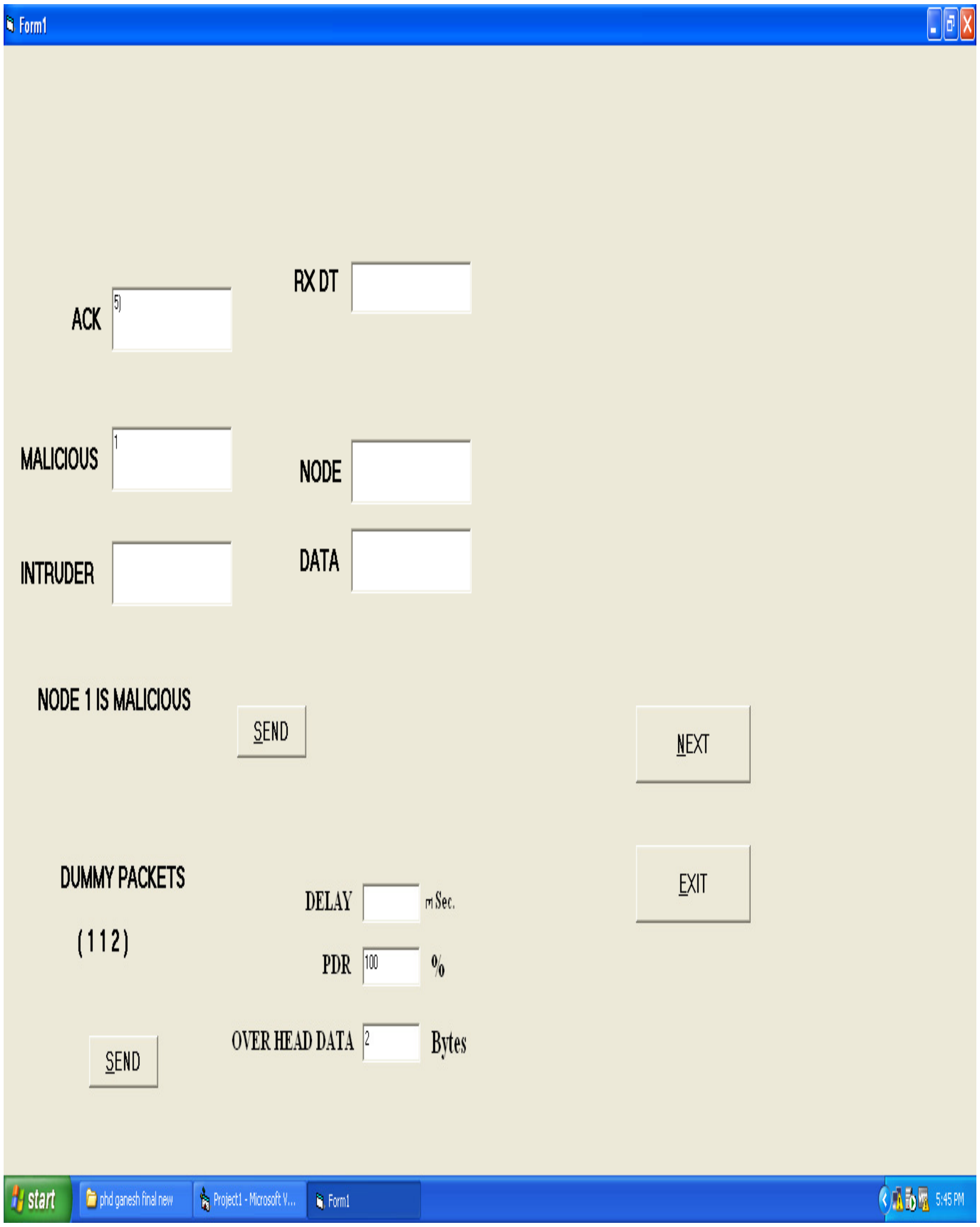

**Figure 4.7 Malicious status identification of node 1**

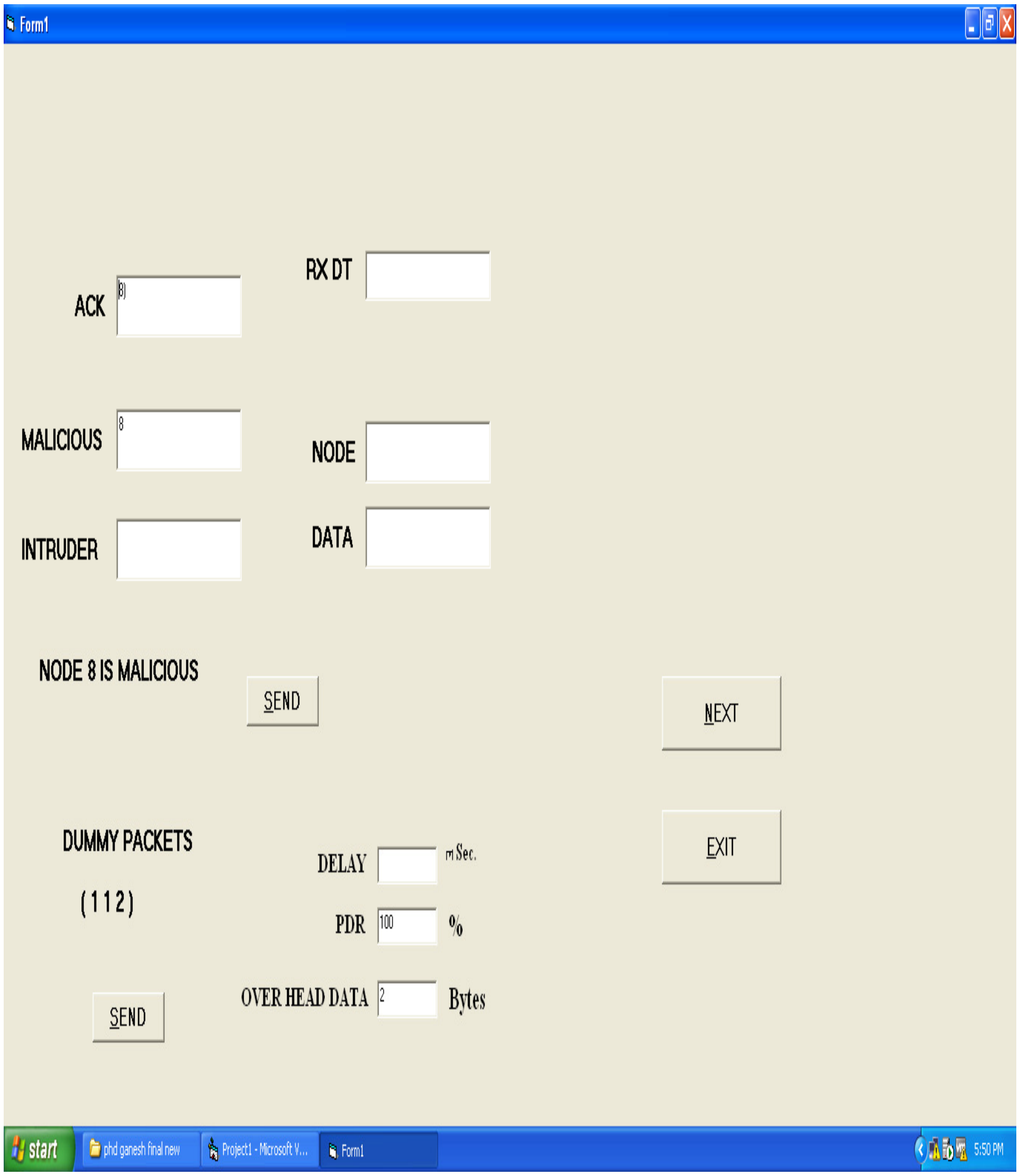

**Figure 4.8 Malicious status identification of node 8**

Node 10, node 12 and node 13 which are specified as node a, node c and node d, respectively as well as the current CM of cluster 3 as shown in Table 4.2, were identified as intruders using the hardware switch as shown in Figure 4.9, Figure 4.10 and Figure 4.11 .

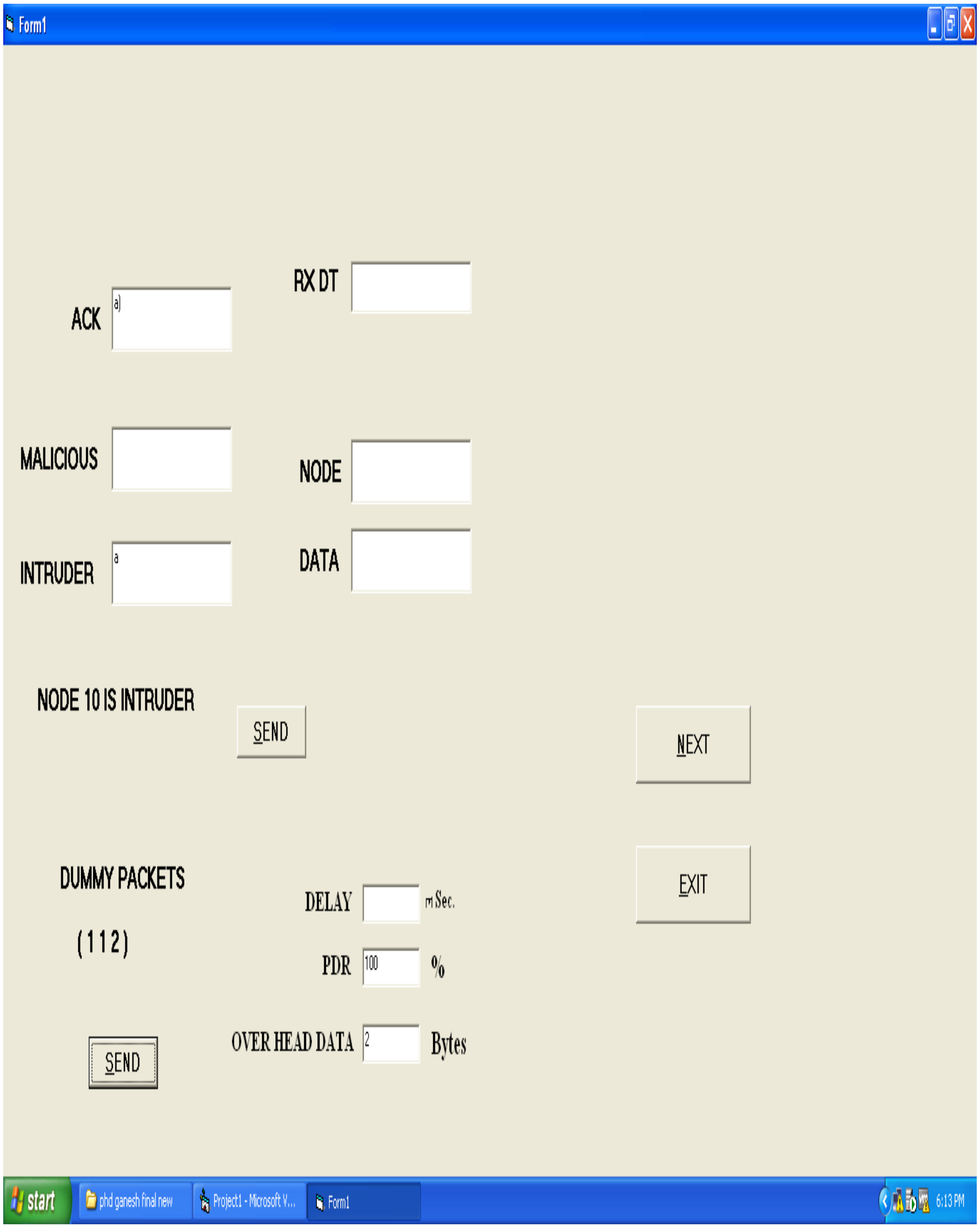

**Figure 4.9 Intruder status identification of node 10**

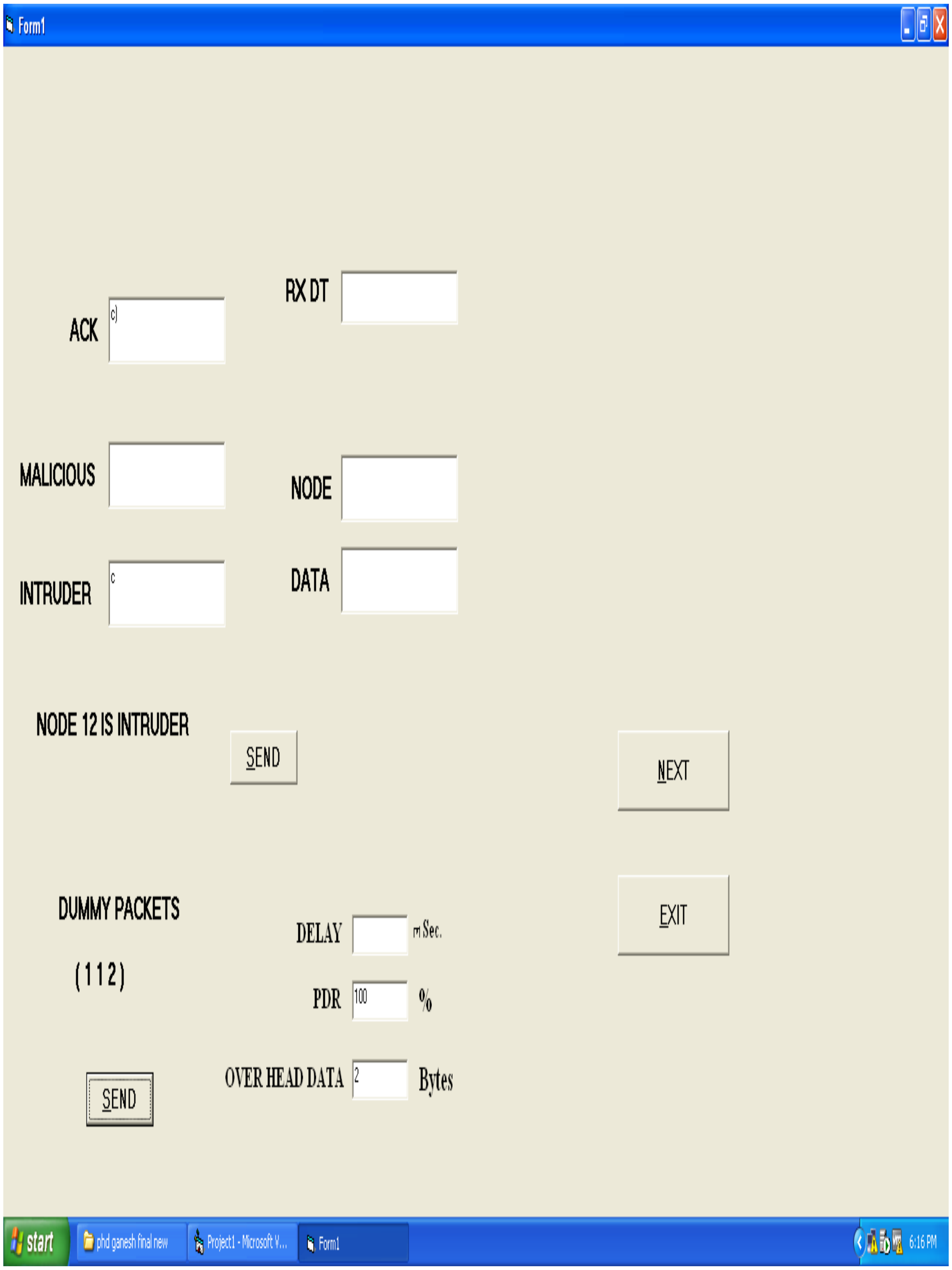

**Figure 4.10 Intruder status identification of node 12**

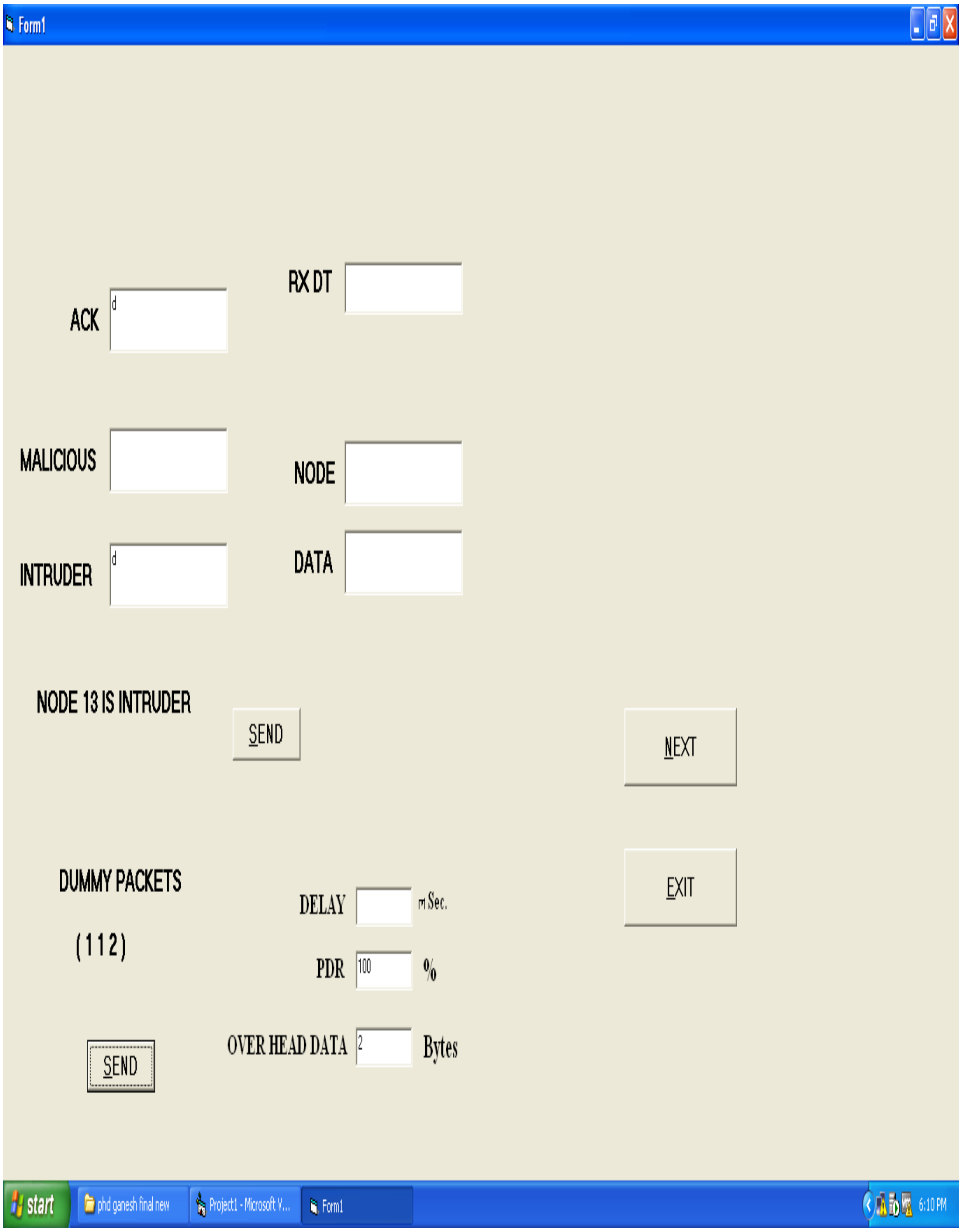

**Figure 4.11 Intruder status identification of node 13**

As shown in Table 4.4, except node 0 which is shown in red color, all other abnormal nodes were advertised a 100% PDR. Hence ESRP algorithm failed to detect node 0, when it was malicious, but could identify the remaining faulty nodes correctly.

**Table 4.4 Abnormal behavior of various nodes**

| Node number | Cluster number | CH | %PDR |
|---|---|---|---|
| 0 | 1 | 2 | 90.94 |
| 1 | 1 | 2 | 100 |
| 8 | 2 | 7 | 100 |
| 10 | 3 | 9 | 100 |
| 12 | 3 | 9 | 100 |
| 13 | 3 | 9 | 100 |

Figure 4.12 shows the cluster reformation phase and Table 4.5 shows the observations.

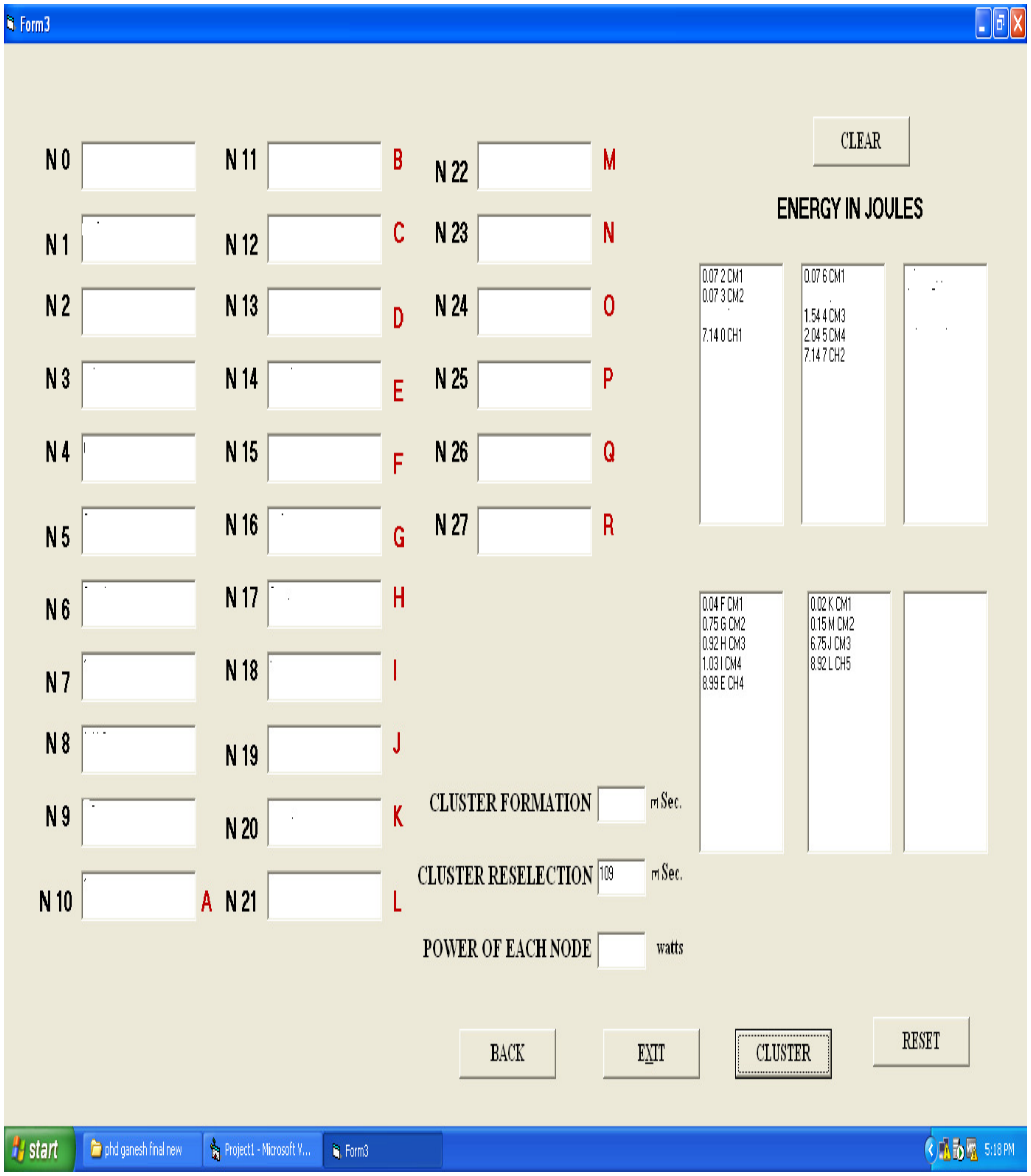

**Figure 4.12 Cluster reformation phase**

As shown in Figure 4.12, ESRP took only 109 ms to reform the cluster, which was lesser than the initial cluster formation time of 136 ms. As stated before, since the malicious status of node 0 was not detected by ESRP, it has been appearing in the reformation phase. In all the clusters, only highest energy nodes, which are marked in green color, were selected as CH Hence the problem faced by ESRP, during the initial cluster formation phase related to the selection of optimal CH has been solved during the cluster reformation phase.

**Table 4.5 Cluster reformation phase**

| Cluster number | Selected CH | Energy consumption of CH in mJ | List of CM nodes | Energy consumption in CM nodes in mJ |
|---|---|---|---|---|
| 1 | 0 | 7.14 | 2,3 | 0.07,0.07 |
| 2 | 7 | 7.14 | 6,4,5 | 0.07,1.54,2.04 |
| 3 | E | 8.99 | F,G,H,I | 0.04,0.75,0.92,1.03 |
| 4 | L | 8.92 | K,M,J | 0.02,0.15,6.75 |

Also, it has been noticed that node 1 and node 8 which are found as malicious nodes and node 10, node 12 and node 13 which are found as intruder nodes, were removed by the ESRP algorithm while reforming the clusters. More over node 9 and node 11 have been switched to a flat routing method as a result of ESRP hybrid routing feature, since no cluster have been formed in their region. Hence the number of clusters has been reduced to four.

Figure 4.13 shows the final stage of the ESRP. Similar to the simulation, the experiment has been carried out for one hour and it has been identified that, at the end of one hour, the WSN is left with only one cluster which includes one CH and four CM as shown in Figure 4.13.

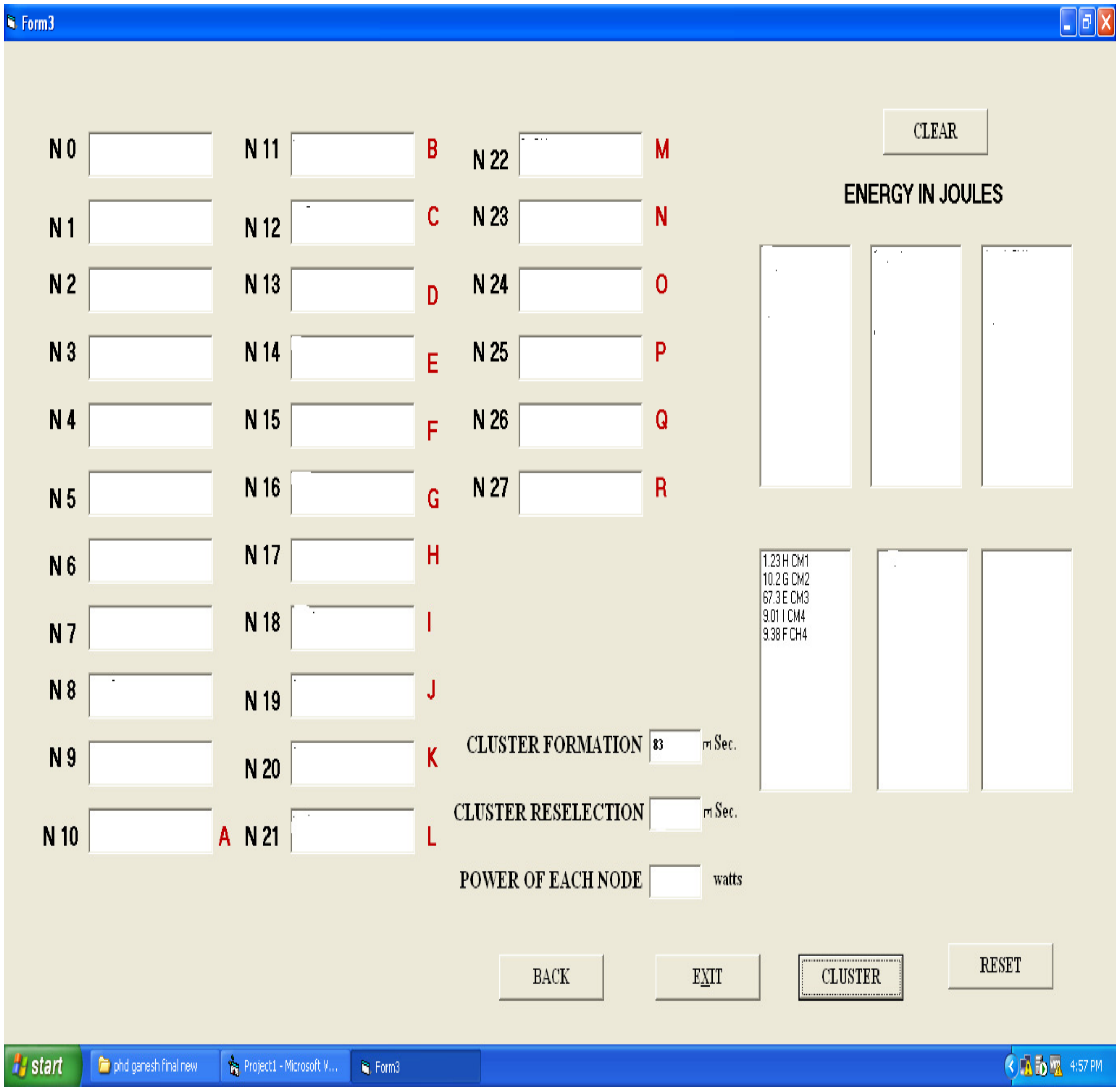

**Figure 4.13 Final stage of ESRP**

The remaining four alive nodes were switched to a flat routing method. Hence out of 28 nodes, the WSN with ESRP had ended up with 9 nodes at the end of the experiment. It can be observed that 32% of nodes were alive at the end of the real time implementation of ESRP as compared to 37% of alive nodes at the end of simulation as specified in CHAPTER 3.

Based on all the above results, it can be stated that except few failures, the proposed ESRP algorithm works well in the real time scenario.

# CHAPTER 5

# CONCLUSION AND FUTURE ENHANCEMENTS

## 5.1 ESRP DRAW BACKS

The drawbacks of the proposed protocol have been mentioned below.

- Mobile nodes have not been considered.
- ESRP implementation limits the scalability of the WSN.
- Idle slots results in to a decreased throughput during lower traffic loads.

## 5.2 CONCLUSION

The deployment nature and limitations of the node resources as well as the wireless communication channel make WSNs susceptible to a variety of new attacks in addition to those attacks which occur in mobile ad hoc networks. A key issue to be solved before a WSN can be conveyed in an extensive variety of uses is the restricted battery limit which obliges the operational time of sensor hubs.

WSN is vulnerable to various attacks such as jamming, battery drainage, routing cycle, Sybil, and cloning. Due to limitation of computation, memory and power resource of sensor nodes, complex security mechanism cannot be implemented in WSN.

Therefore, energy efficient security implementation is an important requirement for WSN.

In this work, an 'Efficient and Secure Routing Protocol' have been created to minimize node level vitality utilization and thereby enhancing the system lifetime.

Because of the vitality mindful directing and element choice about security includes, the ERSP could extensively expand the system lifetime by holding the greatest number of nodes alive, which is 37 out of 100 hubs toward the end of reproduction and thereby empowering the WSN to expend at least 36% of the aggregate vitality.

Other important features such as 'Centralized cluster formation', 'Time driven data reporting', 'Single path routing' and 'Data aggregation using duplicate data suppression', were also played a major role in the superior performance of ESRP.

Based on the simulation results, it has been concluded that the best routing protocol is ESRP, as compared to other similar protocols such as LDTS, SLEACH, SEDR and SERP. Mobility is not an issue in our simulation.

During the real time implementation, Heterogeneous sensor nodes were used where more than one type of node has been integrated in the WSN. While a number of the current regular citizen and military applications of Heterogeneous WSN (H-WSN) do not contrast generously from their homogeneous partners, there are convincing reasons as shown below to incorporate heterogeneity into the network.

The reasons are:

- Improving the scalability of WSNs.
- Addressing the problem of non-uniform energy drainage.
- Taking advantage of the multiple levels of fidelity available in different nodes.
- Diminishing vitality prerequisites without giving up the performance.
- Balancing the cost and functionality of the network.
- Supporting new and higher-bandwidth applications.

## 5.3 FUTURE DIRECTIONS

The future vision of WSN is to embed numerous distributed devices to monitor and interact with physical world phenomena, and to exploit spatially and temporally, dense sensing and actuation capabilities of those sensing devices.

Although extensive efforts have been exerted so far on the routing problem in WSNs, there are still some challenges that confront effective solutions of the routing problem. Despite the fact that the execution of these conventions is promising regarding vitality proficiency, further research would be expected to address issues such as QoS posed by video and imaging sensors and real-time applications.

The ESRP TDMA system obliges some timetable of transmission to be set up either before a framework is sent or constantly. On the occasion that has done logically, the date-book can be set up for the compass of a data trade or for each burst of data. The unused spaces can be dodged by disconnecting the time unequally among the centers sharing the association there by QoS partition can be given.

For example, sensors transmitting low-determination data once in a while are given less time openings than a node that is transmitting high-determination highlight. These structures are limited in cutoff by the measure of time they can administer to each customer.

It can be shown that this TDMA organization is sans crash, however, existing Carrier Sense Multiple Access (CSMA) based systems persevere through basic effects. Further, it can be watched that this service can be used as a piece of a convenient sensor mastermind that gives localization service.

Future examination is to extend this security system to incorporate trust foundation and trust administration in sensor networks. The autonomous mobility management in heterogeneous networks becomes one of future research directions towards seamless mobility. Dynamically repositioning the nodes while the network is operational is necessary to further improve the performance of the network

One fascinating future exploration heading is to study the coordination between the spectrum mobility management and traditional mobility management in Cognitive Radio networks.

Other than this, ESRP can be connected with other steering convention to enhance the vitality effectiveness and can likewise be stretched out with an enhanced interruption discovery framework to enhance the security of WSN.

The ESRP key administration plan can be improved to fathom security issues in mixed media and biometric security, digital security and data affirmation, insurance against wholesale fraud, and criminological figuring. To address these interesting security concerns, it would be basic to study the adjoining innovative advances in disseminating frameworks and omnipresent registering.

Emerging techniques in routing techniques focus on distinctive bearings, all share the regular target of dragging out the system lifetime. Below is the summary of some of these directions.

- **Exploit fault tolerance techniques**: Typically a large number of sensor nodes are implanted inside or beside the phenomenon. Since sensor nodes are prone to failure, fault tolerance techniques come into picture to keep the network operating and performing its tasks. Routing techniques that explicitly employ fault tolerance techniques in an efficient manner are still under investigation.

- **Tiered architectures:** Various leveled directing is an old system to upgrade versatility and effectiveness of the steering convention. On the other hand, novel systems for framework gathering, which extend the framework lifetime, are additionally a hot range of examination in WSN.

- **Exploit spatial diversity and density of sensor nodes**: Nodes will compass a system territory that may be substantial to provide spatial communication between sensor nodes. Accomplishing vitality proficient correspondence in this densely populated environment deserves further investigation. The dense deployment of sensor nodes should allow the network to adapt to unpredictable environment.

- **Time and location synchronization**: Imperativeness profitable systems for accomplice time and spatial bearings with data to support aggregate planning are also required.

- **Self-configuration and Reconfiguration:** This is essential for a lifetime improvement of unattended systems in dynamic and energy obliged environment. This is vital for keeping the system alive. Update and reconfiguration mechanisms should take place when nodes are dying away and leave the network. A feature that is important for every routing protocol is to adapt the topology which changes very quickly and at the same time, to maintain the network functions.

Lastly, one important research direction should receive attention from the researchers, namely the design of routing protocols for Three-Dimensional (3D) sensor fields. Although most of research works on WSN, specifically, on directing, considered Two-Dimensional (2D) settings, where sensors are passed in a planar field, there are a few circumstances where the 2D presumption is not sensible and the utilization of a 3D configuration turns into a need.

Truth be told, 3D settings reflect more exact systematic plan for genuine applications. A valid example, a framework passed on the trees of distinctive statures in a timberland, in a building with different floors, or in submerged, obliges outline in 3D as opposed to in 2D space.

# LIST OF PUBLICATIONS

## International Journals

## National Journals

**International Conferences**

1. Ganesh. S and Amutha. R (2013), “Efficient and Secure Routing protocol for Wireless Sensor Networks”, IEEE-2013 International Conference on Recent Trends in Information Technology (ICRTIT), pp. 631-637.

2. Ganesh. S and Amutha. R (2012), “Efficient and Secure Routing protocol for Wireless Sensor Networks through Selective forwarding”, Eighth International Conference on Wireless Communication and Sensor Networks (WCSN-2012), 6 pages.

3. Ganesh. S and Amutha. R (2012),“Efficient and secure routing protocol for wireless sensor networks using mine detection An extension of triple umpiring system for WSN”, IEEE-ICCM 2012, Eight International Conference on Computing Technology and Information Management (NCM and ICNIT), pp.141 – 145.

**National Conferences**

1. Ganesh. S and Amutha. R (2013)“Efficient and Secure Routing protocol using optimal power management techniques for Wireless Sensor Networks”, NCIEEE2013- 2nd National conference on computational Intelligence in Electrical and Electronics Engineering, pp. 19-26.

2. Ganesh. S and Amutha. R (2012), “Performance of Triple Umpiring System and its Enhancements in Wireless Sensor Networks” First National conference on Machine Intelligence Research and Advancement (NCMIRA, 12), pp. 98-107.

3. Ganesh. S and Amutha. R (2012), “Modified triple umpiring system with optimal power control for Wireless Sensor networks” National Conference on Expanding horizon in Computer, Information Technology, Telecommunication & Electronics (EXCITE 2012), pp.30-34.

# CURRICULUM VITAE

Subramanian Ganesh received the B.E. degree in Electronics and Communications Engineering from Bharadidasan University Trichy, Tamilnadu, India and M.E. degree in Applied Electronics from Anna University, Chennai, Tamilnadu, India, in 1998 and 2008, respectively.

He is working towards his Ph.D. degree in the area of "Efficient and Secure Routing Protocol for Wireless Sensor Networks" as a part-time candidate in Sathyabama University, Chennai, Tamilnadu, India. He is working as an Associate Professor in the Department of Electronics and Communication Engineering at Panimalar Institute of Technology, Chennai.

He has contributed and presented papers at IEEE international conferences in Kerala, Allahabad, China, Korea, and in Thailand. He had published his research works in various reputed international and national journals. The teams led by him have participated in International ROBOSUB competition held at Sandiego, CA,USA, under the sponsorship of Indian Government during the month of July 2012, and in Singapore Autonomous Underwater Vehicle Competition (SAUVC) during the month of March 2013 and March 2015 . He has completed eighteen years in the field of teaching.

## DETAILS OF INDIAN EXAMINER

| | | |
|---|---|---|
| Name of the Candidate | : | S.GANESH |
| Title of the Thesis | : | Efficient and Secure Routing Protocol for Wireless Sensor Network |
| Name of the Indian Examiner | : | Dr. Mrinal Kanti Naskar |
| Designation and Address | : | Professor, Department of ETCE, Jadavpur University, Kolkata-700 032 |

**SATHYABAMA UNIVERSITY**
Jeppiaar Nagar, Chennai – 119.

12/5/16

**EXAMINER'S EVALUATION REPORT FOR PH.D THESIS**

1. Name of the Candidate : S. GANESH
2. Title of the Thesis : Efficient and Secure Routing Protocol for Wireless Sensor Network.
3. Overall Assessment :

| (I) | (II) (a) | (II) (b) | (III) (a) | (III) (b) | (III) ( c ) |
|---|---|---|---|---|---|

(Based on the overall assessment, the examiner shall place the thesis in any one of the three categoies (I), (II) (a), (II) (b), (III) (a), (III) (b) & (III) (c) are given below. (Strike out statements which are not applicable).

(I) I recommend acceptance of the thesis in the present form ~~and further based on the standard attained I classify the work as COMMENDED/HIGHLY COMMENDED.~~ (Strike out which is not applicable)

(II) I recommend acceptance of the thesis subject to the following conditions:
 (a) The candidate shall furnish satisfactory clarification on the queries raised in my detailed report (*) enclosed, during the oral examination.
 (b) The candidate shall Incorporate corrections indicated in my report (*) and place the corrected copy to the Oral Examination Board but the corrected thesis need not be sent me.

(III) I defer my recommendation at this stage and I need the following for further action be taken up enabling me give my recommendation:
 (a) The candidates shall furnish the clarification required on the queries raised in my report (*) and
 (b) The candidate shall incorporate the suggested modifications (*) in the thesis and the corrected thesis along with the candidates clarifications shall be sent to me.
 (c) I reject the thesis for the reasons set out in detail in my report (*).

**DECLARATION**

I declare that I have done the evaluation report for the thesis of the above scholar without the influence of either by supervisor / by the research scholar or by any body else.

Date : 09.05.2016 Signature of the Examiner : [signature]

Place : Jadavpur University Name in BLOCK letters : MRINAL KANTI NASKAR

Address : ETCE Deptt., Jadavpur University, Kol-32.

Note: (*) A detailed report of about 200 or 300 words (or if necessary a longer report) on the thesis, shall be enclosed with the evaluation form, in dictating the standard attained in the case of III (a), the nature and details of the revision of the thesis be made in the case of III (b) and critical points and basis for rejection of the thesis in the case of III (c).

ENCL : Detailed report of the thesis.

[signature]

M. Naskar
Professor
Deptt. of Electronics & Telecommunication Engg.
Jadavpur University, Kolkata-700 032

**Evaluation Report of the Ph. D. thesis entitled "Efficiency and Secure Routing Protocol for Wireless Sensor Network" submitted by Sri S. Ganesh**

The thesis proposed a novel routing protocol named as Efficiency and Secure Routing Protocol (ESRP) for Wireless Sensor Network with the goal to provide an energy efficient routing with dynamic security features.

Extensive literature survey on the topic have been carried out and research gaps in the literature have been identified. The objectives of the thesis are clearly defined and achieved.

The originality of the present work is satisfactory for a Ph. D. dissertation. The work presents a good contribution to research in terms of proposing a new protocol. Various security features are incorporated in Modified Zero-Knowledge Protocol (MZKP). Hardware implementation has also been shown in the work.

Evaluation of the proposed protocol has been carried out through simulation studies and practical experiments. Part of the work have been published in conferences and journals.

The findings of the works are concluded and quantified, contributions are well mentioned and few research problem are highlighted that may be considered as future research directions –

- Exploitation of fault tolerant techniques
- Inclusion of tiered architecture
- Exploitation of spatial diversity and density of sensor nodes
- Time and location synchronization
- Self-configuration and reconfiguration etc.

Overall, the work carried out in the thesis is satisfactory for a Ph. D. Dissertation and this reviewer recommends the acceptance of the thesis for the award of Ph. D. degree.

09.05.16

M. K. Naskar
Professor
Deptt. of Electronics & Telecommunication Engg.
Jadavpur University, Kolkata-700 032

(Signature of the Examiner)

Prof. M. K. Naskar
Professor, Dept. of ETCE,
Jadavpur University, Kolkata – 32
Ph. : 9433276385, email : mrinaletce@gmail.com

# ANSWERS TO THE QUERIES OF INDIAN EXAMINER

The Thesis have been accepted in the present form

## DETAILS OF FOREIGN EXAMINER

| | | |
|---|---|---|
| Name of the Candidate | : | S.GANESH |
| Title of the Thesis | : | Efficient and Secure Routing Protocol for Wireless Sensor Network |
| Name of the Indian Examiner | : | Dr. Abderezak Touzene |
| Designation and Address | : | Associate Professor, Department of Computer Science , Sultan Qaboos University , P.O Box 36, PC 123, Muscat , OMAN. |

# SATHYABAMA UNIVERSITY

Jeppiaar Nagar, Chennai – 119.

22/4/16

(596)

## EXAMINER'S EVALUATION REPORT FOR PH.D THESIS

1. Name of the Candidate : Mr. S. GANESH
2. Title of the Thesis : EFFICIENT AND SECURE ROUTING PROTOCOL FOR WIRELESS SENSOR NETWORK
3. Overall Assessment :

| (I) | (II) (a) | (II) (b) | (III) (a) | (III) (b) | (III) ( c ) |
|---|---|---|---|---|---|

(Based on the overall assessment, the examiner shall place the thesis in any one of the three categories (I), (II) (a), (II) (b), (III) (a), (III) (b) & (III) (c) are given below. (Strike out statements which are not applicable).

(I) I recommend acceptance of the thesis in the present form and further based on the standardattained I classify the work as COMMENDED/~~HIGHLY COMMENDED~~. (Strike out which is not applicalle).

(II) I recommend acceptance of the thesis subject to the following conditions:
- (a) The candidate shall furnish satisfactory clarification on the queries raised in my detailed report (*) enclosed, during the oral examination.
- (b) The candidate shall incorporate corrections indicated in my report (*) and plac the corrected copy to the Oral Examination Board but the corrected thesis need not besent me.

(III) I defer my recommendation at this stage and I need the following for further action be taken up enabling me give my recommendation:
- (a) The candidates shall furnish the clarification required on the queries raised in my report(*) and
- (b) The candidate shall incorporate the suggested modifications (*) in the thesis and he corrected thesis along with the candidates clarifications shall be sent to me.
- (c) I reject the thesis for the reasons set out in detail in my report (*).

## DECLARATION

I declare that I have done the evaluation report for the thesis of the above scholar without the influence of either by supervisor / by the research scholar or by any body else.

Date : 21 – 4 – 2016

Place : MUSCAT OMAN

Signature of the Examiner :

Name in BLOCK letters : DR. ABDEREZAK TOUZENE

Address : COMPUTER SCIENCE DEP. P.O. BOX 36, PC 123, OMAN

Note: (*) A detailed report of about 200 or 300 words (or if necessary a longer report) on the thesis, shall be enclosed with the evaluation form, in dictating the standard attained in the case of III (a), the nature and details of the revision of the thesis be made in the case of III (b) and critical points and basis for rejection of the thesis in the case of III (c).

**Assessment of Ph.D. Thesis SATHYABAMA UNIVERSITY**
**(Candidate Name: S. GANESH)**

**Title of Thesis:**
**"EFFICIENT AND SECURE ROUTING PROTOCOL FOR WIRELESS SENSOR NETWORK".**

**Organization of Thesis**: The objectives of the thesis work have been clearly stated. The overall organization of the report and the flow of the different parts are good.

**Clarity of Language**: The writing style, clarity and quality of the language are very good.

**Methodology**: The Ph.D. student started by reviewing the literature about routing protocols in WSN with a focus on protocols efficiency from one hand and security issues on another hand. He successfully identified the weakness of the existing protocols and he proposed a new protocol ESRP to address the shortcomings. After designing the new protocols an extensive simulation experiments have been conducted to compare the performance of ESRP with the existing ones. Most importantly, real implementation of the protocol has been implemented, tested, and performance results have been obtained. The research methodology presented is good and sound.

**Results & Interpretations**: The student presented a number of plots extracted from the different NS2 simulation runs for the ESRP proposed protocol. He discussed these results and presented elucidations of trends observed on the curves. Finally, he successfully proved the superiority of ESRP protocol compared to the existing ones.

**References**: References have been cited methodically and in the right format.

**In conclusion**: I recommend acceptance of the thesis in the present form and I classify the work as COMMENDED.

Date 21-4-2016

Dr. Abderezak Touzene
Associate Professor, Department of Computer Science.
Sultan Qaboos University
P.O Box 36, PC 123, Muscat, OMAN

## ANSWERS TO THE QUERIES OF FOREIGN EXAMINER

The Thesis have been accepted in the present form and further based on the standard attained, the work have been classified as COMMENDED.